\documentclass[11pt]{article}

\usepackage[usenames,dvipsnames]{pstricks}
\usepackage{epsfig}
\usepackage{pst-grad} 
\usepackage[space]{grffile} 
\usepackage{etoolbox} 
\makeatletter 
\patchcmd\Gread@eps{\@inputcheck#1 }{\@inputcheck"#1"\relax}{}{}
\makeatother

\usepackage{graphicx}
\usepackage{amsmath}
\usepackage{psfrag}
\usepackage{color,soul}
\usepackage{natbib}
\usepackage{siunitx}
\usepackage{float}
\usepackage{overpic}
\usepackage{tikz}
\usepackage{multirow}
\usepackage{bm}
\usepackage{textgreek}

\usepackage{amssymb}
\usepackage{wasysym}
\usetikzlibrary{arrows}

\usepackage[noblocks]{authblk}
\usepackage[margin=1in]{geometry}
\usepackage{float}
\usepackage{enumerate}
\usepackage{caption}
\usepackage[title]{appendix}

  \DeclareTextFontCommand\textsfi{\usefont{OT1}{cmss}{m}{sl}}
  \DeclareMathAlphabet\mathsfi            {OT1}{cmss}{m}{sl}
  \DeclareTextFontCommand\textsfb{\usefont{OT1}{cmss}{bx}{n}}
  \DeclareMathAlphabet\mathsfb            {OT1}{cmss}{bx}{n}
  \DeclareTextFontCommand\textsfbi{\usefont{OT1}{cmss}{m}{sl}}
  \DeclareMathAlphabet\mathsfbi            {OT1}{cmss}{m}{sl}
  \IfFileExists{t1phv.fd}
  {\DeclareTextFontCommand\textsfbi{\usefont{T1}{phv}{b}{it}}
  \DeclareMathAlphabet\mathsfbi            {T1}{phv}{b}{it}
  }{}
  \IfFileExists{ot1phv.fd}
  {\DeclareTextFontCommand\textsfbi{\usefont{OT1}{phv}{b}{it}}
  \DeclareMathAlphabet\mathsfbi            {OT1}{phv}{b}{it}
  }{}

\newcommand\affiliation[1]{\gdef\@affiliation{\let\aff\aff@inst#1}}
\gdef\@affiliation{}

\def\email#1{Email address for correspondence: #1}
\def\aff#1{\ignorespaces\textsuperscript{#1}}
\def\corresp#1{\unskip\thanks{#1}}

\numberwithin{equation}{section}

\renewenvironment{abstract}
{\begin{quote}
\noindent \rule{\linewidth}{.5pt}\par{\bfseries \abstractname.}}
{\medskip\noindent \rule{\linewidth}{.5pt}
\end{quote}
}

\newcommand{\tc}[1]{\mathsfbi{#1}}	
\newcommand{\td}[1]{ {\bf #1} }			
\newcommand{\vc}[1]{\boldsymbol{#1}}
\newcommand{\vd}[1]{ {\bf #1} }			
\newcommand{\ts}[1]{\mathsfi{#1}}	
\newcommand{\ii}{\mathrm{i}}

\newcommand{\qvec}{\chi}
\newcommand{\qvecd}{{\text{\bf \textchi}}}

\newcommand{\rectanglez}[2]{\tikz \draw[very thick,black] (0,0) rectangle (#1,#2);}
\newcommand{\rectanglezRed}[2]{\tikz \draw[thin,red] (0,0) rectangle (#1,#2);}
\newcommand{\linez}[2]{\tikz \draw[thin,red,dashed] (0,0) -- (#1,#2);}
\newcommand{\axez}[2]{\tikz \draw [thick,<->,>=stealth] (0,#2) -- (0,0) -- (#1,0);}

\newcommand{\cmmnt}[1]{\ignorespaces}

\definecolor{blue}{rgb}{0,0.502,1}
\definecolor{darkblue}{rgb}{0,0,0.80}
\definecolor{red}{rgb}{1,0,0}
\definecolor{green}{rgb}{0,0.8824,0}
\definecolor{gray}{rgb}{0.627,0.627,0.627}
\definecolor{black}{rgb}{0,0,0}

\newcommand\solidrule[1][1cm]{\rule[0.5ex]{#1}{.4pt}}

\newcommand{\myRuleSolid}[3][black]{\textcolor{#1}{\rule[0.5ex]{#2}{#3}}}
\newcommand{\myRuleDashed}[3][black]{\textcolor{#1}{\mbox{%
  \solidrule[2mm]\hspace{2mm}\solidrule[2mm]\hspace{2mm}\solidrule[2mm]}}}
\newcommand{\myRuleDashDot}[1]{\textcolor{black}{\mbox{%
  \solidrule[2mm]\hspace{1mm}\solidrule[0.2mm]\hspace{1mm}\solidrule[2mm]\hspace{1mm}\solidrule[0.2mm]\hspace{1mm}\solidrule[2mm]}}}

\usepackage{hyperref}
\hypersetup{
    colorlinks=true,
    linkcolor={darkblue},
    citecolor={darkblue},
    filecolor=black,
    urlcolor={darkblue},
    plainpages=false,
}

\title{\bf Spectral proper orthogonal decomposition \\ and its relationship to dynamic mode \\ decomposition and resolvent analysis}

\author[1]{\bf Aaron Towne\corresp{\email{atowne@stanford.edu}}}
\author[2]{\bf Oliver T. Schmidt}
\author[2]{\bf Tim Colonius}

\affil[1]{\normalsize Center for Turbulence Research, Stanford University, Stanford, CA 94305, USA  }
\affil[2]{\normalsize California Institute of Technology, Pasadena, CA 91125, USA \vspace{-1cm}}

\date{}

\begin{document}
\maketitle

\begin{abstract}

We consider the frequency domain form of proper orthogonal decomposition (POD) called spectral proper orthogonal decomposition (SPOD).  Spectral POD is derived from a space-time POD problem for statistically stationary flows and leads to modes that each oscillate at a single frequency.  This form of POD goes back to the original work of Lumley (\textit{Stochastic tools in turbulence}, Academic Press, 1970), but has been overshadowed by a space-only form of POD since the 1990s.  We clarify the relationship between these two forms of POD and show that SPOD modes represent structures that evolve coherently in space and time while space-only POD modes in general do not.  We also establish a relationship between SPOD and dynamic mode decomposition (DMD); we show that SPOD modes are in fact optimally averaged DMD modes obtained from an ensemble DMD problem for stationary flows.  Accordingly, SPOD modes represent structures that are dynamic in the same sense as DMD modes but also optimally account for the statistical variability of turbulent flows.  Finally, we establish a connection between SPOD and resolvent analysis.  The key observation is that the resolvent-mode expansion coefficients must be regarded as statistical quantities to ensure convergent approximations of the flow statistics.  When the expansion coefficients are uncorrelated, we show that SPOD and resolvent modes are identical. Our theoretical results and the overall utility of SPOD are demonstrated using two example problems: the complex Ginzburg-Landau equation and a turbulent jet.  

\end{abstract}


\section{Introduction}
\label{sec:intro}

Coherent flow-features, or \emph{structures}, play an important role in turbulent flows.  This has lead to efforts to extract these structures from data as well as to model them using simplified equations.  Typically, this involves defining a set of modes that compactly describe the structures.  Some of the most widely used techniques are summarized in recent reviews by \cite{Rowley:2017} and \cite{Taira:2017}.  

Fifty years after its introduction by \cite{Lumley:1967,Lumley:1970}, proper orthogonal decomposition (POD) remains one of the most widely used techniques for educing coherent structures from flow data.  The method is known in other disciplines by a variety of names including principal component analysis and Karhunen-Lo{\`e}ve decomposition.  Proper orthogonal decomposition seeks a set of deterministic modes that optimally capture the energy, or variance, of an ensemble of stochastic flow data.  A rigorous statement of this objective involves defining the ensemble of interest in terms of flow data and choosing an associated inner product and expectation operator; these choices determine the properties of the modes that are obtained.

\begin{sloppypar}
This paper focuses on a specific form of POD called \emph{spectral proper orthogonal decomposition} (SPOD).  To be clear, we are \emph{not} referring to the method recently proposed by \cite{Sieber:2016} that was given the same name.  Instead, we are using the terminology introduced by \cite{Picard:2000} to describe a space-time formulation of POD for statistically stationary flows that goes back to \cite{Lumley:1967,Lumley:1970}.  This terminology is motivated by the fact that the method involves decomposition of the cross-spectral density tensor and leads to modes that each oscillate at a single frequency.  Spectral POD has been applied to a variety of flows including pipes \citep{Hellstrom:2014}, boundary layers \citep{Tutkun:2017}, mixing layers \citep{Delville:1999, Braud:2004}, jets \citep{Glauser:1987b, Arndt:1997, Gordeyev:2000, Citriniti:2000, Gudmundsson:2011,Schmidt:2017a}, wakes  \citep{Tutkun:2008, Araya:2017}, and the flow around an airfoil \citep{Abreu:2017}.   
\end{sloppypar}

Since its introduction and popularization by \cite{Sirovich:1987a}, \cite{Aubry:1988}, and \cite{Aubry:1991b}, a different form of POD has come to dominate the literature.  This form of POD decomposes the \emph{spatial} correlation tensor and leads to spatially orthogonal modes that are modulated in time by expansion coefficients with \emph{random} time dependence.  In what follows, we refer to this form of POD as \emph{space-only POD}.  

Several factors appear to have contributed to the dominance of space-only POD.  First, SPOD requires time-resolved data that were not available using particle image velocimetry until recently.  On the other hand, it is laborious to obtain the cross spectral densities for a large number of spatial points using hot wires, and large arrays of hot wires \citep{Citriniti:2000} become intrusive.  In terms of simulation data, SPOD requires relatively long time integrations that would have been prohibitive using computational resources available at that time.  Finally, space-only POD provides an appealing basis for Galerkin projection of the Navier-Stokes equations, which became popular with the rise of the dynamical systems perspective on turbulence \citep[e.g.,][]{Aubry:1988, Holmes:1997, Noack:2003}.

Perhaps due to the dominance of the space-only variant, there is a lack of clarity in the literature regarding the relationship between space-only and spectral POD.  A survey of some of the most cited and/or recent review articles that address POD reveals a wide-range of perspectives.  Some articles describe POD in purely abstract terms and make no distinction between the two versions of the method \citep{Berkooz:1993,Liang:2002}.  Many papers exclusively present space-only POD \citep{Sirovich:1989, Moin:1989, Holmes:1997, Chatterjee:2000, Rowley:2004, Cordier:2008, Rowley:2017}, while one early review \citep{George:1988} considers only spectral POD.  There are some articles that treat the two variants as separate methods, but the relationship between them is not explored in detail \citep{Aubry:1988, Picard:2000, Chen:2005, Tinney:2008, Holmes:2012, Taira:2017}.  Finally, one recent review describes space-only POD as an approximation of spectral POD \citep{George:2017}.  

Compounding this lack of clarity is the inconsistent and incompatible use of the terms `classical POD' and `snapshot POD'.  Some authors use these terms to refer to the methods we are calling spectral and space-only POD, respectively \citep[e.g.,][]{Hellstrom:2014, Mula:2014, George:2017}.  Others use the name `classical POD' to refer to space-only POD and `snapshot POD' to refer to a particular computational shortcut for computing space-only POD modes \citep[e.g.,][]{Hilberg:1994, Pinier:2007, Cordier:2008}.

The first part of this paper seeks to clarify the relationship between the space-only and spectral formulations of POD.  We will show that they are fundamentally different from one-another -- whereas it is often stated that both methods identify coherent structures, only SPOD modes evolve coherently in space and time.  This suggests that SPOD is better suited for identifying physically meaningful coherent structures in stationary flows.  We also derive formulas relating space-only and spectral POD eigenvalues and eigenvectors.  

The second part of the paper establishes a connection between SPOD and dynamic mode decomposition (DMD).  Dynamic mode decomposition was developed by \cite{Schmid:2010} as an alternative to POD for identifying coherent structures from flow data with the specific aim of obtaining modes that describe the flow dynamics, i.e., the evolution of the flow from one time instant to the next.  This objective was motivated in part by criticism of space-only POD, specifically that the averaging process used to obtain the spatial correlation tensor causes important dynamical information about the flow to be lost.  Our analysis affirms that this criticism is well founded for space-only POD but shows that it does not apply to SPOD.  Moreover, we will show that SPOD modes are in fact optimally averaged DMD modes obtained from an ensemble DMD problem for stationary flows.

Several other methods have been proposed in recent years to try to bridge the gap between the spatial orthogonlization of space-only POD and the temporal orthogonalization of DMD.  \cite{Cammilleri:2013} proposed Cronos-Koopman analysis by treating the projection coefficients from space-ony POD as observables in a Koopman analysis (approximated by DMD).  The aforementioned method of \cite{Sieber:2016} filters the temporal correlation tensor over a time horizon leading to an ad hoc interpolation between space-only POD, which is obtained when the filter width is zero, and the discrete Fourier transform, which is recovered when the filter width is the entire interval of the data.  \cite{Noack:2016} developed a method called recursive dynamic mode decomposition (RDMD) which combines features of space-only POD and DMD. The first RDMD mode is given by the DMD mode that minimizes the time-averaged residual between the modal expansion and the data. Subsequent modes achieve the same objective under the constraint of orthogonality with previous modes. This leads to modes that each oscillate at a single frequency and are spatially orthogonal to all other modes at all frequencies.  The unique properties of each of these methods make them useful for different purposes.  While a detailed comparison is beyond the scope of this paper, our analysis shows that SPOD is optimal by construction for the task of identifying flow structures that evolve coherently in both space and time.  

The third part of the paper shows that SPOD is closely related to resolvent analysis.  Resolvent analysis (also called input/output analysis and frequency response analysis) has its roots in linear systems and control theory.  The resolvent operator is derived from linearized flow equations and constitutes a transfer function between inputs and outputs of interest. It has been used to study the linear response of flows to external body forces and perturbations \citep{Trefethen:1993,Farrell:2001,Schmid:2001,Jovanovic:2005,Bagheri:2009,Sipp:2010} and to forcing from the nonlinear terms in the Navier-Stokes equations \citep{Mckeon:2010, Sharma:2013}.  In the latter context, the method can be derived by partitioning of the Navier-Stokes equations into terms that are linear and nonlinear with respect to perturbations to the turbulent mean flow.  Resolvent analysis then identifies frequency-dependent modes that are optimal in terms of their linear gain between the nonlinear terms and the output.  The idea is to then use a small set of the highest-gain modes as a basis for the output.

The connection we draw between SPOD and resolvent analysis is based on a new statistical interpretation of the resolvent-mode reconstruction of turbulent flows. Due to the sensitivity of the Navier-Stokes equations to small perturbations, each realization of a turbulent flow, e.g., a different run of the same experiment, produces a unique time history that cannot be reliably predicted by knowledge of the time history of a different realization.  In other words, turbulent flows are \emph{random} in the sense defined by \cite{Landahl:1992} and \cite{Pope:2000}, among others.  Accordingly, a statistical description that accounts for many such trajectories provides more information about the likely properties of any specific realization.  This leads to a statistical interpretation of the \emph{expansion coefficients} that are used in the resolvent-mode reconstruction of the flow, which is a departure from past studies that have described them as deterministic quantities described entirely by their amplitude and phase. 

We will show that SPOD and resolvent modes are identical when these expansion coefficients are uncorrelated, which is typically associated with white-noise forcing.  This can be viewed as a statistical counterpart to the relationship between DMD and resolvent analysis recently shown by \cite{Sharma:2016}.  More generally, we will demonstrate the importance of properly accounting for the cross-correlations between the expansion coefficients. We will show that if the expansion coefficients are not treated as statistical quantities, the optimal reconstruction of the flow is always governed by the leading SPOD mode at each frequency; thus the quality of the approximation is dependent first-and-foremost on the low-rank nature of the cross-spectral density tensor rather than the resolvent operator. This limitation can be overcome by using SPOD modes to estimate the statistics of the expansion coefficients. 

The remainder of the paper is organized as follows.  \S\ref{Sec:POD} describes and compares the space-only and spectral formulations of POD.  \S\ref{Sec:SPOD_comp} outlines a procedure for estimating SPOD modes using time-resolved flow data. The relationship between SPOD and DMD is explored in \S\ref{Sec:Compare_DMD}.  \S\ref{Sec:Resolvent} establishes a connection between SPOD and resolvent analysis and shows the key role of the statistics of the resolvent-mode expansion coefficients.  Two example problems that demonstrate the relationships between the various decompositions are given in \S\ref{Sec:Examples}, and \S\ref{Sec:Conclusion} summarizes and concludes the paper.

\section{Proper orthogonal decomposition}
\label{Sec:POD}

The basic objective underlying POD is this: given a zero-mean stochastic process $\{ \vc{q}(\vc{z};\xi) \}$, find the deterministic function $\vc{\phi}(\vc{z})$ that best approximates the stochastic function on average \citep{Lumley:1967,Lumley:1970}.  Here, $\vc{z}$ is a set of independent variables and $\xi$ is an element in the probability space that parameterizes the stochastic variable.  We assume that each realization of the stochastic process belongs to a Hilbert space $\mathcal{H}$ with inner product $\langle \cdot, \cdot \rangle$ and define $E\{ \cdot \}$ to be the expectation operator over the probability space.  With these definitions, this objective is formalized by maximizing the quantity
\begin{equation}
\label{Eq:POD_thoery_opt}
\lambda = \frac{E\{\lvert \langle \vc{q}(\vc{z};\xi),\vc{\phi}(\vc{z}) \rangle \rvert^{2}\}}{\langle \vc{\phi}(\vc{z}),\vc{\phi}(\vc{z}) \rangle}
\end{equation}
over all $\vc{\phi}(\vc{z}) \in \mathcal{H}$.  That is, we wish to find the deterministic function $\vc{\phi}(\vc{z})$ that maximizes the expected value of the normalized projection of the stochastic function.

A standard variational approach can be used to show that the function $\vc{\phi}(\vc{z})$ that maximizes~(\ref{Eq:POD_thoery_opt}) must satisfy the eigenvalue problem 
\begin{equation}
\label{Eq:POD_theory_eig}
\langle \tc{C}(\vc{z},\vc{z}^{\prime}), \vc{\phi}(\vc{z}^{\prime}) \rangle^{*} = \lambda \vc{\phi}(\vc{z}),
\end{equation}
where
\begin{equation}
\label{Eq:POD_theory_R}
\tc{C}\left(\vc{z},\vc{z}^{\prime} \right) = E\{ \vc{q}(\vc{z};\xi) \vc{q}^{*}(\vc{z}^{\prime};\xi) \}
\end{equation}
is the two-point correlation tensor.  Throughout this paper, we use an asterisk superscript to denote both the complex-conjugate of a scalar and the Hermitian-transpose of a vector or tensor.  The properties of the solutions of the eigenvalue problem~(\ref{Eq:POD_theory_eig}) depend critically on the properties of the kernel $\tc{C}$, which in turn depend on the definition of the stochastic ensemble.  Fluid-flows are described by space-time fields $\vc{q}(\vc{x},t)$, and the space-only and spectral variants of POD are obtained by using these flow data to define the stochastic ensemble and the associated inner product and averaging operation in two different ways, as described in the following sections.

\subsection{Space-only POD}

The most commonly employed form of POD generates spatial modes $\vc{\phi}(\vc{x})$.  This is accomplished by defining the stochastic ensemble to consist of snapshots of the flow-field at different time instances.  In other words, the flow at each instant is treated as a realization of a stochastic process.  The appropriate inner product is then 
\begin{equation}
\label{Eq:POD_spacial_innerprod}
\left\langle {\vc{ u}}, {\vc{ v}} \right\rangle_{{x}} =  \int \limits_{\Omega} {\vc{ v}}^{*} ({\vc{ x}},t) {\tc{ W}}({\vc{ x}}) {\vc{ u}} ({\vc{ x}},t) d {\vc{ x}},
\end{equation}
where $\vc{u}$ and $\vc{v}$ are any two elements in $\mathcal{H}$, $\Omega$ denotes the spatial domain over which the flow is defined, and the weight $\tc{W}$ is a positive-definite Hermitian tensor of appropriate dimension.  We will restrict our attention to \emph{bounded} spatial domains, but note that unbounded homogeneous dimensions can be accommodated by transforming those directions to Fourier space \citep{Lumley:1967,Lumley:1970, George:2017}.  The expectation operator for this definition of the stochastic ensemble is simply a time average, so we are restricted to statistically stationary flows.

The quantity to maximize is 
\begin{equation}
\label{Eq:POD_space_opt}
 \lambda = \frac{E\{\lvert \langle \vc{q}(\vc{x},t),\vc{\phi}(\vc{x}) \rangle_{x} \rvert^{2}\}}{\langle \vc{\phi}(\vc{x}),\vc{\phi}(\vc{x}) \rangle_{x}}
\end{equation}
and the resulting Fredholm eigenvalue problem is
\begin{equation}
\label{Eq:POD_space_eig}
\int \limits_{\Omega} \tc{C}(\vc{x},\vc{x}^{\prime}) {\tc{ W}}({\vc{ x}^{\prime}}) \vc{\phi}(\vc{x}^{\prime})  d {\vc{ x}^{\prime}} = \lambda \vc{\phi}(\vc{x}),
\end{equation}
where
\begin{equation}
\label{Eq:POD_space_R}
\tc{C}(\vc{x},\vc{x}^{\prime}) = E\{ \vc{q}(\vc{x},t) \vc{q}^{*}(\vc{x}^{\prime},t) \}
\end{equation}
is the two-point \emph{spatial} correlation tensor.  This tensor is a nuclear kernel, i.e, it is compact and $\int_{\Omega} \tc{C}(\vc{x},\vc{x}) d {\vc{ x}}<\infty$.  As a result, Hilbert-Schmidt theory guarantees that the eigenmodes satisfying~(\ref{Eq:POD_space_eig}) have a number of special properties.  First, there exists a countably infinite set of eigenmodes, $\{\vc{\phi}_{j},\lambda_{j} \}$, that can be ranked according to their eigenvalue, $\lambda_{1} \geq \lambda_{2} \geq \cdots \geq 0$.  Each eigenvalue gives the average energy captured by that mode, measured in the spatial norm induced by the inner product~(\ref{Eq:POD_spacial_innerprod}), and the total energy of the flow is given by the sum of the eigenvalues.  

The eigenvectors are orthogonal, $\langle \vc{\phi}_{j}, \vc{\phi}_{k} \rangle_{x} = \delta_{jk}$, and provide a complete basis for $\vc{q}$.  Accordingly, the flow-field can be expanded as
\begin{equation}
\label{Eq:POD_space_expansion}
\vc{q}(\vc{x},t) = \sum \limits_{j = 1}^{\infty} a_{j}(t) \vc{\phi}_{j}(\vc{x})
\end{equation}
with $a_{j}(t) = \langle \vc{q}(\vc{x},t),\vc{\phi}_{j}(\vc{x}) \rangle_{x}$.  This expansion is optimal in its ability to capture the flow energy; if the expansion is truncated at order $n$, any other orthogonal expansion of the same order will capture less energy.  The expansion coefficients are uncorrelated at zero time lag,
\begin{equation}
\label{Eq:POD_space_a}
E\{a_{j}(t)a_{k}^{*}(t)\} = \lambda_{j}\delta_{jk}.
\end{equation}
Finally, the eigenmodes provide a diagonal representation of the two-point spatial correlation tensor
\begin{equation}
\label{Eq:POD_space_diag}
\tc{C}(\vc{x},\vc{x}^{\prime}) = \sum \limits_{j = 1}^{\infty} \lambda_{j} \vc{\phi}_{j}(\vc{x}) \vc{\phi}_{j}^{*}(\vc{x}^{\prime}),
\end{equation}
and are therefore its principal components.  Accordingly, space-only POD modes optimally represent spatial correlations within the flow.

\subsection{Spectral proper orthogonal decomposition}
\label{Sec:SpectralPOD}

Alternatively, we can seek modes that depend on both space and time.  This is accomplished by defining the stochastic ensemble to consist of a collection of realizations of the time-dependent flow.  For example, different runs of the same experiment are considered to be realizations of a stochastic process.  The appropriate inner product is then
\begin{equation}
\left\langle {\vc{ u}}, {\vc{ v}} \right\rangle_{{x,t}} = \int \limits_{-\infty}^{\infty}  \int \limits_{\Omega} {\vc{ v}}^{*} ({\vc{ x}},t) {\tc{ W}}({\vc{ x}}) {\vc{ u}} ({\vc{ x}},t) d {\vc{ x}} d t
\label{Eq:SPOD_InnerProduct}
\end{equation}
and the expectation operator is an ensemble average over different stochastic realizations of the flow.  The quantity to maximize is then
\begin{equation}
\label{Eq:SPOD_opt}
\lambda = \frac{ E\{\lvert \langle \vc{q}(\vc{x},t),\vc{\phi}(\vc{x},t) \rangle_{x,t} \rvert^{2}\}}{\langle \vc{\phi}(\vc{x},t),\vc{\phi}(\vc{x},t) \rangle_{x,t}},
\end{equation}
which leads to the eigenvalue problem
\begin{equation}
\label{Eq:SPOD_eig}
\int \limits_{-\infty}^{\infty} \int \limits_{\Omega} \tc{C}(\vc{x},\vc{x}^{\prime},t,t^{\prime}) {\tc{ W}}({\vc{ x}^{\prime}}) \vc{\phi}(\vc{x}^{\prime},t^{\prime})  d {\vc{ x}^{\prime}} dt^{\prime} = \lambda \vc{\phi}(\vc{x},t) ,
\end{equation}
where
\begin{equation}
\label{Eq:SPOD_R}
\tc{C}(\vc{x},\vc{x}^{\prime},t,t^{\prime}) = E\{ \vc{q}(\vc{x},t) \vc{q}^{*}(\vc{x}^{\prime},t^{\prime}) \}
\end{equation}
is the two-point space-time correlation tensor.  

Because they persist indefinitely, statistically stationary flows have infinite energy in a space-time norm.  As a result, the space-time correlation tensor~(\ref{Eq:SPOD_R}) is not nuclear and  the eigenmodes of~(\ref{Eq:SPOD_eig}) do not posses the properties generally associated with POD.  To remedy this, a new eigenvalue problem can be obtained in spectral space from which modes with useful properties can be obtained.  In what follows, we focus exclusively on stationary flows.

To derive the spectral eigenvalue problem, we recall that the correlation tensor of a wide-sense stationary flow depends only on the difference between two times,
\begin{equation}
\tc{C}(\vc{x},\vc{x}^{\prime},t,t^{\prime}) \rightarrow \tc{C}(\vc{x},\vc{x}^{\prime},t-t^{\prime}). 
\label{Eq:SPOD_CrossCor_stationary}
\end{equation} 
Then the cross-spectral density tensor $\tc{S}$ can be defined as the Fourier transform pair of the correlation tensor,
\begin{equation}
{\tc{ S}}({\vc{ x}},{\vc{ x}}^{\prime}, f) = \int \limits_{-\infty}^{\infty} {\tc{ C}}({\vc{ x}},{\vc{ x}}^{\prime}, \tau) e^{-\ii 2\pi f \tau} d \tau
\label{Eq:SPOD_CrossCor_stationary_FT}
\end{equation}
and
\begin{equation}
{\tc{C}}({\vc{ x}},{\vc{ x}}^{\prime}, \tau) = \int \limits_{-\infty}^{\infty} {\tc{S}}({\vc{ x}},{\vc{ x}}^{\prime}, f) e^{\ii 2\pi f \tau} d f.
\label{Eq:SPOD_CrossCor_stationary_IFT}
\end{equation}

Using these definitions, the following result can be derived: for any frequency $f^{\prime}$, the function ${ \vc{\phi}} ({\vc{ x}},t) = {\vc{\psi}} ({\vc{ x}},f^{\prime}) e^{\ii 2\pi f^{\prime} t}$ is a solution of the eigenvalue problem~(\ref{Eq:SPOD_eig}) with eigenvalue $\lambda(f^{\prime})$, where ${ \vc{\psi}} ({\vc{ x}},f^{\prime})$ and $\lambda(f^{\prime})$ satisfy the spectral eigenvalue problem
\begin{equation}
\int \limits_{\Omega} {\tc{ S}}({\vc{ x}},{\vc{ x}}^{\prime}, f^{\prime})   {\tc{ W}}(\vc{x}^{\prime}) \vc{\psi}({\vc{ x}^{\prime}},f^{\prime}) d {\vc{ x}}^{\prime}  = 
\lambda(f^{\prime})   \vc{\psi}({\vc{ x}},f^{\prime}).
\label{Eq:SPOD_eigSpectral}
\end{equation}
This result was first given by \cite{Lumley:1967,Lumley:1970}; we offer an alternative derivation in Appendix A.  

The cross-spectral density tensor is nuclear, so the eigenmodes of~(\ref{Eq:SPOD_eigSpectral}) \emph{at each frequency} inherit properties analogous to those of space-only POD modes. There is a countably infinite set of eigenfunctions $\vc{\psi}_{j}(\vc{x},f)$ at each frequency that are orthogonal to all other modes at the same frequency in the \emph{spatial} inner product~(\ref{Eq:POD_spacial_innerprod}), i.e., $\langle \boldsymbol{\psi}_{j},\boldsymbol{\psi}_{k} \rangle_{x} = \delta_{jk}$.  The Fourier modes of each flow realization are optimally expanded as
\begin{equation}
\label{Eq:SPOD_expansion_freq}
\hat{\vc{q}}(\vc{x},f) = \sum \limits_{j = 1}^{\infty} a_{j}(f) \vc{\psi}_{j}(\vc{x},f)
\end{equation}
with $a_{j}(f) = \langle \hat{\vc{q}}(\vc{x},f), \vc{\psi}_{j}(\vc{x},f) \rangle_{x}$ and the expansion coefficients are uncorrelated,
\begin{equation}
\label{Eq:SPOD_a_freq}
E\{a_{j}(f)a_{k}^{*}(f)\} = \lambda_{j}(f)\delta_{jk}.
\end{equation}
The cross-spectral density tensor has the diagonal representation
\begin{equation}
\label{Eq:SPOD_diag_freq}
\tc{S}(\vc{x},\vc{x}^{\prime},f) = \sum \limits_{j = 1}^{\infty} \lambda_{j}(f) \vc{\psi}_{j}(\vc{x},f) \vc{\psi}_{j}^{*}(\vc{x}^{\prime},f)
\end{equation}
so the SPOD modes are its principal components.  Furthermore, the modes $\vc{\psi}_{j}(\vc{x},f) e^{\ii 2\pi f t}$ are orthogonal in the space-time inner product~(\ref{Eq:SPOD_InnerProduct}), so each mode at each frequency can be viewed as a distinct space-time mode.  The space-time correlation tensor can be written as
\begin{equation}
\label{Eq:SPOD_correlation_expansion}
\tc{C}(\vc{x},\vc{x}^{\prime},\tau) = \int \limits_{-\infty}^{\infty}  \sum \limits_{j = 1}^{\infty} \lambda_{j}(f) \vc{\psi}_{j}(\vc{x},f) \vc{\psi}_{j}^{*}(\vc{x}^{\prime},f) e^{\ii 2\pi f \tau}  d f.
\end{equation}
In summary, for stationary flows the space-time POD formulation leads to spectral POD modes that each oscillate at a single frequency and optimally represent the second-order space-time flow statistics.

\subsection{Spectral vs. space-only POD}
\label{Sec:Compare_POD}

The essential difference between spectral and space-only POD is that the former yields modes that are coherent in space and time, whereas the later gives modes that are only spatially coherent.  First consider space-only POD.  The time dependence of the flow field $\vc{q}(\vc{x},t)$ is treated as a stochastic parameter, with different time instances taken to represent different members in an ensemble of spatially-dependent fields.  Once interpreted in this way, the flow snapshots that make up the ensemble lose any concept of sequential ordering, so the time-dependent evolution of the flow has no impact on the definition of the POD modes.  Therefore, the POD modes are impervious to temporal correlation within the data, which is an essential feature of physical coherent structures.  Because of this, POD modes do not necessarily represent structures that evolve coherently in space and time.  This can be shown explicitly by writing the space-time correlation tensor in terms of space-only POD modes,
\begin{subequations}
\label{Eq:SPOD_comparePOD_POD_cor}
\begin{align}
\tc{C}(\vc{x},\vc{x}^{\prime},\tau) &= E\left\{  \left( \sum\limits_{j=1}^{\infty} a_{j}(t) \vc{\phi}_{j}(\vc{x})\right) \left( \sum\limits_{k=1}^{\infty} a_{k}(t+\tau) \vc{\phi}_{k}(\vc{x}^{\prime}) \right)^{*} \right\} \\ &= \sum\limits_{j=1}^{\infty} \sum\limits_{k=1}^{\infty} \ts{C}^{\mathrm{POD}}_{a_{j}a_{k}}(\tau) \vc{\phi}_{j}(\vc{x}) \vc{\phi}_{k}^{*}(\vc{x}^{\prime})
\end{align}
\end{subequations}
with 
\begin{equation}
\label{Eq:SPOD_comparePOD_CaaPOD}
\ts{C}_{a_{j}a_{k}}^{\mathrm{POD}}(\tau) = E\{ a_{j}(t) a_{k}^{*}(t+\tau)\}.
\end{equation}
When $\tau = 0$, (\ref{Eq:POD_space_a}) ensures that $\ts{C}^{\mathrm{POD}}_{a_{j}a_{k}}(\tau) = \lambda_{j}\delta_{jk}$ and~(\ref{Eq:SPOD_comparePOD_POD_cor}) reduces to the spatial correlation tensor. On the other hand, (\ref{Eq:POD_space_a}) is not applicable when $\tau \neq 0$, and, as a result, POD theory does not guarantee any special properties for $\ts{C}^{\mathrm{POD}}_{a_{j}a_{k}}(\tau)$; thus, the temporal correlation of two terms in the POD expansion is not known \emph{a priori}.  This means that the part of the flow described by a given POD mode is not necessarily correlated with the part of the flow described by the same POD mode at a later time, nor is it necessarily uncorrelated with the part of the flow described by a different mode at a later time.  Therefore, contrary to some previous statements, space-only POD modes \emph{do not} necessarily represent flow structures that evolve coherently.  A recent analysis of space-only POD by \cite{George:2017} erroneously assumed the expansion coefficients to be uncorrelated at different times.  

In contrast, SPOD modes do represent structures that evolve coherently.  The space-time correlation tensor is written in terms of SPOD modes in~(\ref{Eq:SPOD_correlation_expansion}).  This form of the space-time correlation tensor is the result of special properties of the correlations
\begin{equation}
\label{Eq:SPOD_comparePOD_CaaSPOD}
\ts{C}_{a_{j}a_{k}}^{\mathrm{SPOD}}(f,f^{\prime}) \triangleq E\{ a_{j}(f) a_{k}^{*}(f^{\prime})\} = \lambda_{j} \delta_{jk} \delta(f-f^{\prime}),
\end{equation}
which govern the correlation between the part of the flow described by individual SPOD modes.  Specifically, (\ref{Eq:SPOD_correlation_expansion}) can be obtained by inserting (\ref{Eq:SPOD_comparePOD_CaaSPOD}) along with the inverse Fourier transform of (\ref{Eq:SPOD_expansion_freq}) into the definition of the space-time correlation tensor given by (\ref{Eq:SPOD_R}) and (\ref{Eq:SPOD_CrossCor_stationary}).  The first two terms in the final form of~(\ref{Eq:SPOD_comparePOD_CaaSPOD}) follow from~(\ref{Eq:SPOD_a_freq}) and ensure that two terms in the SPOD expansion at the same frequency are uncorrelated.  The final term is a consequence of the fact that the frequency components from the Fourier transform of a stationary random process are uncorrelated \citep{Lumley:1970, George:1988}.  

In sum, (\ref{Eq:SPOD_correlation_expansion})~and~(\ref{Eq:SPOD_comparePOD_CaaSPOD}) show that the part of the flow described by a particular SPOD mode is perfectly correlated with the part of the flow described by that same mode at all times and entirely uncorrelated with the part of the flow described by all other modes at all times.  In other words, each SPOD mode describes a structure that evolves coherently in space and time.

The preceding analysis does not imply that space-only POD modes can \emph{never} exhibit space-time coherence.  For example, \cite{Rowley:2004} observed that some of the leading space-only POD modes of a compressible cavity flow capture single-frequency Rossiter modes.  Rather, our analysis shows that space-only POD modes do not have this property by construction, in contrast to SPOD modes which evolve coherently in space and time by construction.  

It is also possible to derive equations relating space-only and spectral POD modes and eigenvalues.  Using the fact that the spatial correlation tensor is equivalent to the zero-time-lag space-time correlation tensor, (\ref{Eq:SPOD_CrossCor_stationary_IFT}) implies that
\begin{equation}
\label{Eq:SPOD_comparePOD_RS}
\tc{C}(\vc{x},\vc{x}^{\prime}) =  \int \limits_{-\infty}^{\infty} \tc{S}(\vc{x},\vc{x}^{\prime},f) d f.
\end{equation}
Expanding ${\tc{C}}$ and ${\tc{S}}$ in terms of POD and SPOD modes, respectively, gives
\begin{equation}
\label{Eq:SPOD_comparePOD_RS_expand}
\sum \limits_{j = 1}^{\infty} \lambda_{j} \vc{\phi}_{j}(\vc{x}) \vc{\phi}_{j}(\vc{x}^{\prime})^{*} = \int \limits_{-\infty}^{\infty} \sum \limits_{k = 1}^{\infty} \lambda_{k}(f) \vc{\psi}_{k}(\vc{x},f) \vc{\psi}_{k}(\vc{x}^{\prime},f)^{*} d f.
\end{equation}
Applying the operation $\langle \vc{\phi}_{j} , \cdot \rangle_{x}$ to both sides of~(\ref{Eq:SPOD_comparePOD_RS_expand}) and dividing by $\lambda_{j}$ leads to the expression
\begin{equation}
\label{Eq:SPOD_comparePOD_Phi}
\vc{\phi}_{j}(\vc{x})  =  \int \limits_{-\infty}^{\infty} \sum \limits_{k = 1}^{\infty} \frac{\lambda_{k}(f)}{\lambda_{j}} \, c_{jk}(f) \vc{\psi}_{k}(\vc{x},f) \, d f,
\end{equation}
where $c_{jk}(f) = \langle \vc{\phi}_{j}(\vc{x}) , \vc{\psi}_{k}(\vc{x},f) \rangle_{x}$.  Taking the same inner product again and moving $\lambda_{j}$ back to the left-hand-side yields
\begin{equation}
\label{Eq:SPOD_comparePOD_Lam}
\lambda_{j} = \int \limits_{-\infty}^{\infty}  \sum \limits_{k=1}^{\infty}  \lambda_{k}(f) \, \lvert  c_{jk}(f) \rvert^{2} \,d f.
\end{equation}
Equations~(\ref{Eq:SPOD_comparePOD_Phi}) and~(\ref{Eq:SPOD_comparePOD_Lam}) show that each space-only POD mode is potentially made up of many SPOD modes.  Physically, this means that the spatially coherent structures represented by space-only POD are composed of contributions from spatiotemporal coherent structures at many frequencies.  Practically, this is manifested as broadband frequency content within the coefficients $a_{j}(t)$.  This highlights the fact that each space-only POD mode typically represents flow phenomena at many different time scales, which muddies their interpretation.  In contrast, SPOD modes decouple flow phenomena at different time scales, which can be helpful for understanding the flow dynamics and deriving non-empirical models.

\section{Computing SPOD modes from data}
\label{Sec:SPOD_comp}

An efficient algorithm for computing space-only POD modes from snapshots of discrete flow data using the method of snapshots \citep{Sirovich:1987a} is well-known and described in detail in numerous  publications \citep[e.g.,][]{Rowley:2017,Taira:2017}.  Techniques for computing SPOD modes from snapshots of the flow are not as well documented. Here, we outline a procedure similar to the one used by \cite{Gordeyev:2000} and \cite{Citriniti:2000} but with an additional simplification that reduces the computational cost in most cases.

Let the vector $\vd{q}_{k} \in \mathbb{R}^{N}$ represent the instantaneous state of $\vc{q}(\vc{x},t)$ at time $t_{k}$ on a discrete set of points in the spatial domain $\Omega$.  The total length $N$ of the vector is equal to the number of grid points times the number of flow variables, since all of these values have been stacked into the vector $\vd{q}_{k}$, which we call a snapshot of the flow.  Now suppose that these data are available for $M$ equally spaced time instances, $t_{k+1} = t_{k} + \Delta t$.  This data set can be compactly represented by the data matrix
\begin{equation}
\label{Eq:SPOD_comput_dataMat}
\td{Q} = \left[ \vd{q}_{1} , \vd{q}_{2}, \dots, \vd{q}_{M} \right] \in \mathbb{R}^{N \times M}.
\end{equation}

\begin{figure}
 \centering
 \begin{overpic}[trim=1.0cm 3.5cm 0.5cm 3.5cm, clip=true,width=5.0in]{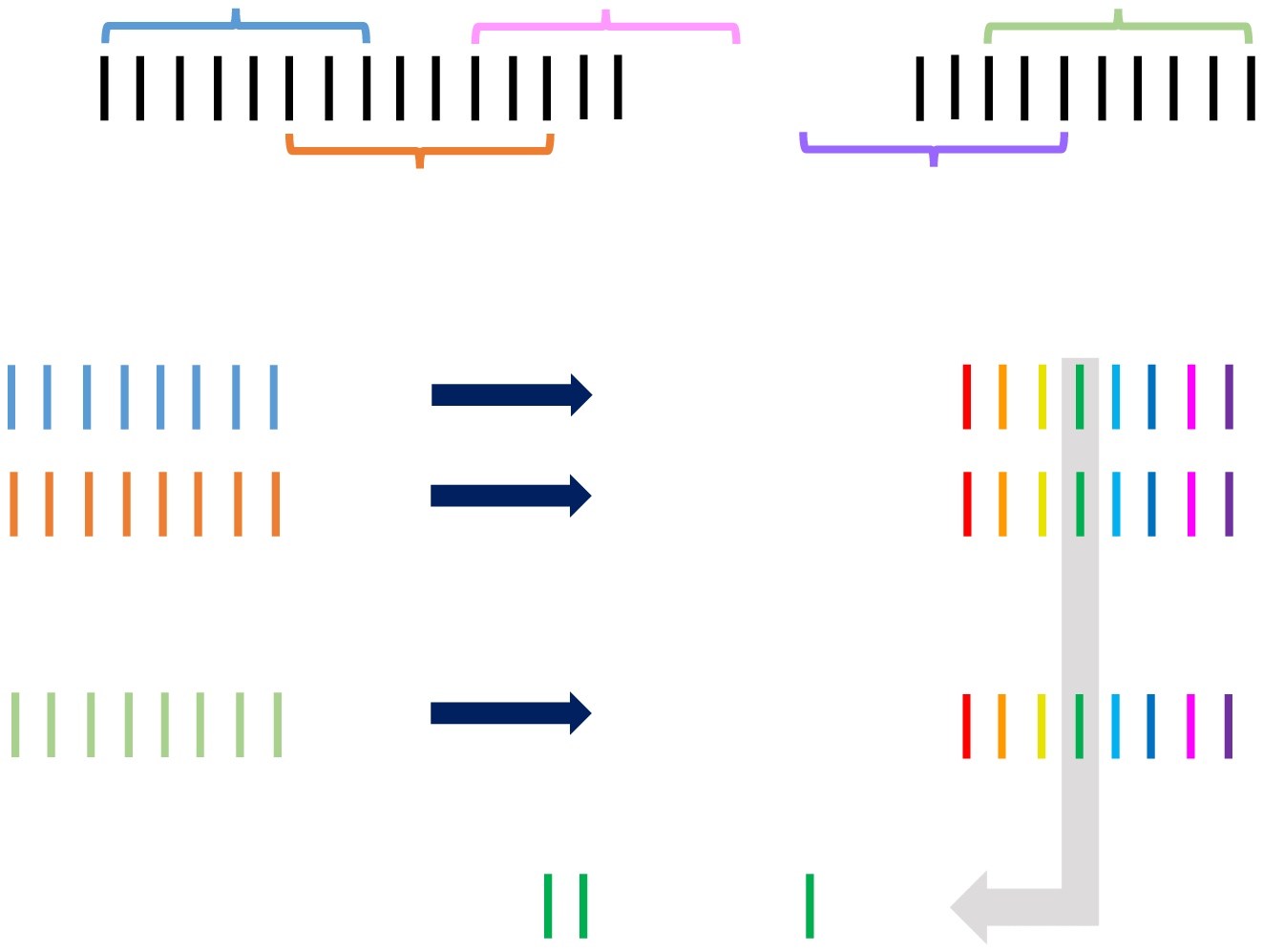}
 \put (7.5,0){\rectanglez{4.0in}{2.45in}}
 \put (20.25,39.875) {\large $\td{Q} = [$}
 \put (82.25,39.875) {\large $]$}
 \put (57.00,39.875) {\large $\boldsymbol{\cdots}$}
 \put (27.25,44) {\small $t \rightarrow$}
 \put (33.625,45.3) {$(1)$}
 \put (50.125,45.3) {$(3)$}
 \put (72.15,45.3) {$(N_{b})$}
 \put (41.85,34.3) {$(2)$}
 \put (61.55,34.3) {$(N_{b}-1)$}
 \put (13.00,26.10) {\large $\td{Q}^{(1)} = [$}
 \put (38.50,26.10) {\large $]$}
 \put (13.00,21.30) {\large $\td{Q}^{(2)} = [$}
 \put (38.50,21.30) {\large $]$}
 \put (11.25,11.45) {\large ${\td{Q}^{(N_{b})} = [}$}
 \put (38.50,11.45) {\large $]$}
 \put (23.25,29.25) {\small $t \rightarrow$}
 \put (20.625,16.0) {\large \bf $\vdots$}
 \put (44.5,29.5) {DFT}
 \put (47,16.0) {\large \bf $\vdots$}
 \put (55.50,26.10) {\large $\hat{\td{Q}}^{(1)} = [$}
 \put (81.00,26.10) {\large $]$}
 \put (55.50,21.30) {\large $\hat{\td{Q}}^{(2)} = [$}
 \put (81.00,21.30) {\large $]$}
 \put (53.75,11.45) {\large ${\hat{\td{Q}}^{(N_{b})} = [}$}
 \put (81.00,11.45) {\large $]$}
 \put (65.75,29.25) {\small $f \rightarrow$}
 \put (63.125,16.0) {\large \bf $\vdots$}
 \put (32.25,3.4) {\large $\hat{\td{Q}}_{f_k} = \sqrt{\kappa}\,[$}
 \put (62.325,3.4) {\large $]$}
 \put (53.75,3.4) {\large  $\boldsymbol{\cdots}$}
 \end{overpic}
 \caption{Schematic depiction of Welch's method for estimating SPOD modes.  A detailed description of each step is given in the text.  }
 \label{fig:Welch}
\end{figure}

The cross-spectral density tensor could be naively estimated using the discrete Fourier transform (DFT) of the data matrix $\td{Q}$.  However, it is well known that spectral estimates obtained in this way do not converge as the number of snapshots $M$ is increased.  In fact, the uncertainty in the estimate at each frequency is as large as the magnitude of the estimate itself \citep{George:1978,Bendat:1990}.  To obtain convergent estimates of the spectral densities, it is necessary to appropriately average the spectra over multiple realizations of the flow.  This can be accomplished using Welch's (\citeyear{Welch:1967}) method, which is represented schematically in Figure~\ref{fig:Welch}.  The first step is to partition the data matrix into a set of smaller, possibly overlapping blocks.  Precisely, if we write each block as
\begin{equation}
\label{Eq:SPOD_comput_dataMat_block}
\td{Q}^{(n)} = \left[ \vd{q}_{1}^{(n)} , \vd{q}_{2}^{(n)}, \dots, \vd{q}_{N_{f}}^{(n)} \right] \in \mathbb{R}^{N \times N_{f}},
\end{equation}
then the $k$-th entry in the $n$-th block is $\vd{q}_{k}^{(n)} = \vd{q}_{k+(n-1)(N_{f}-N_{o})}$, where $N_{f}$ is the number of snapshots in each block, $N_{o}$ is the number of snapshots by which the blocks overlap, and $N_{b}$ is the total number of blocks.  By the ergodicity hypothesis, each of these blocks can be regarded as a member of an ensemble of realizations of the flow.

Next, the DFT is computed for each block,
\begin{equation}
\label{Eq:SPOD_comput_DFT_Qhat}
\hat{\td{Q}}^{(n)} =  \left[ \hat{\vd{q}}_{1}^{(n)} , \hat{\vd{q}}_{2}^{(n)}, \dots, \hat{\vd{q}}_{N_{f}}^{(n)} \right]
\end{equation}
with
\begin{equation}
\label{Eq:SPOD_comput_DFT}
\hat{\vd{ q}}_{k}^{(n)} = \frac{1}{\sqrt{N_{f}}} \sum \limits_{j=1}^{N_{f}} w_{j} \vd{q}_{j}^{(n)} e^{-\ii 2\pi (k-1) \frac{j-1}{N_{f}}}
\end{equation}
for $k = 1,\dots, N_{f}$ and $n = 1,\dots,N_{b}$.  The scalar weights $w_{j}$ are nodal values of a window function that can be used to reduce spectral leakage due to non-periodicity of the data in each block \citep[e.g.,][]{Heinzel:2002}.  We have included the $1/\sqrt{N_{f}}$ factor to make the discrete transform unitary for a rectangular window ($w_{j}=1$ for all $j$), which will be convenient later in \S\ref{Sec:Compare_DMD}.  Here, $\hat{\vd{q}}_{k}^{(n)}$ is the Fourier component at frequency $f_{k}$ in the $n$-th block and the resolved frequencies are
\begin{equation}
    f_{k} =
\begin{cases}
    \frac{k-1}{N_{f} \Delta t} & \text{for} \quad k\leq N_{f}/2, \\[6pt]
    \frac{k-1-N_{f}}{N_{f} \Delta t} & \text{for} \quad k > N_{f}/2.
\end{cases}
\label{Eq:SPOD_comput_fk}
\end{equation}

The cross-spectral density tensor $\tc{S}(\vc{x},\vc{x}^{\prime},f)$ can be estimated at frequency $f_{k}$ by the average
\begin{equation}
\label{Eq:SPOD_comput_CSD_sum}
\td{S}_{f_{k}} = \frac{\Delta t}{s N_{b}} \sum \limits_{n = 1}^{N_{b}} \hat{\vd{q}}_{k}^{(n)} \left(\hat{\vd{ q}}_{k}^{(n)}\right)^{*}, 
\end{equation}
where $s = \sum_{j=1}^{N_{f}} w_{j}^{2}$.  This can be written compactly by arranging the Fourier coefficients at frequency $f_{k}$ from each block into the new data matrix
\begin{equation}
\label{Eq:SPOD_comput_dataMat_freq}
\hat{\td{Q}}_{f_{k}} = \sqrt{\kappa} \left[ \hat{\vd{q}}_{k}^{(1)} , \hat{\vd{q}}_{k}^{(2)}, \dots, \hat{\vd{q}}_{k}^{(N_{b})} \right] \in \mathbb{R}^{N \times N_{b}},
\end{equation}
where $\kappa = {\Delta t/ (s N_{b}) }$.
Then the estimated cross-spectral density tensor at frequency $f_{k}$ can be written as
\begin{equation}
\label{Eq:SPOD_comput_CSD_mats}
\td{S}_{f_{k}} = \hat{\td{Q}}_{f_{k}} \hat{\td{Q}}_{f_{k}}^{*}.
\end{equation}
This estimate converges as the number of blocks $N_{b}$ and the number of snapshots in each block $N_{f}$ are increased together \citep{Welch:1967,Bendat:1990}.

Using this estimate, the infinite dimensional SPOD eigenvalue problem~(\ref{Eq:SPOD_eigSpectral}) reduces to an $N \times N$ matrix eigenvalue problem
\begin{equation}
\label{Eq:SPOD_comput_eigDisc}
\td{S}_{f_{k}}\td{W}{\text{\bf \textPsi}}_{f_{k}} =  {\text{\bf \textPsi}}_{f_{k}} {\text{\bf \textLambda}}_{f_{k}}
\end{equation}
at each frequency.  Here, the spatial inner product~(\ref{Eq:POD_spacial_innerprod}) is approximated as $\langle \hat{\vd{q}}_{1}, \hat{\vd{q}}_{2}  \rangle_{\vd{ x}} = \hat{\vd{q}}_{2}^{*} \td{W} \hat{\vd{q}}_{1}$; the positive-definite Hermitian matrix $\td{W}  \in \mathbb{C}^{N \times N}$ accounts for both the weight $\tc{W}(\vc{x})$ and the numerical quadrature of the integral on the discrete grid.  The approximate SPOD modes are given by the columns of ${\text{\bf \textPsi}}_{f_{k}}$ and are ranked according to their corresponding eigenvalues given by the diagonal matrix ${\text{\bf \textLambda}}_{f_{k}}$.  Note that at most $N_{b}$ nonzero eigenvalues can be obtained.  The approximate modes mimic the properties of the continuous modes.  For example, they are discretely orthogonal, $ {\text{\bf \textPsi}}_{f_{k}}^{*} \td{W} {\text{\bf \textPsi}}_{f_{k}} = \td{I}$, and the estimated cross-spectral density tensor can be expanded as $\td{S}_{f_{k}} = {\text{\bf \textPsi}}_{f_{k}} {\text{\bf \textLambda}}_{f_{k}} {\text{\bf \textPsi}}_{f_{k}}^{*}$.

In practice, the number of blocks $N_{b}$ is typically much smaller than the discretized problem size $N$. Using the definition of $\td{S}_{f_{k}}$ from~(\ref{Eq:SPOD_comput_CSD_mats}), it is possible to show that the $N_{b} \times N_{b}$ eigenvalue problem
\begin{equation}
\label{Eq:SPOD_comput_eigDisc_MOS}
\hat{\td{Q}}_{f_{k}}^{*}\td{W}\hat{\td{Q}}_{f_{k}}\text{\bf \textTheta}_{f_{k}} =  \text{\bf \textTheta}_{f_{k}} \tilde{\text{\bf \textLambda}}_{f_{k}}
\end{equation}
supports the same non-zero eigenvalues as~(\ref{Eq:SPOD_comput_eigDisc}).  The eigenvectors corresponding to these non-zero eigenvalues can be exactly recovered as
\begin{equation}
\label{Eq:SPOD_comput_ModeRecover}
\tilde{\text{\bf \textPsi}}_{f_{k}} = \hat{\td{Q}}_{f_{k}} \text{\bf \textTheta}_{f_{k}} \tilde{\text{\bf \textLambda}}_{f_{k}}^{-1/2}.
\end{equation}

The complete procedure for computing SPOD modes from data snapshots is outlined in the following algorithm.  Variables that are assigned using the ``$\gets$'' operator can be deleted or overwritten after each iteration in their respective loop to reduce memory usage.  A \textsc{Matlab} implementation is available at \url{https://github.com/SpectralPOD/spod_matlab}.  We also note that \cite{Schmidt:2017b} recently formulated a streaming version of the algorithm that can reduce computational cost for large data sets.  

\clearpage
\medskip
\renewcommand{\labelenumi}{\arabic{enumi}. }
{\bf\noindent Algorithm} (SPOD)
\begin{enumerate}
\item For each data block $n = 1,2,\dots, N_{b}$: 
\begin{enumerate}
    \item Assemble the data matrix \[\td{Q}^{(n)} \gets [\vd{q}_{1+(n-1)(N_f -N_o)}, \vd{q}_{2+(n-1)(N_f-N_o)}, \dots, \vd{q}_{N_f+(n-1)(N_\omega- N_o)}].\] 
    \item Using a (windowed) fast Fourier transform, calculate and store the row-wise DFT \[\hat{\td{Q}}^{(n)} = \mathrm{FFT}(\td{Q}^{(n)}) = [\hat{\vd{q}}_1^{(n)}, \hat{\vd{q}}_2^{(n)}, \dots, \hat{\vd{q}}_{N_\omega}^{(n)}].\] The column $\hat{\vd{q}}_k^{(n)}$ contains the $n^{\text{th}}$ realization of the Fourier mode at the $k^{\text{th}}$ discrete frequency $f_k$.
    \end{enumerate}
\item For each frequency $k = 1,2,\dots, N_{f}$ (or some subset of interest):
\begin{enumerate}
    \item Assemble the matrix $\hat{\td{Q}}_{f_{k}} \gets \sqrt{\kappa}\,[\hat{\vd{q}}_k^{(1)}, \hat{\vd{q}}_k^{(2)}, \dots, \hat{\vd{q}}_k^{(N_b)}]$ of Fourier realizations from the $k^{\text{th}}$ column of each $\hat{\td{Q}}^{(n)}$.
    \item Calculate the matrix $\td{M}_{f_{k}} \gets \hat{\td{Q}}_{f_{k}}^* \td{W} \hat{\td{Q}}_{f_{k}}$.
    \item Compute the eigenvalue decomposition $\td{M}_{f_{k}} = \text{\bf \textTheta}_{f_{k}} \tilde{\text{\bf \textLambda}}_{f_{k}} \text{\bf \textTheta}_{f_{k}}^*$.
    \item Obtain and store the SPOD modes $\tilde{\text{\bf \textPsi}}_{f_{k}} = \hat{\td{Q}}_{f_{k}} \text{\bf \textTheta}_{f_{k}} \tilde{\text{\bf \textLambda}}_{f_{k}}^{-1/2}$ and modal energies $\tilde{\text{\bf \textLambda}}_{f_{k}}$ for the $k^{\text{th}}$ discrete frequency.
  \end{enumerate}
\end{enumerate}

\section{Spectral POD and DMD}
\label{Sec:Compare_DMD}

In this section, we investigate the relationship between SPOD and dynamic mode decomposition (DMD) and show that SPOD can be understood as an optimal form of DMD for statistically stationary turbulent flows.  Dynamic mode decomposition was developed as an alternative to POD for identifying coherent structures from flow data \citep{Schmid:2008,Rowley:2009,Schmid:2010}.  The method approximates the eigenmodes of a linear operator that maps the state of the flow from one time instant to the next.  Since the operator is linear, the temporal evolution of each mode is described by a single frequency and growth/decay rate and the modes are in general spatially non-orthogonal.  Just as space-only POD can be described as a spatial orthogonalization of the flow data, DMD can be understood as a temporal orthogonalization of the data \citep{Schmid:2010}.

\subsection{DMD and the DFT}
\label{Sec:DMD_DFT_connect}

For zero-mean data that are uniformly sampled in time, \cite{Chen:2012} showed that DMD is formally equivalent to the DFT (the flow snapshots must also be linearly independent for this to hold, which is true in most applications).  As a result, each DMD mode has zero growth/decay rate.  This formal DMD-DFT equivalence does not hold for data with a nonzero mean, but in practice the physically-relevant DMD modes tend to have have nearly zero growth/decay rates for stationary flows \cite[e.g.,][]{Rowley:2009,Chen:2012,Schmid:2012,Semeraro:2012}, which is logical since stationary flows are persistent by definition.  

This tendency is explained more rigorously by the connection between DMD and Koopman operator theory \citep{Mezic:2005, Rowley:2009}.  The Koopman operator is an infinite-dimensional \emph{linear} operator that describes the evolution of scalar observables of a \emph{nonlinear} dynamical system on a finite manifold.  The eigenvectors of the Koopman operator can be used to decompose vector-valued observables  (equivalent to our $\vc{q}$) into modes that evolve with a single frequency and growth/decay rate.  \cite{Mezic:2005} showed that for any dynamical system with a Borel probability measure, the growth/decay rate is zero and Koopman modes are equivalent to Fourier modes.  Stationary flows possess an ergodic measure by definition, so their Koopman modes are simply Fourier modes.

\cite{Rowley:2009} showed that DMD modes approximate Koopman modes when the flow snapshots used to compute the modes are linearly independent, and \cite{Tu:2013} showed that this holds under a slightly weaker condition on the data termed \emph{linear consistency}.  In light of this, it is not surprising that DMD modes tend to be similar to DFT modes for stationary flows.  Moreover, deviations of the DMD modes from Fourier modes can be viewed as artifacts of the DMD approximation of the underlying Koopman operator.  This suggests that, contrary to prevailing wisdom, it is advantageous to subtract the mean when applying DMD to stationary flows to ensure that the DMD modes mimic the zero-growth-rate property of the underlying Koopman modes.

\subsection{An ensemble DMD problem for stationary flows}
\label{Sec:DMD_ensemble}

To conceptually relate SPOD and DMD, imagine that we have an ensemble of $N_{e}$ data sets each representing a stochastic realization of the same stationary flow.  There are at least two approaches one could adopt for applying DMD to this problem.  The first would be to simply perform separate DMD computations for each realization of the flow.  If we agree to subtract the mean to ensure zero growth rates as described previously, each DMD mode will exactly correspond to a DFT mode, but in general the mode at a given frequency will be different for each realization.  

A more sophisticated approach for applying DMD to this problem would be to use the approach of \cite{Tu:2013} to combine multiple flow realization into a single DMD calculation.  Their variant of DMD, which they call `exact DMD', is defined in terms of the operator
\begin{equation}
\label{Eq:DMD_A_def}
\td{A} \triangleq \td{Y} \td{X}^{+},
\end{equation}
where $\td{X}^{+}$ is the pseudo-inverse of $\td{X}$ and the columns of the matrices $\td{X}$ and $\td{Y}$ are input-output data pairs that are related by a linear operator that is to be approximated by the matrix $\td{A}$.  The DMD modes and eigenvalues are then given by the eigenvectors and eigenvalues of $\td{A}$, respectively.  

For a standard application in which the flow data consist of sequential snapshots of the flow, the input and output matrices are
\begin{subequations}
\label{Eq:DMD_XY_def}
\begin{align}
\td{X} \triangleq& \left[ \vd{q}_{1} , \vd{q}_{2}, \dots, \vd{q}_{M-1} \right] = \td{Q} \, \td{T}_{\mathrm{X}}, \\
\td{Y} \triangleq& \left[ \vd{q}_{2} , \vd{q}_{3}, \dots, \vd{q}_{M} \right] = \td{Q} \, \td{T}_{\mathrm{Y}},
\end{align}
\end{subequations}
where each $\vd{q}_{j} \in \mathbb{R}^{N}$ is a snapshot of the flow as defined in \S\ref{Sec:SPOD_comp} and $\td{Q} \in \mathbb{R}^{N \times M}$ is the data matrix given by~(\ref{Eq:SPOD_comput_dataMat}).  The matrices $\td{T}_{\mathrm{X}}, \td{T}_{\mathrm{Y}} \in \mathbb{R}^{M \times M-1}$ select the appropriate columns of $\td{Q}$ and are given by
\begin{equation}
\label{Eq:DMD_SS_def}
\td{T}_{\mathrm{X}} \triangleq {\begin{bmatrix}1&0&\cdots&0 \\ 0&1 & \ddots & \vdots \\ \vdots &\ddots&\ddots&0  \\0&\cdots&0&1 \\0&0&\cdots&0 \end{bmatrix}}, \qquad \td{T}_{\mathrm{Y}} \triangleq {\begin{bmatrix} 0&0&\cdots&0 \\  1&0&\cdots&0 \\ 0&1 & \ddots & \vdots \\ \vdots &\ddots&\ddots&0  \\0&\cdots&0&1  \end{bmatrix}}.
\end{equation}

As described by \cite{Tu:2013}, multiple realizations of a flow can be accommodated within the exact DMD framework by arranging the realizations together into ensemble input and output matrices
\begin{subequations}
\label{Eq:DMD_XY_def_ensemble}
\begin{align}
\td{X} \triangleq& \left[ \td{Q}^{(1)} \td{T}_{\mathrm{X}}  , \dots, \td{Q}^{(N_{e})} \td{T}_{\mathrm{X}}  \right], \\
\td{Y} \triangleq& \left[ \td{Q}^{(1)} \td{T}_{\mathrm{Y}} ,  \dots, \td{Q}^{(N_{e})} \td{T}_{\mathrm{Y}}  \right],
\end{align}
\end{subequations}
where each $\td{Q}^{(n)} \in \mathbb{R}^{N \times M}$ contains snapshots from the $n$-th realization of the flow, as in~(\ref{Eq:SPOD_comput_dataMat_block}).

The DMD/DFT equivalence for zero-mean data proven by \cite{Chen:2012} was derived within the context of the original Arnoldi-based formulation of DMD given by \cite{Schmid:2010} and was restricted to the case of sequential, linearly independent snapshots.  Accordingly, it does not apply to the ensemble DMD problem defined by~(\ref{Eq:DMD_A_def}) and~(\ref{Eq:DMD_XY_def_ensemble}).

In Appendix~\ref{Sec:DMDensembleProof}, we prove that the DMD modes obtained from this ensemble formulation are precisely the DFT modes of each realization of the flow if each realization has zero mean (i.e., that we've subtracted the mean in line with our earlier arguments for stationary flows) and the ensemble input/output data matrices are linearly consistent. Under these conditions, (\ref{Eq:DMD_A_def}) and~(\ref{Eq:DMD_XY_def_ensemble}) reduce to 
\begin{equation}
\label{Eq:DMD_eigprob_final}
\td{A}\left[ \tilde{\td{Q}}^{(1)}, \dots, \tilde{\td{Q}}^{(N_{e})}\right]  = \left[ \tilde{\td{Q}}^{(1)}, \dots, \tilde{\td{Q}}^{(N_{e})} \right] 
 \begingroup 
\setlength\arraycolsep{2pt} 
\renewcommand{\arraystretch}{1}
 \begin{bmatrix}
{\text{\bf \textLambda}}_{\mathrm{DMD}}  &  & \\
 & \ddots &  \\
 &  & {\text{\bf \textLambda}}_{\mathrm{DMD}} \end{bmatrix}
 \endgroup,
\end{equation}
where the columns of $\tilde{\td{Q}}^{(n)}$ contain the DFT modes of ${\td{Q}}^{(n)}$ excluding the mean component,
\begin{equation}
\label{Eq:DMD_LAMBDA}
{\text{\bf \textLambda}}_{\mathrm{DMD}}  = \begin{bmatrix}
z &  & & \\
 & z^{2} & &  \\
 & & \ddots &  \\
 & & & z^{M-1} \end{bmatrix},
\end{equation}
and $z = e^{-\mathrm{i} 2 \pi /M}$ is the primitive $M$-th root of unity.  Therefore, each DFT mode $\hat{\vd{q}}^{(n)}_{k}$, $k = 2, \dots M$ is an eigenvector of $\td{A}$ with eigenvalue $z^{k-1}$.  These eigenvalues have unit magnitude so the growth/decay rate is zero, as expected.  The frequency corresponding to each DFT mode is recovered from the eigenvalue as $f_{k} = \mathrm{Re}\left[ \mathrm{ln}(z^{k-1}) / (-\mathrm{i} 2 \pi \Delta t  )\right]$, and it is easy to verify that these frequencies match those given by~(\ref{Eq:SPOD_comput_fk}).  Therefore, each DFT mode from each realization of the flow is a DMD mode that oscillates at the corresponding DFT frequency.  Since each eigenvalue is repeated $N_{e}$ times, any linear combination of the DFT modes at a given frequency is also a DMD mode.

\subsection{DMD and SPOD}
\label{Sec:DMD_SPOD_connect}

Overall, for either the simple approach in which a separate $\td{A}$ operator is defined for each flow realization or the more sophisticated approach in which multiple realizations are incorporated into a single ensemble DMD computation, the end result is a \emph{set} of DMD modes at each frequency consisting of the DFT mode of each realization (or a linear combination thereof).  Because of the random nature of turbulent flows and the uncertainty inherent to the DFT, the mode for a given frequency obtained from each flow realization will in general be different.  This statistical variability may seem undesirable, but in fact it exposes an important characteristic of turbulent flows -- the behavior at a given frequency cannot be described by a single deterministic mode, i.e, the cross-spectral density tensor is not rank one.  In light of this, how can useful information be extracted from this set of differing DMD modes?  A sensible approach is to search for functions that best represent the ensemble of DMD modes at each frequency; these are given precisely by the SPOD modes at that frequency. In other words, SPOD modes provide the optimal basis, as defined by (\ref{Eq:SPOD_opt}), for describing the variability within the ensemble of DMD modes.  Said the other way around, each DMD mode in the ensemble at a particular frequency represents one of the possible ways the coherent structures represented by the SPOD modes can coexist in one particular realization of the turbulent flow.

The preceding analysis of the ensemble DMD problem enables us to make a still stronger connection between DMD and SPOD for stationary flows.  As already mentioned, any linear combination of the set of DFT modes at a given frequency is also a DMD mode.  Recall from~(\ref{Eq:SPOD_comput_ModeRecover}) that each SPOD mode can be written as a linear combination of the DFT modes used to estimate the cross-spectral density tensor.  Therefore, each SPOD mode is itself a DMD mode.  Specifically, SPOD modes are given by the particular linear combination, or weighted average, of DFT/DMD modes that yields uncorrelated modes that optimally capture the flow energy, as defined by~(\ref{Eq:SPOD_opt}).  Therefore, SPOD modes are ensemble DMD modes that have been \emph{optimally averaged} to provide the best possible representation of the second-order space-time flow statistics.  This means that SPOD modes are dynamically significant in the same sense as standard DMD modes, but at the same time optimally describe the random nature of turbulent flows.  This connection between SPOD and DMD suggests that SPOD could be directly related to the concept of the stochastic Koopman operator \citep{Mezic:2005}, but this remains to be explored in detail.


\section{Spectral POD and resolvent analysis}
\label{Sec:Resolvent}

In this section, a connection is made between Spectral POD and resolvent analysis for stationary flows.  Resolvent analysis \citep{Jovanovic:2005, Bagheri:2009, Sipp:2010, Mckeon:2010} identifies modes that optimally describe the linear growth/amplification mechanisms within the Navier-Stokes equations and is based on analysis of the linearized Navier-Stokes equations rather than time-resolved data.  This approach has been previously related to other modal decomposition techniques.  \cite{Dergham:2013} showed that resolvent modes can be used to approximate the controllability Gramian of a linear time-invariant system, the eigenvectors of which are equivalent to space-only POD modes if the system is forced by white noise \citep{Rowley:2005, Bagheri:2009}.  \cite{Sharma:2016} recently suggested a connection between DMD/Koopman operator theory and resolvent analysis.  Specifically, they showed that the resolvent operator relates Koopman modes of the input and output for a flow on an attractor.

The connection that we make between SPOD and resolvent modes is based on a statistical perspective on the resolvent-mode expansion of stationary turbulent flows.  In the following sections, we provide a brief development of the basic resolvent theory to introduce our notation and terminology, make a connection between SPOD and resolvent modes, and discuss the implications of this connection.


\subsection{Resolvent analysis}
\label{Sec:Resolvent_formulation}

We begin with nonlinear flow equations of the form
\begin{equation}
\label{Eq:resolvent_F}
\frac{\partial \vc{\qvec}}{\partial t} = \vc{\mathcal{F}}\left( \vc{\qvec} \right),
\end{equation}
where $\vc{\qvec}$ is a state-vector of flow variables.  The compressible Navier-Stokes equations are naturally written in the form of~(\ref{Eq:resolvent_F}), whereas the incompressible Navier-Stokes equations can be written in this form by eliminating the pressure by projecting the velocity field onto a divergence-free basis.  Substituting the standard Reynolds decomposition
\begin{equation}
\label{Eq:resolvent_reynolds}
\vc{\qvec}\left(\vc{x},t \right) = \bar{\vc{\qvec}}\left(\vc{x}\right) + \vc{\qvec}^{\prime}\left(\vc{x},t \right) 
\end{equation}
into~(\ref{Eq:resolvent_F}) and isolating the terms that are linear in $\vc{\qvec}^{\prime}$ yields an equation of the form
\begin{equation}
\label{Eq:resolvent_LNS_time}
\frac{\partial \vc{\qvec}^{\prime}}{\partial t} - \mathcal{A}\left(\bar{\vc{\qvec}}\right) \vc{\qvec}^{\prime} = \vc{h}\left(\bar{\vc{\qvec}},\vc{\qvec}^{\prime}\right),
\end{equation}
where 
\begin{equation}
\label{Eq:resolvent_A}
\mathcal{A}\left(\bar{\vc{\qvec}}\right) = \frac{\partial \vc{\mathcal{F}}}{\partial \vc{\qvec}}\left(\bar{\vc{\qvec}}\right)
\end{equation}
is the linearized flow operator and $\vc{h}$ contains the remaining nonlinear terms as well as any exogenous inputs such as incoming fluctuations at boundaries or environmental disturbances.  

To make the ensuing analysis as flexible as possible, it is useful to introduce two new variables: an input variable $\vc{\eta}$ and an output variable $\vc{y}$ defined by the relations
\begin{equation}
\label{Eq:resolvent_input_time}
\vc{h}(\vc{x},t) = \mathcal{B} \vc{\eta}(\vc{x},t)
\end{equation}
and
\begin{equation}
\label{Eq:resolvent_output_time}
\vc{y}(\vc{x},t) = \mathcal{C} \vc{\qvec}^{\prime}(\vc{x},t),
\end{equation}
respectively.  The linear operator $\mathcal{C} = \mathcal{C}(\vc{x})$ can be used to select certain flow variables and/or spatial regions that are of particular interest, or in general any linear function of the state perturbation.  Similarly, the linear operator $\mathcal{B} = \mathcal{B}(\vc{x})$ limits the allowable form of $\vc{h}$, which can be useful for enforcing known properties of the nonlinearity in certain flows, e.g., that the incompressible continuity equation is linear so the corresponding component of $\vc{h}$ must be zero.  Since the flow is stationary, all quantities can be Fourier decomposed and (\ref{Eq:resolvent_LNS_time}), (\ref{Eq:resolvent_input_time}), and (\ref{Eq:resolvent_output_time}) can be written in the frequency domain as
\begin{equation}
\label{Eq:resolvent_LNS_freq}
\left(\ii 2\pi f \mathcal{I} - \mathcal{A} \right) \hat{\vc{\qvec}} = \hat{\vc{h}} =  \mathcal{B} \hat{\vc{\eta}}
\end{equation}
and
\begin{equation}
\label{Eq:resolvent_output_freq}
\hat{\vc{y}}= \mathcal{C} \hat{\vc{\qvec}}.
\end{equation}

Following \cite{Mckeon:2010}, the basic objective of resolvent analysis is to find pairs of inputs and outputs at each frequency that are optimal in terms of their linear gain
\begin{equation}
\label{Eq:resolvent_gain}
\sigma^{2} = \frac{\langle \hat{\vc{y}},\hat{\vc{y}}\rangle_{y}}{\langle \hat{\vc{\eta}},\hat{\vc{\eta}}\rangle_{\eta}},
\end{equation}
where
\begin{equation}
\label{Eq:resolvent_ip_y}
\langle \hat{\vc{y}}_{1}, \hat{\vc{y}}_{2}  \rangle_{y} = \int \limits_{\Omega} \hat{\vc{y}}_{2}^{*}(\vc{x},f) \tc{W}_{y}(\vc{x}) \hat{\vc{y}}_{1}(\vc{x},f) d \vc{x}
\end{equation}
and
\begin{equation}
\label{Eq:resolvent_ip_f}
\langle \hat{\vc{\eta}}_{1}, \hat{\vc{\eta}}_{2}  \rangle_{\eta} = \int \limits_{\Omega} \hat{\vc{\eta}}_{2}^{*}(\vc{x},f) \tc{W}_{\eta}(\vc{x}) \hat{\vc{\eta}}_{1}(\vc{x},f) d \vc{x}
\end{equation}
are inner products on the output and input spaces, respectively.

Using~(\ref{Eq:resolvent_LNS_freq}) and~(\ref{Eq:resolvent_output_freq}), the input and output are related as
\begin{equation}
\label{Eq:intro_model_IO}
\hat{\vc{y}}(\vc{x},f) = \mathcal{R}(\vc{x},f) \hat{\vc{\eta}}(\vc{x},f),
\end{equation}
where
\begin{equation}
\label{Eq:intro_model_R}
\mathcal{R}(\vc{x},f) = \mathcal{C}(\vc{x}) \left(\ii 2 \pi f \mathcal{I} - \mathcal{A}(\vc{x}) \right)^{-1} \mathcal{B}(\vc{x})
\end{equation}
is termed the resolvent operator.  Note that $\left(\ii 2\pi f \mathcal{I} - \mathcal{A} \right)$ is invertible for all $f$ if $\mathcal{A}$ is stable; i.e., the real part of all of its eigenvalues is negative. 

The optimal inputs and outputs are defined by the Schmidt decomposition of the resolvent operator
\begin{equation}
\label{Eq:resolvent_schmidt}
\mathcal{R}(\vc{x},f) = \sum \limits_{j = 1}^{\infty} \sigma_{j}(f) \hat{\vc{u}}_{j}(\vc{x},f) \otimes \hat{\vc{v}}_{j}(\vc{x},f),
\end{equation}
defined in terms of the inner products~(\ref{Eq:resolvent_ip_y}) and~(\ref{Eq:resolvent_ip_f}).  The input modes $\hat{\vc{v}}_{j}$ and output modes $\hat{\vc{u}}_{j}$ provide complete bases for their respective spaces and are orthogonal in their respective inner products (i.e., $\langle \hat{\vc{u}}_{j}, \hat{\vc{u}}_{k}  \rangle_{y} = \langle \hat{\vc{v}}_{j}, \hat{\vc{v}}_{k}  \rangle_{\eta} = \delta_{jk}$).  The modes are ordered by their singular value $\sigma_{j}$, and the gain between $\hat{\vc{v}}_{j}$ and $\hat{\vc{u}}_{j}$ is $\sigma_{j}^{2}$.  The resolvent operator is said to be \emph{low rank} if the magnitude of the singular values falls off rapidly with increasing $j$.  

Since the bases are complete, the output can be expanded as
\begin{equation}
\label{Eq:resolvent_expansion}
\hat{\vc{y}}(\vc{x},f)  = \sum \limits_{j=1}^{\infty} \hat{\vc{u}}_{j}(\vc{x},f) \sigma_{j}(f) \beta_{j}(f),
\end{equation}
where
\begin{equation}
\label{Eq:resolvent_beta}
\beta_{j}(f) = \langle  \hat{\vc{\eta}}(\vc{x},f), \hat{\vc{v}}_{j}(\vc{x},f) \rangle_{\eta}
\end{equation}
is the projection of the forcing onto the $j$-th input mode \citep{Mckeon:2010}.  In some flows, reduced-order models have been obtained by retaining a limited number of terms in the expansion when the resolvent operator is low rank \citep[e.g.,][]{Beneddine:2016, Gomez:2016b}.


\subsection{Spectral densities in terms of resolvent modes}
\label{Sec:Resolvent_stats}

A connection between resolvent analysis and SPOD can be obtained by writing the cross-spectral density tensor in terms of resolvent modes.  To do so, it is convenient to write the cross-spectral density tensor as
\begin{equation}
\label{Eq:resolvent_Syy_def}
\tc{S}_{yy}(\vc{x},\vc{x}^{\prime},f) = E\{ \hat{\vc{y}}(\vc{x},f) \hat{\vc{y}}^{*}(\vc{x}^{\prime},f) \},
\end{equation}
which can be proven to be equivalent to the previous definition given by~(\ref{Eq:SPOD_CrossCor_stationary_FT}) \citep{Bendat:1990}.  Inserting the resolvent-mode expansion~(\ref{Eq:resolvent_expansion}) leads to the expression
\begin{equation}
\tc{S}_{yy}(\vc{x},\vc{x}^{\prime},f) = \sum \limits_{j=1}^{\infty} \sum \limits_{k=1}^{\infty} \hat{\vc{u}}_{j}(\vc{x},f) \hat{\vc{u}}^{*}_{k}(\vc{x}^{\prime},f) \sigma_{j}(f)     \sigma_{k}(f) \ts{S}_{\beta_{j} \beta_{k}}(f), \label{Eq:resolvent_Syy_4}
\end{equation}
where
\begin{equation}
\label{Eq:resolvent_Sbb_def}
\ts{S}_{\beta_{j} \beta_{k}}(f)  = E\{ \beta_{j}(f) \beta^{*}_{k}(f)  \}
\end{equation}
is the scalar-valued cross-spectral density between the $j$-th and $k$-th expansion coefficients.  In obtaining~(\ref{Eq:resolvent_Sbb_def}), the output resolvent modes and singular values were moved outside of the expectation operator, which is permitted because they are deterministic quantities, depending only on the resolvent operator and the inner products.  On the other hand, the expansion coefficients depend on the forcing term $\hat{\vc{\eta}}$, which is stochastic due the the random nature of turbulent flows, so the expansion coefficients must be described by their cross-spectral density.  This fact is central to the remainder of this section.

A connection between between SPOD and resolvent analysis is obtained by choosing the SPOD quantity of interest $\vc{q}$ and the resolvent output $\vc{y}$ to be the same, with the same inner products, $\langle \cdot , \cdot \rangle_{x} = \langle \cdot , \cdot \rangle_{y}$.  Then the cross-spectral density tensors $\tc{S}$ and $\tc{S}_{yy}$ are identical and the SPOD expansion~(\ref{Eq:SPOD_diag_freq}) and resolvent-mode expansion~(\ref{Eq:resolvent_Syy_4}) can be equated,
\begin{subequations}
\label{Eq:resolvent_Syy_SPODandRES}
\begin{align}
\label{Eq:resolvent_Syy_SPODandRES1}
\tc{S}_{yy}(\vc{x},\vc{x}^{\prime},f) &= \sum \limits_{j = 1}^{\infty} \lambda_{j}(f) \vc{\psi}_{j}(\vc{x},f) \vc{\psi}_{j}^{*}(\vc{x}^{\prime},f)  
\\ &= \sum \limits_{j=1}^{\infty} \sum \limits_{k=1}^{\infty} \hat{\vc{u}}_{j}(\vc{x},f) \hat{\vc{u}}^{*}_{k}(\vc{x}^{\prime},f) \sigma_{j}(f)     \sigma_{k}(f) \ts{S}_{\beta_{j} \beta_{k}}(f).\label{Eq:resolvent_Syy_SPODandRES2}
\end{align}
\end{subequations}
The implications of~(\ref{Eq:resolvent_Syy_SPODandRES}) will be explored in the following two subsections.


\subsection{Special case: uncorrelated expansion coefficients}
\label{Sec:Resolvent_uncor}

The resolvent-mode expansion of the cross-spectral density tensor can be simplified in the special case in which the $\beta_{j}$ expansion coefficients are uncorrelated from one another, which is expressed as $\ts{S}_{\beta_{j} \beta_{k}}(f) = \mu_{j}(f) \delta_{jk}$.  Substituting this into~(\ref{Eq:resolvent_Syy_SPODandRES2}) leads to the simplified expression
\begin{subequations}
\label{Eq:resolvent_Syy_SPODandRES_white}
\begin{align}
\tc{S}_{yy}(\vc{x},\vc{x}^{\prime},f) &= \sum \limits_{j = 1}^{\infty}  \vc{\psi}_{j}(\vc{x},f) \vc{\psi}_{j}^{*}(\vc{x}^{\prime},f) \lambda_{j}(f)
\\ &= \sum \limits_{j=1}^{\infty} \hat{\vc{u}}_{j}(\vc{x},f) \hat{\vc{u}}^{*}_{j}(\vc{x}^{\prime},f) \sigma_{j}^{2}(f) \mu_{j}(f). 
\end{align}
\end{subequations}
Since the diagonalization of a normal operator via a basis that is orthogonal in a given inner product is unique, (\ref{Eq:resolvent_Syy_SPODandRES_white}) shows that \emph{the sets of SPOD and resolvent modes are identical}.  Specifically, the resolvent mode with maximum $\sigma_{j}^{2} \mu_{j}$ corresponds to the first SPOD mode, and so on.  If $\mu_{j} = 1$ for every $j$, then the ordering of the two sets of modes is the same and $\sigma_{j}^{2}(f) = \lambda_{j}(f)$, $\hat{\vc{u}}_{j}(\vc{x},f) = \vc{\psi}_{j}(\vc{x},f)$.

The conditions for which the expansion coefficients are uncorrelated can be elucidated by manipulating the definition of $\ts{S}_{\beta_{j} \beta_{k}}$ using properties of inner products, 
\begin{equation}
\label{Eq:resolvent_Sbb}
\ts{S}_{\beta_{j} \beta_{k}}= E\{ \langle \hat{\vc{\eta}}, \hat{\vc{v}}_{j} \rangle_{\eta} \langle \hat{\vc{\eta}}, \hat{\vc{v}}_{k} \rangle_{\eta}^{*} \} = E\{ \langle \langle \hat{\vc{\eta}}, \hat{\vc{v}}_{j} \rangle_{\eta} \hat{\vc{\eta}}^{*}, \hat{\vc{v}}_{k}  \rangle_{\eta}^{*} \} =  \langle \langle \tc{S}_{\eta\eta} ,\hat{\vc{v}}_{j} \rangle_{\eta}^{*}, \hat{\vc{v}}_{k}  \rangle_{\eta}^{*}, 
\end{equation}
where $\tc{S}_{\eta\eta}(\vc{x},\vc{x}^{\prime},f) = E\{ \hat{\vc{ \eta}}({\vc{ x}},f) \hat{\vc{ \eta}}^{*}({\vc{ x}^{\prime}},f)\}$ is the cross-spectral density tensor of the input $\vc{\eta}$.  Recalling the orthogonality of the resolvent output modes, the last form of~(\ref{Eq:resolvent_Sbb}) shows that $\ts{S}_{\beta_{j} \beta_{k}} = \mu_{j} \delta_{jk}$ only if $\langle \tc{S}_{\eta\eta}(\vc{x},\vc{x}^{\prime},f), \hat{\vc{ v}}_{j} ({\vc{ x}^{\prime}},f) \rangle_{\eta}^{*} = \mu_{j}\hat{\vc{ v}}_{j} ({\vc{ x}},f)$.  Using the definition of the inner product, this can be written as
\begin{equation}
\label{Eq:resolvent_Sff_eigprob}
\int \limits_{\Omega} {\tc{ S}}_{\eta\eta}({\vc{ x}},{\vc{ x}}^{\prime}, f^{\prime})   {\tc{ W}_{\eta}}(\vc{x}^{\prime}) \vc{v}_{j}({\vc{ x}^{\prime}},f^{\prime}) d {\vc{ x}}^{\prime}  = 
\mu(f^{\prime})   \vc{v}_{j}({\vc{ x}},f^{\prime}).
\end{equation}
This is precisely the eigenvalue problem that defines the SPOD modes of the \emph{input}.  Thus, uncorrelated expansion coefficients imply that the resolvent input modes are the same as the SPOD modes of the input (if there are repeated eigenvalues/singular values then the corresponding modes must span the same subspace).  Conversely, if the SPOD modes of the input are the same as the input resolvent modes, inserting the SPOD expansion of $\tc{S}_{\eta\eta}$ into~(\ref{Eq:resolvent_Sbb}) shows that the expansion coefficients are uncorrelated.  Therefore, the expansion coefficients are uncorrelated if and only if the resolvent input modes correspond exactly with the SPOD modes of the input.

On inspection of~(\ref{Eq:resolvent_Sff_eigprob}), the stronger condition that $\mu_{j} = 1$ for every $j$ requires that 
\begin{equation}
\label{Eq:resolvent_SWeqIdelta}
\tc{S}_{\eta\eta}(\vc{x},\vc{x}^{\prime},f){\tc{ W}}_{\eta}({\vc{ x}^{\prime}}) = \tc{I} \delta(\vc{x} - \vc{x}^{\prime}). 
\end{equation} 
When ${\tc{ W}}_{\eta} = \tc{I}$, this condition takes on physical significance: $\tc{S}_{\eta\eta}(\vc{x},\vc{x}^{\prime},f) = \tc{I} \delta(\vc{x} - \vc{x}^{\prime})$ corresponds to an input that is completely uncorrelated in space and time and has unit amplitude everywhere, i.e., unit-amplitude white noise.  While the nonlinear forcing terms in any real flow are unlikely to be white, this approximation has been shown to be reasonable in some flows and is frequently used for the construction of low-order models \citep{Farrell:1993,Farrell:1996,Farrell:2001}, including resolvent-based models \citep{Jovanovic:2001,Bagheri:2009,Sipp:2010,Moarref:2012,Dergham:2013}.  In these models, then, resolvent modes can be understood as approximations of SPOD modes.  The equivalence of SPOD and resolvent modes in the case of white-noise forcing has been recently pointed out by multiple authors working in the area of jet-noise modeling \citep{Towne:2015,Towne:2016b,Semeraro:2016}.


\subsection{General case: flow reconstruction with correlated expansion coefficients}
\label{Sec:Resolvent_recon}

The nonlinear forcing terms in real flows are not white \citep{Zare:2017,Towne:2017a}, which leads to correlated resolvent-mode expansion coefficients.  In this case, a more general relationship between SPOD and resolvent modes can be derived that provides insight into how the resolvent-mode expansion coefficients can be chosen to optimally reconstruct the flow.

A number of strategies for choosing the resolvent-mode expansion coefficients have been proposed and employed in the last few years.  Perhaps the most straight-forward approach, used for example by \cite{Jeun:2016}, is to use simulation data to compute $\vc{\eta}$, estimate $\hat{\vc{\eta}}$ using a single DFT of the time series, and calculate the expansion coefficients using their definition in~(\ref{Eq:resolvent_beta}).  However, this method suffers from the uncertainty inherent in the DFT, as discussed earlier.  Methods using converged statistics of the flow must be used to eliminate this uncertainty.

\cite{Moarref:2014} formulated a convex optimization problem to find a vector of expansion coefficients that optimally reproduce experimentally-measured streamwise energy spectra, averaged in the other directions, for a turbulent channel flow.  The coefficients that were optimal in the sense they defined led to qualitative agreement with the statistics, but notably the error did not tend toward zero as the number of resolvent modes was increased.

\cite{Gomez:2016a} derived an equation for the power spectral densities of the expansion coefficients, which define their magnitude.   These are obtained by projecting the time-varying forcing term $\vc{\eta}$ onto the resolvent input basis for a range of frequencies, and computing the power spectral densities of these terms.  In a separate paper \citep{Gomez:2016b}, the same authors suggested a somewhat different method using a time-series of data at one location in the flow.  In this approach, the resolvent expansion is inverse Fourier transformed back into the time domain and the coefficients for all frequencies are simultaneously determined by matching the time series in a least-squares sense.

\cite{Beneddine:2016} used two different approaches for finding the coefficient for (only) the leading resolvent mode at each frequency for the flow over a backward-facing step.  In the first approach, the amplitudes of the expansion coefficients were chosen so that the rank-one resolvent-mode reconstruction matched the power spectral density of the spanwise-averaged streamwise velocity at a single location in the flow.  In the second approach, the expansion coefficient for the leading mode at each frequency was chosen so that the rank-one resolvent expansion matched the leading SPOD mode, again at a single location.  They found this second method to provide improved estimates of the power spectral density at other spatial locations in the flow.  Their results were further improved by attempting to match the SPOD mode at several locations using a least-squares approach.

All of these methods assume that each expansion coefficient can be described by a single deterministic amplitude and phase.  In contrast, (\ref{Eq:resolvent_Syy_4}) shows that the cross-spectral densities of the expansion coefficients are required to properly reconstruct the second-order statistics of the flow.  Notably, this includes the power spectral density, which is obtained by setting $\vc{x}^{\prime} = \vc{x}$.  Treating the expansion coefficients as deterministic quantities imposes a fundamental limitation on the quality of the flow-approximation that can be achieved by the resolvent model.  For clarity, we will use the symbol $b_{j}$ in place of $\beta_{j}$ to denote these deterministic expansion coefficients.  Regardless of how the $b_{j}$ values are chosen, the approximation of $\ts{S}_{\beta_{j}\beta_{k}}$ implied by these coefficients is $\ts{S}_{b_{j}b_{k}} = b_{j}b_{k}^{*}$.  Making this substitution in~(\ref{Eq:resolvent_Syy_4}), the resulting cross spectral density tensor can be factored into the form
\begin{equation}
\label{Eq:resolvent_Syy_factored}
\tc{S}_{yy}(\vc{x},\vc{x}^{\prime},f) = \left( \sum \limits_{j=1}^{\infty} \hat{\vc{u}}_{j}(\vc{x},f) \sigma_{j}(f)   b_{j}(f) \right)\left( \sum \limits_{k=1}^{\infty} \hat{\vc{u}}_{k}(\vc{x}^{\prime},f) \sigma_{k}(f)   b_{k}(f) \right)^{*}.
\end{equation}
Since the right-hand-side is the product of two vectors, this approximation of the cross-spectral density tensor has rank equal to one, whereas the correct cross-spectral density tensor has rank equal to the number of nonzero SPOD eigenvalues at each frequency.  This limits the quality of the flow-approximation that can be achieved with the resolvent model using deterministic expansion coefficients, no matter how their amplitudes and phases are chosen.  Specifically, we will see that convergent approximations of the power spectral and cross-spectral densities are not possible when the statistical nature of the expansion coefficients is not respected.

A few specific choices of $b_{j}$ are worth a closer look.  First, we consider the case where the $b_{j}$ values are chosen to have have the correct magnitude, so that their power spectral densities $\ts{S}_{b_{j}b_{j}} = |b_{j}|^{2}$ match the correct values $\ts{S}_{\beta_{j}\beta_{j}}$.  This could be accomplished, for example, using the approach of \cite{Gomez:2016a}.  Equation~(\ref{Eq:resolvent_Syy_4}) shows that the cross-spectral and power spectral densities depend also on off-diagonal terms $\ts{S}_{\beta_{j}\beta_{k}}$.  These will be replaced by the values $\ts{S}_{b_{j}b_{k}} = b_{j}b_{k}^{*}$, and there is no reason to expect this to provide a good approximation.  For example, in the case of uncorrelated expansion coefficients, the off-diagonal terms of $\ts{S}_{\beta_{j}\beta_{k}}$ are zero, but $b_{j}b_{k}^{*}$ is nonzero unless either mode $j$ or mode $k$ has zero magnitude.  Thus, even though the power spectral densities of the expansion coefficients are correct, the power spectral densities of the output will be incorrect.  To make the error explicit, (\ref{Eq:resolvent_Syy_factored}) can be manipulated into the form
\begin{multline}
\label{Eq:resolvent_Syy_RightWrong}
\tc{S}_{yy}(\vc{x},\vc{x}^{\prime},f) = \sum \limits_{j=1}^{\infty} \hat{\vc{u}}_{j}(\vc{x},f) \hat{\vc{u}}^{*}_{j}(\vc{x}^{\prime},f) \sigma_{j}^{2}(f)     |b_{j}(f)|^{2}  \\ + \sum \limits_{j=1}^{\infty} \sum\limits_{{\substack{k = 1 \\ k \neq j}}}^{\infty} \hat{\vc{u}}_{j}(\vc{x},f) \hat{\vc{u}}^{*}_{k}(\vc{x}^{\prime},f) \sigma_{j}(f)     \sigma_{k}(f) b_{j}(f) b_{k}^{*}(f).
\end{multline}
Using the power-spectral-density-based expansion coefficients, the first term in~(\ref{Eq:resolvent_Syy_RightWrong}) is correct, while the second term is incorrect no matter how many terms are included in the expansion.  

An alternative approach would be to choose the $b_{j}$ coefficients such that the resolvent-mode expansion optimally matches the data, as defined by the inner product $\langle \cdot,\cdot \rangle_{y}$, in the limit of high $N_{r}$.  Spectral POD theory tells us that the optimal representation of the data under the rank-one limitation is given by the leading SPOD mode.  Therefore the optimal $b_{j}$ values are those that reconstruct the leading SPOD mode at each frequency.  These values can be obtained by directly projecting the leading SPOD mode onto the resolvent modes, giving 
\begin{equation}
\label{Eq:resolvent_bj_opt}
b_{j}(f) = \frac{\sqrt{\lambda_{1}(f)}}{\sigma_{j}(f)} \langle {\vc{\psi}}_{1},\hat{\vc{u}}_{j} \rangle_{y}.
\end{equation}
By construction, these expansion coefficients guarantee that the first SPOD mode at each frequency is recovered when enough terms are retained in the resolvent-mode expansion.  Even though this is the optimal expansion possible using a deterministic vector of expansion coefficients, it will provide a good approximation of the flow only if $\lambda_{1}(f) \gg \lambda_{2}(f)$ over the range of relevant frequencies.  Therefore, the quality of the resolvent-mode expansion using optimal (or any other) deterministic coefficients is dependent first and foremost on the low-rank nature of the cross-spectral density tensor rather than of the resolvent operator.  

In contrast, convergent approximations of the flow statistics can be achieved if the statistical nature of the resolvent-mode expansion coefficients is incorporated at the outset.  Accordingly, we propose a practical method for approximating the required cross-spectral densities $\ts{S}_{\beta_{j}\beta_{k}}$ using the leading SPOD modes.  The resolvent mode expansion coefficients are related to SPOD modes by comparing the SPOD and resolvent expansions of the cross-spectral density tensor given in~(\ref{Eq:resolvent_Syy_SPODandRES}).  By taking two successive inner products of both expansions with respect to $\hat{\vc{u}}_{j}$ and $\hat{\vc{u}}_{k}$ and dividing by $\sigma_{j}(f)$ and $\sigma_{k}(f)$, we obtain the expression
\begin{equation}
\label{Eq:resolvent_Sbb_SPOD}
\ts{S}_{\beta_{j}\beta_{k}} = \sum \limits_{n = 1}^{\infty} \frac{\lambda_{n}(f)}{\sigma_{j}(f)\sigma_{k}(f)} \gamma_{nj}(f) \gamma_{nk}^{*}(f),
\end{equation}
where
\begin{equation}
\label{Eq:resolvent_gamma}
\gamma_{jk}(f) = \langle {\vc{\psi}}_{j},\hat{\vc{u}}_{k} \rangle_{y}
\end{equation}
is the projection between the $j$-th SPOD mode and the $k$-th resolvent output mode.  

Use of the first $N_{s}$ terms in~(\ref{Eq:resolvent_Sbb_SPOD}) to approximate $\ts{S}_{\beta_{j}\beta_{k}}$ and a resolvent-mode expansion with $N_{r}$ terms leads to the approximation of the second-order flow statistics 
\begin{equation}
\label{Eq:resolvent_Syy_Sbb_SPOD}
\tc{S}_{yy}(\vc{x},\vc{x}^{\prime},f) \approx \sum \limits_{n = 1}^{N_{s}}  \lambda_{n}(f)  \left(\sum \limits_{j=1}^{N_{r}} \gamma_{nj}(f) \hat{\vc{u}}_{j}(\vc{x},f) \right)  \left( \sum \limits_{k=1}^{N_{r}}   \gamma_{nk}(f)  \hat{\vc{u}}_{k}(\vc{x}^{\prime},f) \right)^{*}.
\end{equation}
From the definition of $\gamma_{jk}$ and the fact that the input resolvent modes are complete, we have that 
\begin{equation}
\label{Eq:resolvent_highNrLimit}
\sum \limits_{j=1}^{N_{r}} \gamma_{nj}(f) \hat{\vc{u}}_{j}(\vc{x},f) \to \vc{\psi}_{n}(\vc{x},f) \quad \mathrm{as} \quad N_{r} \to \infty.
\end{equation}
Therefore, using the first $N_{s}$ SPOD modes to approximate $\ts{S}_{\beta_{j}\beta_{k}}$ leads to a resolvent-mode expansion capable of capturing the first $N_{s}$ SPOD modes, if enough resolvent modes are retained.  The approximation is therefore convergent as $N_{s}$ and $N_{r}$ are increased, and if the SPOD spectrum falls off rapidly, accurate approximations of the flow are possible using only a few SPOD modes to compute the expansion coefficients.  If a sufficiently high value of $N_{s}$ is used, then the quality of the approximation that can be obtained with a limited number of resolvent modes is determined by the convergence of the limit in~(\ref{Eq:resolvent_highNrLimit}).  It is notable that the resolvent gains do not appear explicitly in~(\ref{Eq:resolvent_Syy_Sbb_SPOD}) or~(\ref{Eq:resolvent_highNrLimit}).  Instead, the efficacy of the resolvent expansion is determined by the efficiency of the leading resolvent modes in providing a basis for the leading SPOD modes.


\section{Examples}
\label{Sec:Examples}

This section contains two example problems that demonstrate our theoretical results as well as the overall utility of SPOD.


\subsection{Complex Ginzburg-Landau equation}
\label{Sec:Examples_GL}

Our first example consists of the linearized complex Ginzburg-Landau equation, which is frequently used as a model for instabilities in spatially-evolving flows.  The equation can be written in the form of~(\ref{Eq:resolvent_LNS_time}) with
\begin{equation}
\label{Eq:exGL:A}
\mathcal{A} = -\nu \frac{\partial}{\partial x} + \gamma \frac{\partial^{2}}{\partial x^{2}} + \mu(x).
\end{equation}
The spatial dependence of the solution is a consequence of the parameter $\mu(x)$, for which we adopt the quadratic form
\begin{equation}
\label{Eq:exGL_mu}
\mu(x) = (\mu_{0} - c_{\mu}^2) + \frac{\mu_{2}}{2} x^{2}
\end{equation}
used previously by several authors \citep{Hunt:1991,Bagheri:2009, Chen:2011}.  We take $\mu_{0} = 0.23$ and all other parameters in~(\ref{Eq:exGL:A}) and~(\ref{Eq:exGL_mu}) are set to the values used by \cite{Bagheri:2009}.  The resulting model is globally stable (all eigenvalues of $\mathcal{A}$ have negative real part) but is susceptible to non-modal growth due to the non-normality of $\mathcal{A}$, which within the context of resolvent analysis leads to gains much larger than one.  The input and output operators and inner-product matrices ($\mathcal{B}$, $\mathcal{C}$, $\tc{W}_{y}$, and $\tc{W}_{\eta}$, respectively) are all set to unity.  The value of $\mu_{0}$ was chosen so that the gain of the leading resolvent mode at its peak frequency is 100 times larger than the gain of the second mode, which is a typical value for real flows.

The equations are discretized with a pseudo-spectral approach using Hermite polynomials, as in \cite{Bagheri:2009} and \cite{Chen:2011}.  The collocation points correspond to the roots of the first $N$ Hermite polynomials; following \cite{Bagheri:2009} we use $N = 220$.  This leads to a computational domain $x \in [-85,85]$, which is large enough to mimic an infinite domain.  Since the collocation points are unevenly distributed, the discretized form of the inner product would contain matrices different from the identity.  To avoid this, we define the input $\eta$ and output $y$ on a different, uniformly-spaced grid.  This is accomplished by setting the discretized forms of the input and output matrices $B$ and $C$ so that they represent an interpolation from the uniform grid to the Hermite grid and from the Hermite grid to the uniform grid, respectively.  On the uniform input/output grid, the discrete inner-product matrices $\td{W}_{\vd{y}}$ and $\td{W}_{{\text{\bf \texteta}}}$ are the identity matrix.

To generate data for our analysis, the discretized equations are stochastically excited in the time domain using forcing terms with prescribed statistics.  In the following two subsections, we consider cases of white-noise and correlated forcing, respectively.

\subsubsection{White-noise forcing}

We begin by forcing the linearized Ginzburg-Landau equations with band-limited white noise that is spatially uncorrelated.  This forcing is realized by setting the value at each grid point and at each discrete time instance to a random complex number with uniformly distributed phase (between $0$ and $2\pi$) and normally distributed amplitude with unit variance.  This signal is then low-pass filtered in time using a 10th-order finite-impulse-response filter with a cut-off frequency equal to $60\%$ of the Nyquist frequency.  Additionally, the forcing is spatially limited to the interior portion of the domain using an exponential envelope of the form $\exp[ (x/L)^{p} ]$ with $L = 60$ and $p = 10$.  

\begin{sloppypar}
The equations are integrated using a fourth-order embedded Runge-Kutta method \citep{Shampine:1997}, and a total of $40000$ snapshots of the solution are collected with spacing $\Delta t = 0.5$, leading to a Nyquist frequency of $f_{\mathrm{Nyquist}} = 1$.  A large number of snapshots are required for this problem because of the slow statistical convergence of the white-noise forcing; problems with correlated forcing, as is typical in real flows, tend to require fewer snapshots.  Spectra and SPOD modes are computed as outlined in \S\ref{Sec:SPOD_comp} with a Hann window function, $75\%$ overlap, and $N_{f} = 384$.   
\end{sloppypar}


\begin{figure}[!t]
\input{fig2.tex}
\centering
\includegraphics[trim=0cm 0.0cm 0cm 0.0cm, clip=true,width=0.8\textwidth]{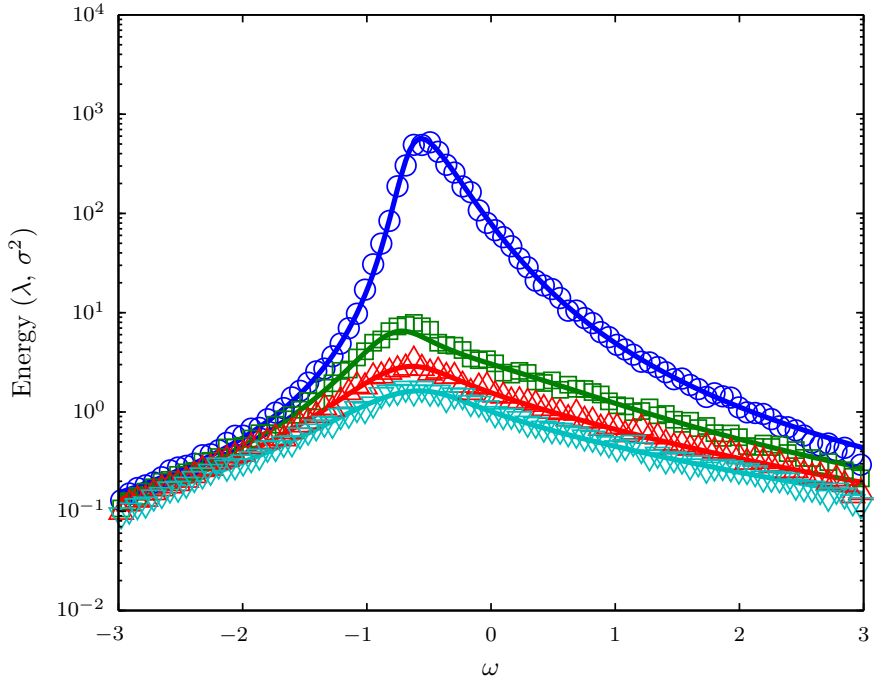}
\caption{Comparison of the first four resolvent gains $\sigma_{j}^{2}$ (lines) and SPOD eigenvalues $\lambda_{j}$ (symbols) as a function of frequency for the Ginzburg-Landau model forced with white noise: ($\Circle$, blue) mode 1; ($\square$, green) mode 2; ($\triangle$, red) mode 3; ($\triangledown$, cyan) mode 4. For each mode, every other SPOD eigenvalue has been omitted from the plot to improve readability. }
\label{fig:LamVsSig2White}
\end{figure}

Since the forcing is white and the input inner product has unit weight, the conditions are met under which resolvent and SPOD modes are theoretically equivalent.  The first four SPOD eigenvalues and resolvent gains are shown in Figure~\ref{fig:LamVsSig2White} as a function of frequency.  For consistency with previous publications, frequencies are reported in terms of the angular frequency $\omega = 2 \pi f$.  Overall, the SPOD eigenvalues agree with the resolvent gains, but two minor differences can be observed which are related to the finite level of convergence of the spectral estimates used to approximate the SPOD modes.  First, the SPOD eigenvalues are not completely smooth as a function of frequency.  This is indicative of the remaining uncertainty in the spectral estimates, which can be further reduced by averaging over more data ensembles.  Second, there is a small overshoot in the value of the second SPOD eigenvalue at the peak frequency of the first mode.  This overshoot is related to spectral leakage, which can be reduced by increasing the frequency resolution of the spectral estimates by increasing the length of the data ensembles.  When estimating the cross-spectral density using a data set of finite length, increasing the number of ensembles reduces the length of each ensemble and vice versa.  Therefore, decreasing one of the two types of errors tends to increase the other.  The proper compromise will in general be problem-dependent.  As a general rule, SPOD modes are more difficult to converge for flows that exhibit sharp spectral peaks.


\begin{figure}[!t]
\input{fig3.tex}
\centering
\includegraphics[trim=0cm 0.0cm 0cm 0.0cm, clip=true,width=0.95\textwidth]{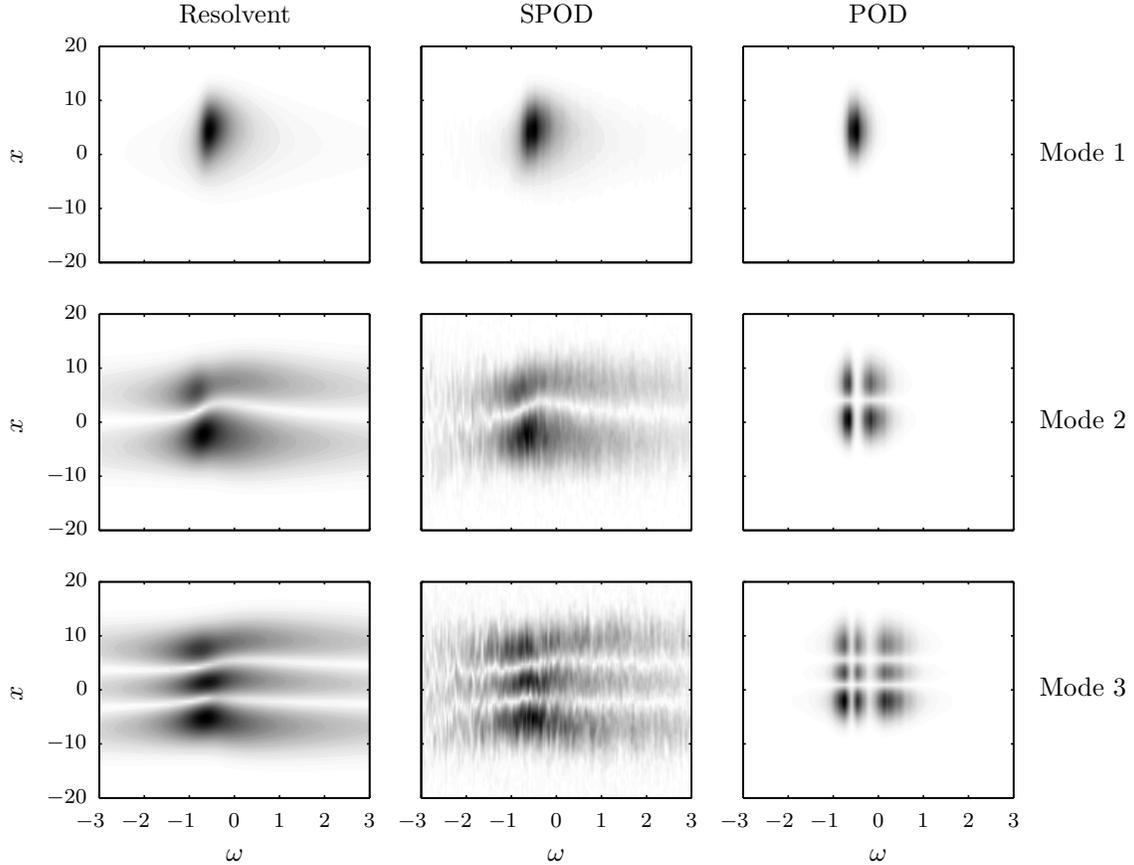}
\caption{Weighted mode shapes in $\omega$--$x$ space for the Ginzburg-Landau equation forced with white noise.  First column: resolvent modes $\sigma_{j}(\omega)|\hat{\vc{u}}_{j}(x,\omega)|$. Second column: SPOD modes $\sqrt{\lambda_{j}(\omega)}|\vc{\psi}_{j}(x,\omega)|$.  Third column: space-only POD modes $|\hat{a}_{j}(\omega)|\,|\vc{\phi}_{j}(x)|$. }
\label{fig:ModeShapesWhite}
\end{figure}

The structure of the first three resolvent and SPOD modes is depicted in Figure~\ref{fig:ModeShapesWhite} as a function of $x$ and $\omega$ (space-only POD results are also shown and will be discussed later).  Specifically, we plot the weighted quantities $\sigma_{j}(\omega)|\hat{\vc{u}}_{j}(x,\omega)|$ and $\sqrt{\lambda_{j}(\omega)}|\vc{\psi}_{j}(x,\omega)|$ to clearly show where each mode is active in $\omega$-$x$ space.  The contour levels in each plot are distributed between zero and the maximum value of the weighted mode.  


\begin{figure}[!t]
\input{fig4.tex}
\centering
\includegraphics[trim=0cm 0.0cm 0cm 0.0cm, clip=true,width=0.9\textwidth]{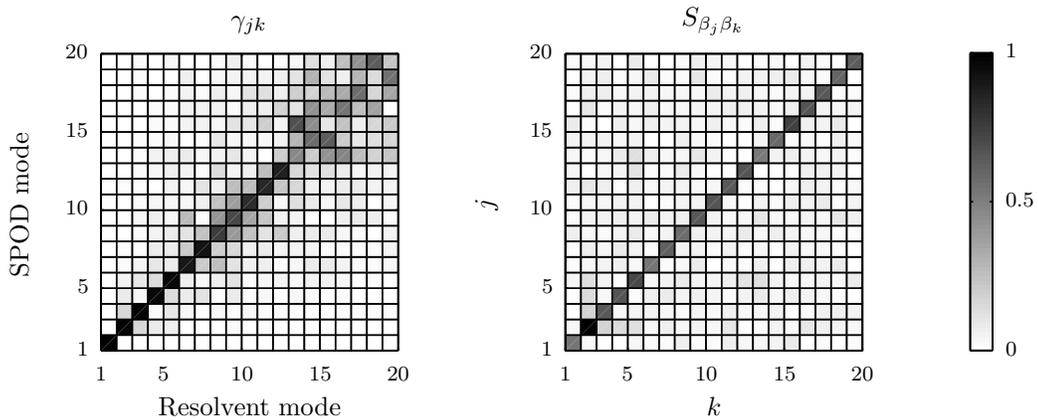}
\caption{SPOD/resolvent mode projection coefficient (left) and resolvent-mode expansion coefficient cross-spectral density (right) at the frequency of highest gain for the Ginzburg-Landau model forced with white noise. The first twelve SPOD and resolvent modes are nearly identical (diagonal $\gamma_{jk}$) because the resolvent-mode expansion coefficients are almost completely uncorrelated (diagonal $\ts{S}_{\beta_{j}\beta_{k}}$).   }
\label{fig:GamSbbWhite}
\end{figure}

The first resolvent and SPOD modes are nearly indistinguishable.  The sub-optimal SPOD modes likewise closely mimic their corresponding resolvent modes, although the limited smoothness as a function of frequency is again evident.  The match between the resolvent and SPOD modes can be quantified using the projection $\gamma_{jk}$, defined in~(\ref{Eq:resolvent_gamma}).  This is shown for the peak frequency of the leading resolvent mode, $\omega = -0.6$, in Figure~\ref{fig:GamSbbWhite}(a).   The diagonal ($j=k$) entries dominate up to $j=k=12$, indicating that the first 12 resolvent and SPOD modes are nearly the same.  Finally, Figure~\ref{fig:GamSbbWhite}(b) shows the expansion coefficient cross-spectral density tensor $\ts{S}_{\beta_{j}\beta_{k}}$ for the same frequency, normalized by its maximum value.  As expected, it is nearly diagonal, which is the root cause of the approximate equivalence of the resolvent and SPOD modes for this problem.

We also briefly compare the SPOD results with those obtained using space-only POD.  The rightmost column of Figure~\ref{fig:ModeShapesWhite} shows the $\omega$-$x$ distribution of the first three POD modes.  The frequency distribution was obtained by computing the power spectral densities of the time-dependent expansion coefficients $a_{j}(t)$ using Welch's method, which is similar to the method of \cite{Cammilleri:2013} except that we use proper spectral averaging rather than a single DFT to obtain frequency information.  The quantity plotted is then $|\hat{a}_{j}(\omega)|\,|\vc{\phi}_{j}(x)|$  As expected, each space-only POD mode represents a range of frequencies.  The first space-only POD mode is relatively similar to the leading SPOD mode, but subsequent POD modes represent structures that are quite different.  Two things stand out.  First, since the spatial shape of each space-only POD mode does not depend on frequency, they are unable to capture the frequency-dependent shape of the resolvent modes that is captured by the SPOD modes.  Second, the frequency-dependent magnitude of each sub-optimal space-only POD mode exhibits a clear dip at the frequencies at which previous modes peak.  Overall, the SPOD modes provide a better representation than the space-only POD modes of the underlying dynamics and non-normal growth mechanisms, which are completely described by the resolvent modes for this problem since the forcing is white.

\begin{figure}[!t]
\centering
\begin{overpic}[trim=-0.5cm 0.0cm 0cm -0.25cm, clip=true,width=0.9\textwidth]{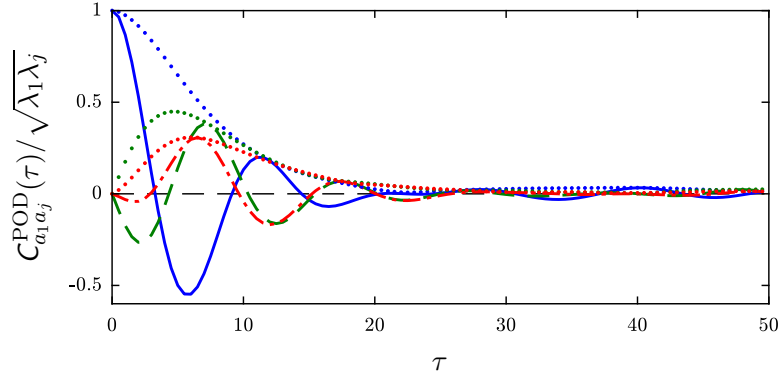}
 \put (13,10){\rotatebox{90}{$\ts{C}^{\mathrm{POD}}_{a_{1}a_{j}}(\tau) / \sqrt{\lambda_{1}\lambda_{j} }$}}
 \put (50.5,0.5){$\tau$}
\end{overpic}
\caption{Temporal correlation of the first three space-only POD expansion coefficients with the first coefficient (see~\ref{Eq:SPOD_comparePOD_CaaPOD}) for the Ginzburg-Landau model forced with white noise. Real part of the correlation for:  (\myRuleSolid[black]{1cm}{0.4pt}, blue) mode 1; (\myRuleDashed[black]{1cm}{0.4pt}, green) mode 2; (\myRuleDashDot{}, red) mode 3.  The light dotted lines show the magnitude of each correlation.    Since the autocorrelation of the first mode decays and the cross-correlation with the sub-optimal modes becomes non-zero for $\tau>0$, the space-only POD modes do not represent structures that evolve coherently in space and time.}
\label{fig:aCorrelationWhite}
\end{figure}

Finally, we demonstrate the result from \S\ref{Sec:Compare_POD} that space-only POD modes do \emph{not} represent structures that are uncorrelated from one another in time.  Figure~\ref{fig:aCorrelationWhite} shows the correlation $\ts{C}^{\mathrm{POD}}_{a_{1}a_{j}}(\tau) = E\{ a_{1}(t) a_{j}^{*}(t+\tau)\}$ of the first POD expansion coefficient with the first three expansion coefficients ($j = 1,2,3$) as a function of the temporal lag $\tau$, scaled by $\sqrt{\lambda_{1}\lambda_{j}}$.  The solid and broken lines show the real part and magnitude of the correlations, respectively.  At zero time lag, the scaled autocorrelation $\ts{C}^{\mathrm{POD}}_{a_{1}a_{1}}(0) / \lambda_{1}$ is one and the cross-correlations $\ts{C}^{\mathrm{POD}}_{a_{1}a_{k}}(0)$ are zero.  Both of these values follow theoretically from~(\ref{Eq:POD_space_a}).  However, for $\tau >0$ the autocorrelation quickly drops and the cross-correlations become nonzero.  Accordingly, the space-only POD modes do not represent structures that evolve coherently in time.


\subsubsection{Correlated forcing}

Next, we force the linearized Ginzburg-Landau equations with forcing terms that are spatially correlated.  This forcing is generated by convolving the same band-limited white noise signal used previously with a kernel of the form
\begin{equation}
\label{Eq:ex:filter}
g(x,x^{\prime}) = \frac{1}{\sqrt{2\pi} \sigma_{\eta}} \exp \left[ -\frac{1}{2} \left( \frac{x-x^{\prime}}{\sigma_{\eta}} \right)^{2} \right] \exp \left[ \ii 2\pi \frac{x - x^{\prime}}{\lambda_{\eta}} \right],
\end{equation}
where $\sigma_{\eta}$ is the standard deviation of the envelope and $\lambda_{\eta}$ is the wavelength of the filter.  This leads to a forcing that is white-in-time up to the cut-off frequency but that has non-zero spatial correlation in the form of~(\ref{Eq:ex:filter}) but with $\sigma_{\eta}$ replaced with $\sqrt{2}\sigma_{\eta}$.  This form of the forcing statistics is qualitatively similar to those of the nonlinear forcing terms in real flows, such as a turbulent jet \citep{Towne:2017a}.  We use $\sigma_{\eta} = 4$ and $\lambda_{\eta} = 20$.  The statistics of the response converge more rapidly in this case due to the correlated forcing, so we use a smaller number of snapshots, $N = 10000$, than in the uncorrelated case.


\begin{figure}[!t]
\input{fig6.tex}
\centering
\includegraphics[trim=0cm 0.0cm 0cm 0.0cm, clip=true,width=0.9\textwidth]{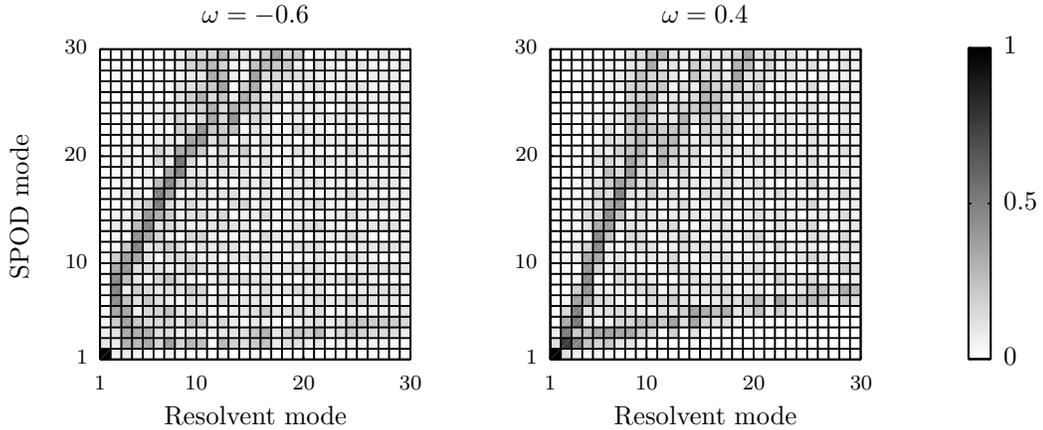}
\caption{SPOD/resolvent projection coefficients $\gamma_{jk}$ at two frequencies for the Ginzburg-Landau model forced by spatially correlated input. Since the tensors are not diagonal, the resolvent and SPOD modes are not the same in this case.}
\label{fig:GammaCorrelated}
\end{figure}

The correlation of the forcing leads to differences between the output resolvent and SPOD modes.  This is clearly demonstrated by the $\gamma_{jk}$ projection coefficients, which are shown for $\omega = -0.6$ and $0.4$ in Figure~\ref{fig:GammaCorrelated}.  The first of these frequencies is near the peak frequency of the leading resolvent mode where $\sigma_{1}^{2} \approx 560$ and $\sigma_{1}^{2}/\sigma_{2}^{2} \approx 100$, while the second is away from the peak and $\sigma_{1}^{2} \approx 20$ and $\sigma_{1}^{2}/\sigma_{2}^{2} \approx 10$.  In both cases, the diagonal character of $\gamma_{jk}$ observed for the white-noise forcing (Figure~\ref{fig:GamSbbWhite}a) is no longer observed, indicating that the SPOD and resolvent modes are no longer the same.  


\begin{figure}[!t]
\input{fig7.tex}
\centering
\includegraphics[trim=0cm 0.0cm 0cm 0.0cm, clip=true,width=0.9\textwidth]{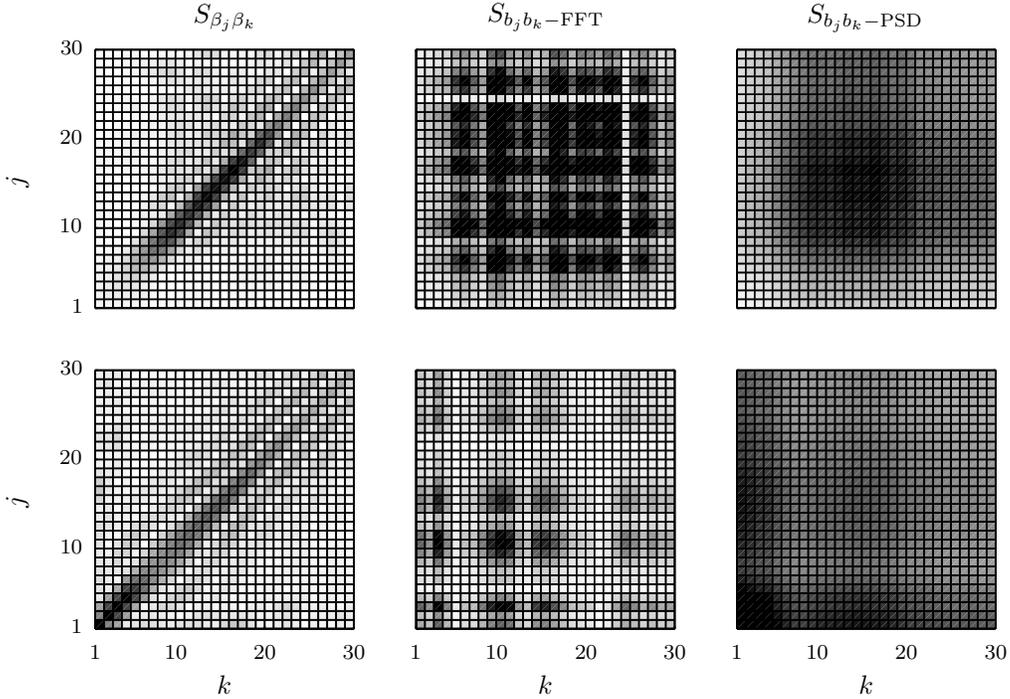}
\caption{Cross-spectral density of the resolvent-mode expansion coefficients for (top row) $\omega = -0.6$ and (bottom row) $\omega = 0.4$.  First column: correct values.  Second column: values implied by deterministic coefficients based on a single long-time DFT of the solution.  Third column: values implied by deterministic coefficients with correct power spectral density.}
\label{fig:SbbCorrelated}
\end{figure}

As discussed in \S\ref{Sec:Resolvent}, this can be attributed to the fact that the correlated forcing produces correlated resolvent-mode expansion coefficients.  The magnitude of the cross-spectral density of the expansion coefficients $\ts{S}_{\beta_{j}\beta_{k}}$ is shown in the first column of Figure~\ref{fig:SbbCorrelated} for the same two frequencies.   The strictly diagonal correlation tensor observed for the white-noise forcing (Figure~\ref{fig:GamSbbWhite}b) has been replaced by a more complicated banded structure, indicating substantial correlations between different expansion coefficients.

The expansion-coefficient correlations are not captured by standard approaches to computing the expansion coefficients that treat them as deterministic quantities.  For example, the second and third columns of Figure~\ref{fig:SbbCorrelated} show the correlations implied using the DFT- and power spectral density (PSD)-based methods discussed in \S\ref{Sec:Resolvent_recon}, both of which lead to large errors away from the main correlation bands.  

\begin{figure}[!t]
\input{fig8.tex}
\centering
\includegraphics[trim=0cm 0.0cm 0cm 0.0cm, clip=true,width=0.95\textwidth]{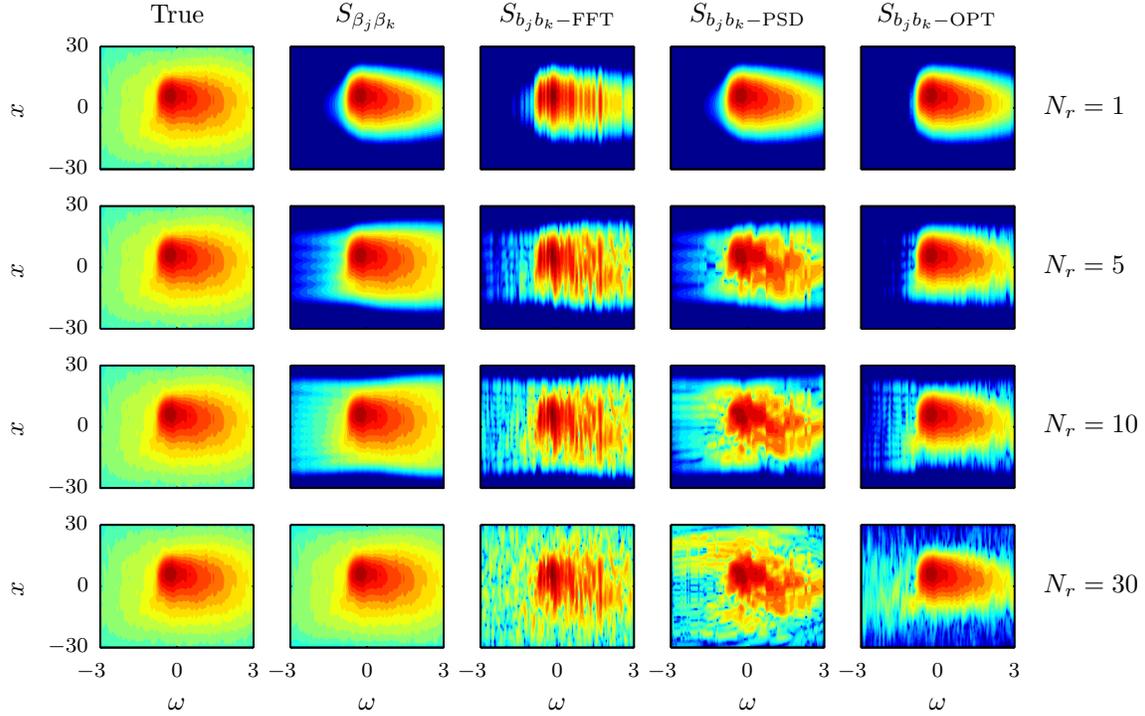}
\caption{Resolvent-mode reconstructions of the power spectral density of the solution of the Ginzburg-Landau for spatially correlated forcing using different expansion coefficients.  Rows: number of resolvent modes in the expansion increases from top to bottom.  First column: true PSD (same in every row).  Second column: reconstruction with correct statistical expansion coefficients.  Third column: reconstruction with deterministic expansion coefficients based on a single long-time DFT.  Fourth column: reconstruction with deterministic PSD-based expansion coefficients.  Fifth column: reconstruction with optimal deterministic expansion coefficients.}
\label{fig:PSDreconstructCorrelated}
\end{figure}

As indicated by~(\ref{Eq:resolvent_Syy_4}), these errors in the expansion coefficient cross-spectral density lead to errors in the resolvent-mode reconstruction of the output statistics.  Figure~\ref{fig:PSDreconstructCorrelated} compares the true power spectral density of the output as a function of $\omega$ and $x$ with the power spectral densities obtained from resolvent-mode reconstructions of the flow with $N_{r} = 1$, $5$, $10$, and $30$ modes and with the expansion coefficients specified in four different ways.  The first column shows the true power spectral density, which does not depend on $N_{r}$.  The second column shows the power spectral density obtained using the correct values of $\ts{S}_{\beta_{j}\beta_{k}}$, computed using~(\ref{Eq:resolvent_Sbb_SPOD}).  The third and fourth columns show the power spectral densities obtained using the DFT- and PSD-based deterministic expansion coefficients, respectively, and the final column shows results for the optimal $b_{j}$ values given by~(\ref{Eq:resolvent_bj_opt}).  The contour levels in all of the panels are the same and span six orders-of-magnitude with the upper bound set to the maximum value of the true power spectral density.  


All four resolvent-mode reconstructions produce similar power spectral densities for $N_{r} = 1$, but the DFT-based expansion is noticeably noisy due to the uncertainty implicit in this approach.  As the number of resolvent modes included in the expansion increases, the power spectral density obtained using the proper $\ts{S}_{\beta_{j}\beta_{k}}$ values converge to the true power spectral density.  In contrast, none of the three methods using deterministic expansion coefficients converges to the true PSD due to their incorrect implied correlations.  For the DFT- and PSD-based methods, the reconstructed power spectral densities improve with increasing $N_{r}$ in low-energy regions of $\omega$-$x$ space but do not improve in high-energy regions.  In fact, the reconstruction of the high-energy regions actually worsens with increasing $N_{r}$ for the PSD-based expansion coefficients due to accumulation of the error terms from~(\ref{Eq:resolvent_Syy_RightWrong}) as more modes are included in the expansion.  On the other hand, the inclusion of more terms in the resolvent-mode expansion using the optimal $b_{j}$ values has less effect on the low-energy regions but does not degrade the approximation of the high-energy regions.  This is because the resolvent-mode reconstruction using these expansion coefficients is converging toward the leading SPOD mode at each frequency, and these modes effectively represent high-energy regions by construction.  The expansion with optimal $b_{j}$ values is well-converged using 30 resolvent modes, so this is the best approximation of the second-order output statistics that can be obtained using deterministic expansion coefficients.


\subsection{Turbulent jet}
\label{Sec:Jet}

Our second example is a subsonic turbulent jet issued from a round convergent-straight nozzle.  The Mach number, temperature ratio, and Reynolds number of the jet are $M = U_{j}/c_{j} = 0.4$, $T_{j}/T_{\infty} = 1$, and $Re = \rho_{j} U_{j} D / \mu_{j} \approx 10^{6}$, respectively, where $U$ is the velocity, $c$ is the speed of sound, $T$ is the temperature, $\rho$ is the density, $D$ is the nozzle diameter, $\mu$ is the dynamic viscosity, and the subscripts $j$ and $\infty$ indicate mean conditions at the nozzle exit and in the ambient far-field, respectively.  

We use data from a large-eddy simulation (LES) as input for the various empirical modal decomposition methods discussed in this paper.  The simulation was performed using the compressible flow solver ``Charles'' developed at Cascade Technologies \citep{Bres:2017}, which solves the spatially-filtered compressible Navier-Stokes equations on unstructured grids using a finite-volume method and third-order Runge-Kutta time integration.  The simulation used approximately sixteen million control volumes and was run for a duration of 2000 acoustic time units ($t c_{\infty}/D$).  The numerical methods, geometry, and grid are identical to those used in a previous simulation at a higher Mach number, which was validated using double-blind comparisons with an extensive set of experimental measurements \citep{Bres:2015,Bres:2016,Bres:2017b}.

\begin{figure}[!t]
 \centering
 \begin{overpic}[trim= -2cm 0cm 0cm -0.75cm, clip, width=0.75\textwidth]{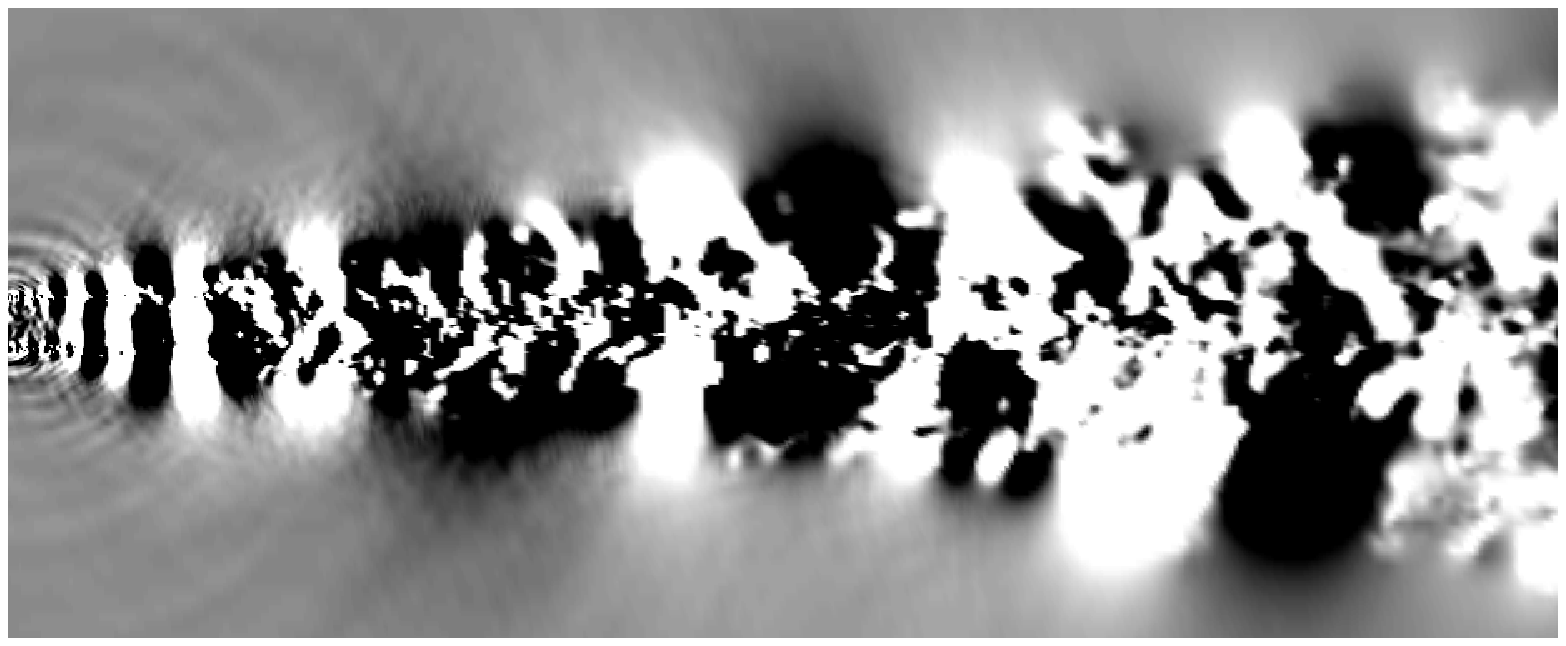}
 \put (0,2.25){\includegraphics[trim=-2cm 0cm 0cm -0.75cm, clip=true, width = 0.75\textwidth]{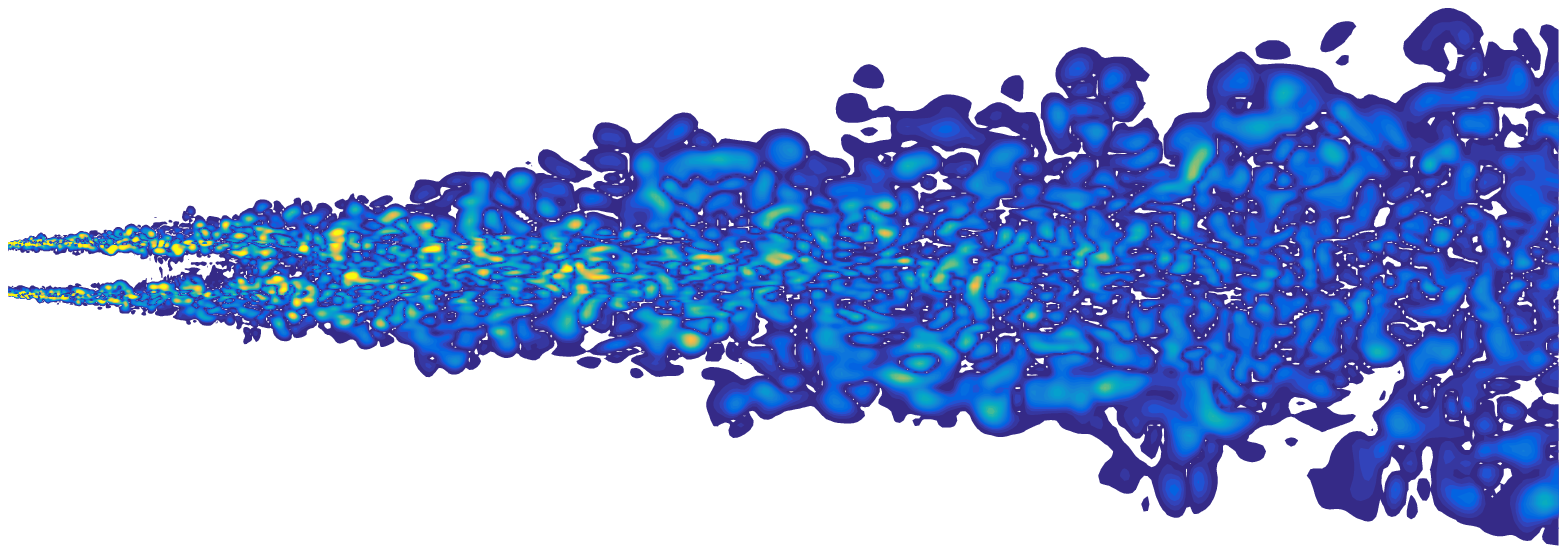}}
 \put (-5.725,3.95){\includegraphics[trim= 0cm 5cm 39.175cm 5cm, clip, width=2.10cm]{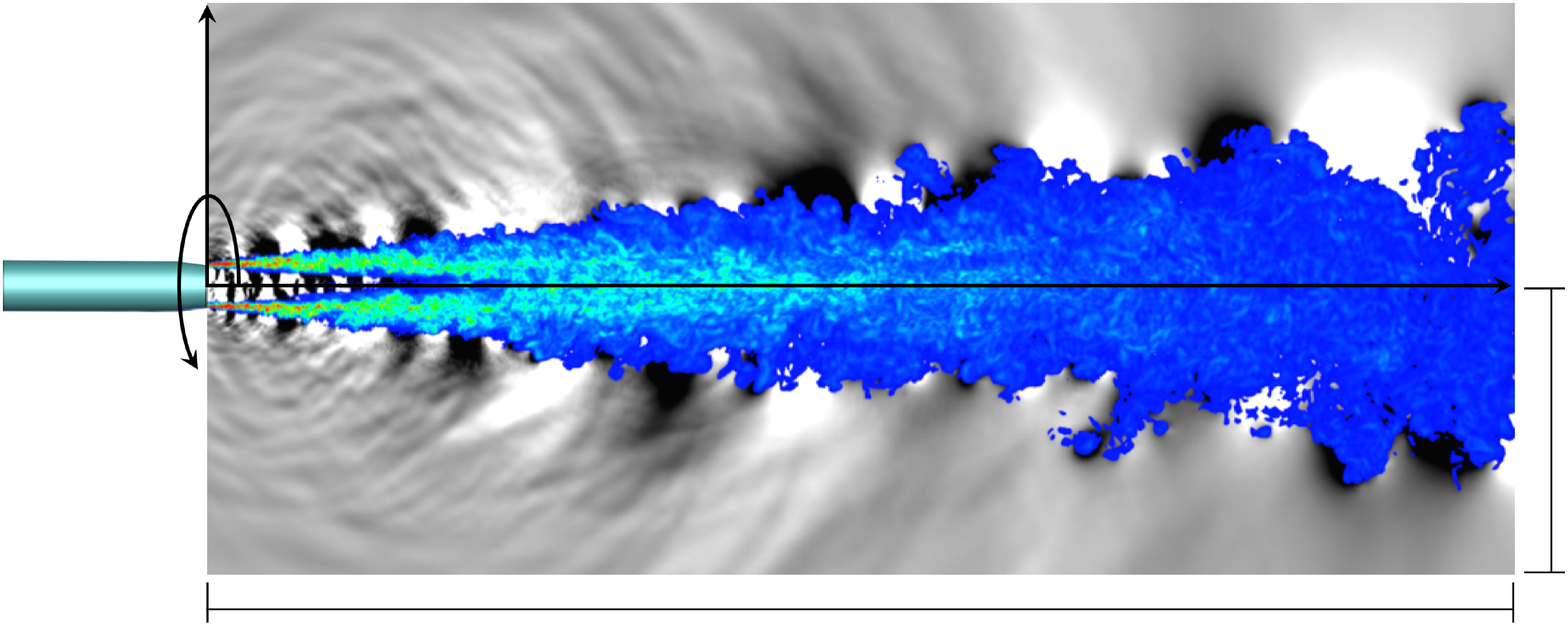}}
  \put (-4,21) {nozzle}
 \put (7,11) {$\theta$}
 \put (11.25,17.75) {\axez{11.25}{2.5}}
 \put (8.75,35.25) {$r$}
 \put (100.5,19) {$x$}
 \end{overpic}
 \vspace{0.25cm}
 \caption{Instantaneous snapshot of the Mach 0.4 turbulent jet. Grayscale: pressure fluctuations.  Color: vorticity magnitude.}
 \label{fig:jet_snapshot}
\end{figure}

The available LES database consists of $10000$ snapshots of the jet (velocities, density, and pressure) sampled every $0.2$ acoustic time units on a structured cylindrical output grid that approximately mirrors the underlying LES resolution and extends a distance of thirty jet diameters in the streamwise direction and six jet diameters in the radial direction, respectively.  The azimuthal vorticity magnitude and pressure fluctuations from one of the snapshots are shown in Figure~\ref{fig:jet_snapshot}.  These snapshots are used to compute SPOD, space-only POD, DMD, and DFT modes.  Since the azimuthal coordinate is periodic, the snapshots can be decomposed into azimuthal Fourier modes, and each azimuthal mode can treated independently within the context of the modal decompositions.  For the sake of brevity, we exclusively present results for the axi-symmetric component, and all modes are visualized using the pressure field.

\subsubsection{Spectral POD modes}
\label{Sec:Jet_SPOD}

First, we study the SPOD modes of the jet.  Spectral POD has been applied to jets before using data from both experiments \citep{Glauser:1987b, Arndt:1997, Citriniti:2000, Gudmundsson:2011, Sinha:2014, Semeraro:2016} and simulations \citep{Towne:2015,Schmidt:2017}.  These studies have used a variety of different definitions of the quantity of interest $\vc{q}$, spatial domain $\Omega$, and inner-product weight $\tc{W}$.  Here, we use the full state vector $\vc{q} = \left[ \rho, u_{x}, u_{r}, u_{\theta}, p \right]$ in the domain described previously, and we define the inner-product weight such that the induced norm is equivalent to the compressible energy norm proposed by \cite{Chu:1965}.  The SPOD modes are estimated using the procedure outlined in \S\ref{Sec:SPOD_comp} using blocks containing $N_{f} = 256$ snapshots, leading to $N_{b} = 78$ realizations of the jet.  A standard Hann window is used to reduce spectral leakage. Frequencies are reported in terms of the Strouhal number $St = f D / U_{j}$.


\begin{figure}[!t]
\input{fig10.tex}
\centering
\includegraphics[trim=0cm 0.0cm 0cm -0.25cm, clip=true,width=0.95\textwidth]{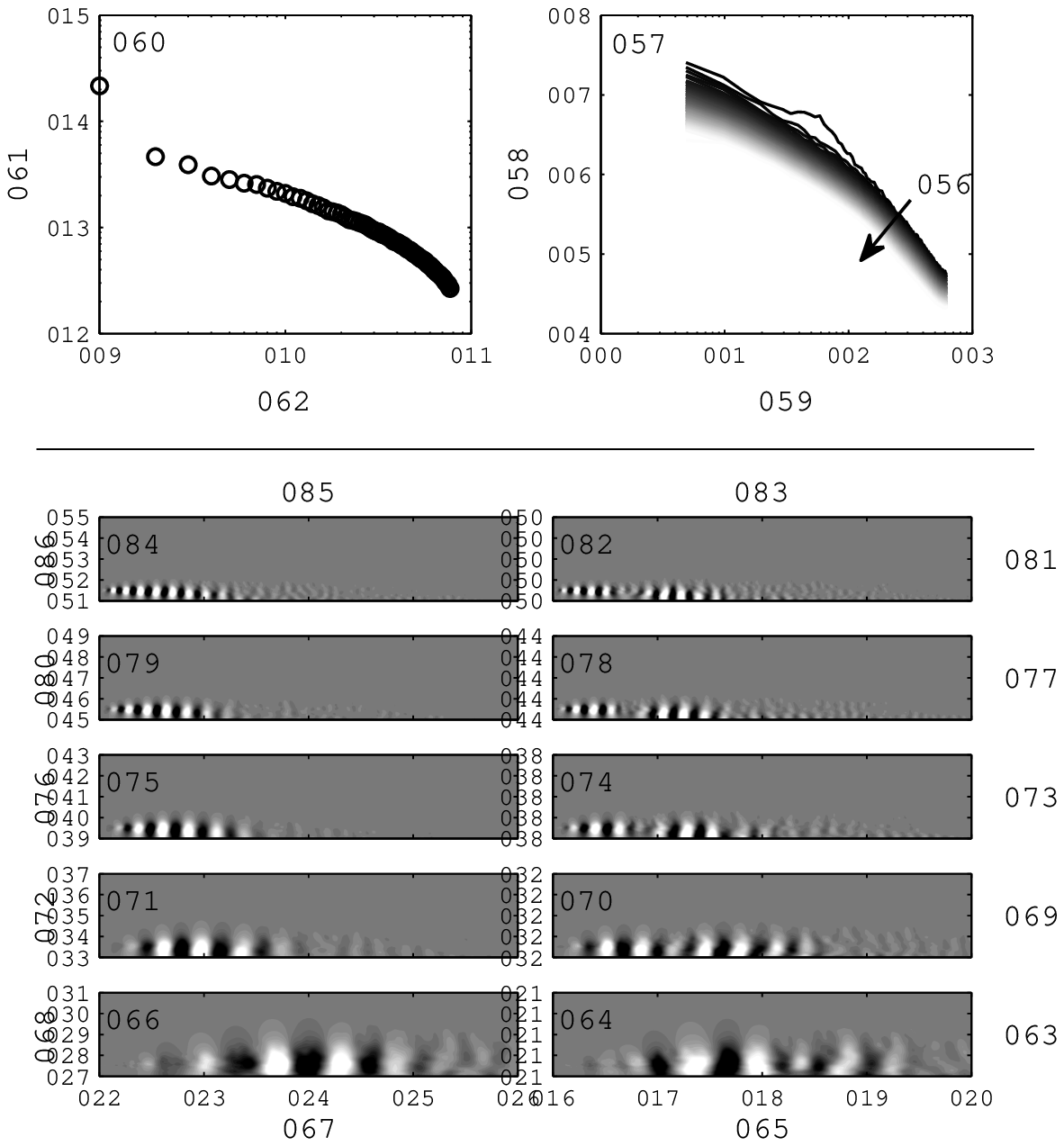}
\caption{Spectral POD modes of the Mach 0.4 turbulent jet: (a) SPOD eigenvalues at $St = 0.6$, normalized by the total energy at that frequency. (b) SPOD eigenvalues as a function of frequency, normalized by total flow energy. (c-l) Pressure field of the first (left column) and second (right column) SPOD modes at the five indicated frequencies.}
\label{fig:Jet_SPOD}
\end{figure}

The SPOD modes are summarized in Figure~\ref{fig:Jet_SPOD}.  The eigenvalues are depicted in two different ways in Figure~\ref{fig:Jet_SPOD}(a,b).  Figure~\ref{fig:Jet_SPOD}(a) shows the eigenvalues as a function of the mode number for $St = 0.6$.  The eigenvalues are normalized by the total energy at this frequency, so each scaled eigenvalue represents the fraction of the energy at $St = 0.6$ described by that mode.  The leading mode is substantially more energetic than the suboptimal modes and captures $\sim 22 \%$ of the flow energy at this frequency.  This is indicative of low-rank dynamics at this frequency \citep{Schmidt:2017}.  The full spectrum of SPOD eigenvalues is shown in Figure~\ref{fig:Jet_SPOD}(b) as a function of frequency.  All eigenvalues have been normalized by the total flow energy integrated over all frequencies, so each one can be interpreted as the fraction of the total flow energy described by that mode.  The shading of the curves varies linearly from black to white as the mode number increases from 1 to $N_{b} = 78$.  The low-rank behavior observed at $St = 0.6$, indicated by a large gap between the first and second SPOD eigenvalues, persists in the range $0.3\lesssim St \lesssim 1$.  

The pressure fields of the first two SPOD modes are plotted for several frequencies in Figure~\ref{fig:Jet_SPOD}(c-l).  All of the modes take the form of coherent wavepacket structures.  The wavelength and spatial support of the wavepackets are strongly dependent on the frequency; both quantities increase with decreasing frequency.  The wavepacket described by the first mode at each frequency has a single streamwise maximum.  These wavepackets have been linked to the Kelvin-Helmholtz instability of the annular jet shear layer for $St \gtrsim 0.3$ \citep{Suzuki:2006,Gudmundsson:2011} and to a superposition of non-normal modes for $ St \lesssim 0.3$ \citep{Jordan:2017,Schmidt:2017}.  This link between the real turbulent jet and these simple concepts from linear stability theory provides a starting point for the construction of non-empirical reduced-order models.  The wavepackets described by the second mode at each frequency have two streamwise maxima, and subsequent modes have increasingly complex structure.  These modes optimally describe the variability in the shape of the wavepackets observed in the jet at different times (see \S\ref{Sec:Jet_DMD}).  These variations are central to the acoustic characteristics of the jet \citep{Cavalieri:2011, Cavalieri:2014}.

Overall, the SPOD modes provide valuable insights that enhance physical understanding of the jet dynamics and motivate reduced-order models.  In particular, the wavepackets represented by the SPOD modes have been shown to play a key role in generating acoustic radiation and provide a rigorous starting point for modeling and mitigating jet noise.

\subsubsection{Space-only POD modes}

\begin{figure}[!t]
\input{fig11.tex}
\centering
\includegraphics[trim=0cm 0.0cm 0cm -0.25cm, clip=true,width=0.95\textwidth]{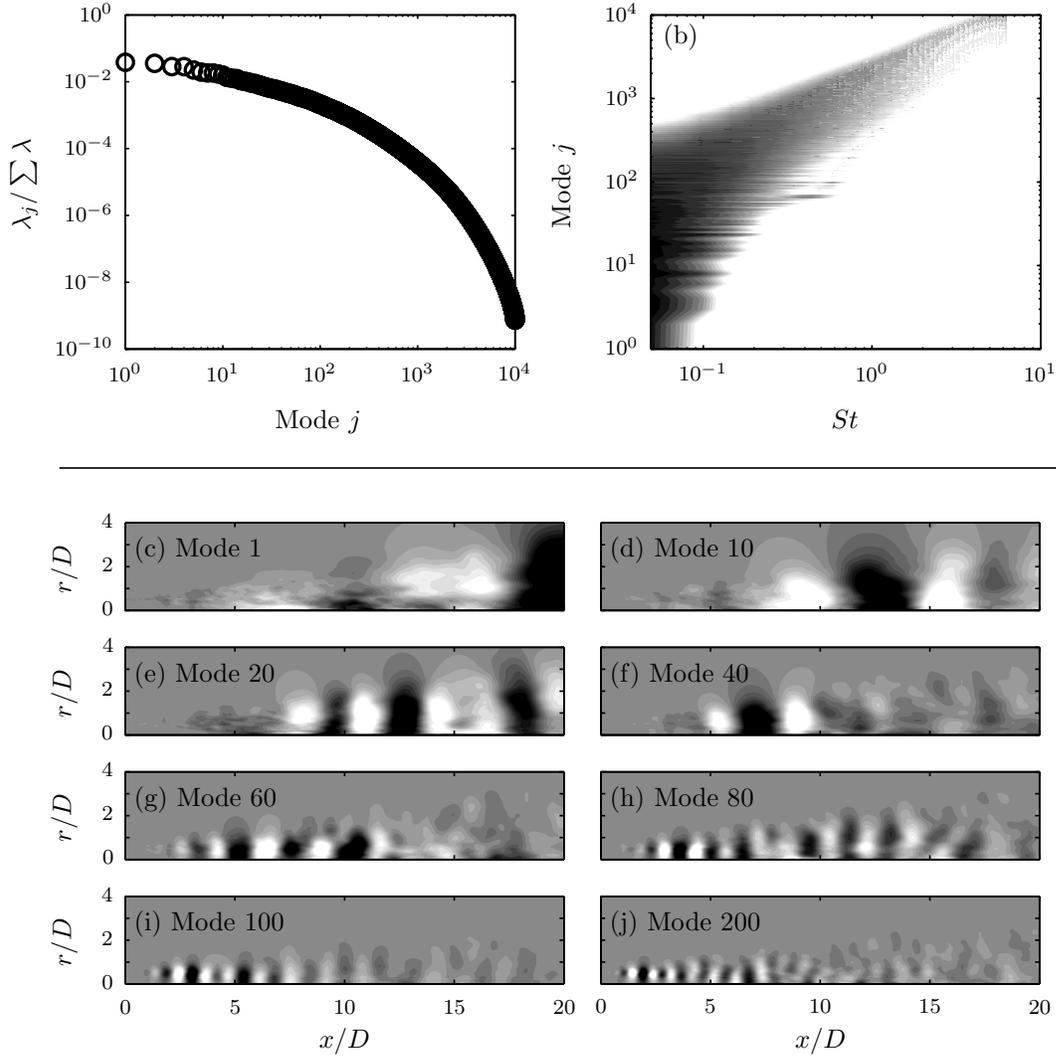}
\caption{Space-only POD modes of the Mach 0.4 turbulent jet: (a) space-only POD eigenvalues, normalized by the total flow energy. (b) Power spectral density of the space-only POD expansion coefficients; (c-j) Pressure field of space-only POD modes 1, 10, 20, 40, 60, 80, 100, 200.}
\label{fig:Jet_POD}
\end{figure}

Second, we examine the space-only POD modes of the jet.  We will see that, in contrast to SPOD, the space-only POD modes provide little insight into the jet dynamics.  The space-only POD eigenvalues are shown in Figure~\ref{fig:Jet_POD}(a).  The eigenvalues decay slowly and give no indication of the low-rank behavior of the jet revealed by SPOD.  The power spectral density of each expansion coefficient is shown in Figure~\ref{fig:Jet_POD}(b).  The contour levels are logarithmically distributed over three orders of magnitude with the upper limit equal to the maximum value.  The first approximately $10$ modes are dominated by low frequencies that are below the relevant range for jet noise research; even though these are the highest-energy structures in the jet they are of little interest.  All of the modes, and especially for $j \gtrsim 10$, contain contributions from a wide range of frequencies -- each mode represents behavior at many different time scales.  


Because of this mixing of different time scales, the space-only POD modes cannot reproduce the orderly wavepacket structures identified by SPOD.  The pressure fields of several modes are shown in Figure~\ref{fig:Jet_POD}(c-h).  Modes 1 and 10 are dominated by large structures associated with the low frequencies that these modes represent.  The higher modes have a disorganized appearance caused by the superposition of many different spatial scales associated with the wide range of time scales contained in each mode.  Fragments of different wavepackets can be observed in these modes, but a clean separation of distinct structures is not achieved; as indicated by~(\ref{Eq:SPOD_comparePOD_Phi}), the space-only POD modes are made up of a combination of many SPOD modes at different frequencies.  This scale mixing makes the space-only POD modes far less helpful for understanding and modeling the jet dynamics.  This is perhaps unsurprising in light of the analysis in \S\ref{Sec:Compare_POD} showing that space-only POD modes do not represent physically meaningful coherent structures.

\subsubsection{DMD modes}
\label{Sec:Jet_DMD}

Third, we compare four different types of DMD modes.  In all cases, we use the formulation given by~(\ref{Eq:DMD_A_def}) with different choices of the matrices $\td{X}$ and $\td{Y}$.  The first set of DMD modes is obtained using all 10000 snapshots of the jet to define $\td{X}$ and $\td{Y}$ as in~(\ref{Eq:DMD_XY_def}) \emph{without} subtracting the mean -- this is the standard way in which DMD would be applied to the jet database.  The second set of DMD modes is obtained again using all 10000 snapshots in the same way, but the mean is subtracted -- this is equivalent to taking a single long-time DFT of the database.  The third set of DMD modes is obtained from the ensemble DMD problem defined by~(\ref{Eq:DMD_XY_def_ensemble}), and the mean is subtracted from each flow realization -- the DMD modes are thus the short-time DFT modes of each realization, as shown in \S\ref{Sec:Compare_DMD}.  The details of the ensemble definition are the same as described in \S\ref{Sec:Jet_SPOD} for the SPOD computation.  The fourth and final set of DMD modes is the SPOD modes constructed from the ensemble of short-time DFT modes.


\begin{figure}[!t]
\input{fig12.tex}
\centering
\begin{overpic}[trim=0cm 0.0cm 0cm -0.25cm, clip=true, width=0.95\textwidth]{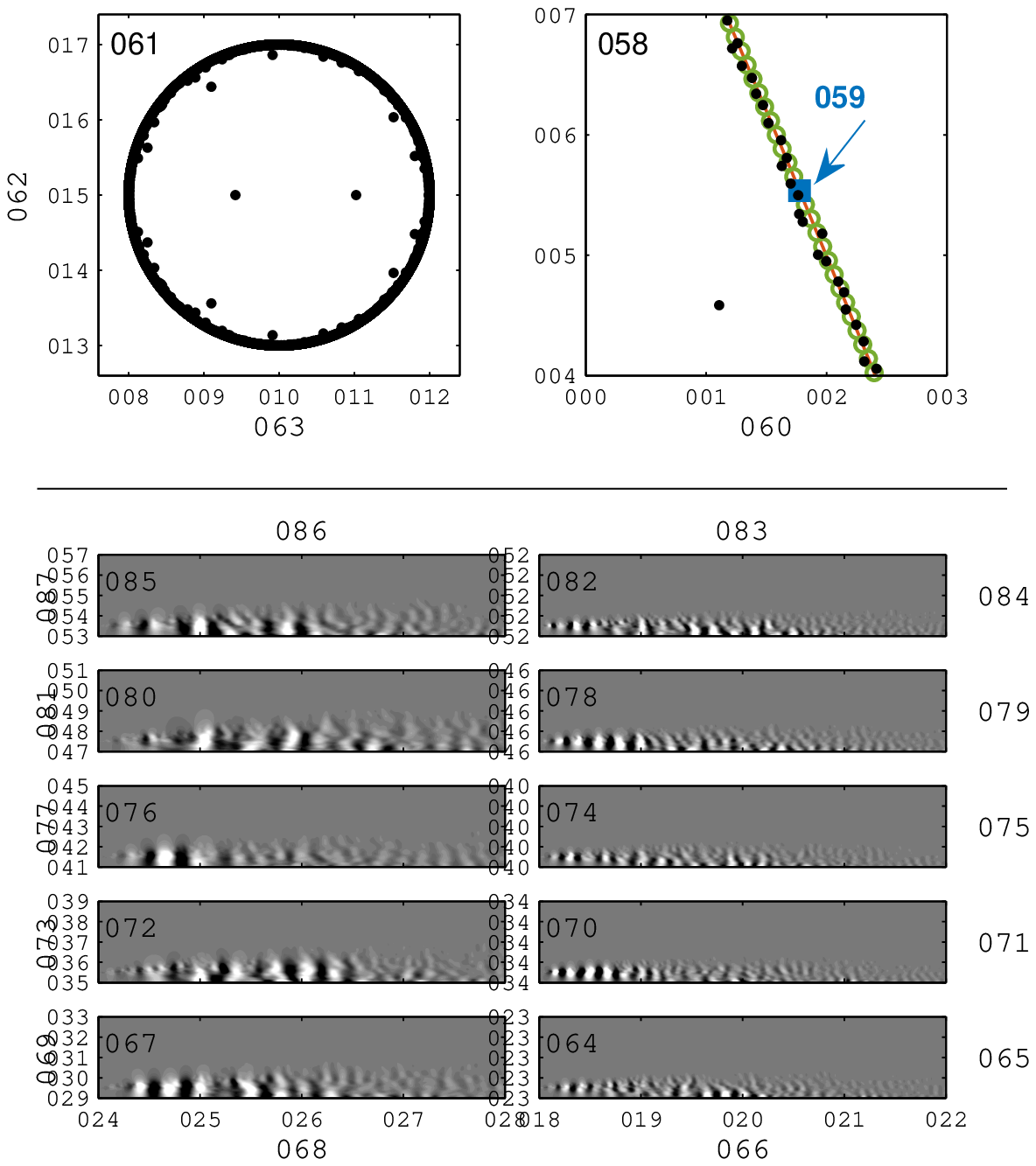}
 \put (33.575,85){\rectanglezRed{0.22cm}{0.22cm}}
 \put (33.575,86.225){\linez{2.51cm}{1.63cm}}
 \put (33.575,66.675){\linez{2.51cm}{-3.28cm}}
\end{overpic}
\caption{DMD/DFT modes of the Mach 0.4 turbulent jet: (a) Eigenvalues for DMD without mean subtraction; (b) Eigenvalues in the vicinity of $St = 0.8$ for: ({\bf\CIRCLE}) DMD without mean subtraction; ({\bf\Circle}, green) DMD with mean subtraction, i.e., a long-time DFT; ($\blacksquare$, blue) ensemble DMD, i.e., short-time DFT; (c-l) Pressure field of the different types of DMD modes: (c-d) DMD without mean subtraction; (e-f) long-time DFT; (g-l) three realization of the short-time DFT.  The left and right columns show the modes nearest $St = 0.4$ and 0.8, respectively. }
\label{fig:Jet_DMD}
\end{figure}

The DMD eigenvalues associated with the different sets of DMD modes are depicted in Figure~\ref{fig:Jet_DMD}(a,b).  Figure~\ref{fig:Jet_DMD}(a) shows the full set of eigenvalues for the standard DMD calculation in which the mean is not removed.  With the exception of a small number of outliers, the eigenvalues are tightly clustered along the unit circle and therefore have nearly zero growth/decay rates.  This is expected since the jet is a stationary flow.  Figure~\ref{fig:Jet_DMD}(b) focuses on the portion of the unit circle corresponding to $St \approx 0.8$ and confirms that the growth/decay rates of the standard DMD modes are indeed small.  Also shown in this plot are the DMD eigenvalues associated with the long-time and short-time DFT (only one short-time eigenvalue is visible).  Most of the eigenvalues from the standard DMD calculation lie near one of the eigenvalues from the long-time DFT, which shows that the standard DMD eigenvalues are nearly evenly spaced and suggests that each standard DMD mode is probably similar to one of the long-time DFT modes, even without mean subtraction.  

The pressure fields of some of the DMD modes are shown in Figure~\ref{fig:Jet_DMD}(c-l).  Specifically, we show the modes from each different type of DMD that are nearest to the frequencies $St = 0.4$ and 0.8.  The standard DMD and long-time DFT modes are shown in Figure~\ref{fig:Jet_DMD}(c,d) and Figure~\ref{fig:Jet_DMD}(e,f), respectively.  We see that these modes are quite similar, which is expected both from examination of the associated DMD eigenvalues and from the theoretical connection between DMD modes, Koopman modes, and Fourier modes discussed in \S\ref{Sec:Compare_DMD}.  Three of the short-time DFT modes associated with the ensemble DMD problem are shown for the two frequencies in Figure~\ref{fig:Jet_DMD}(g-l).  Significant variability is observed between the different realizations, which is a consequence of the physical flow variability in different short-time intervals and the uncertainty inherent to the DFT.  The SPOD modes shown in Figure~\ref{fig:Jet_SPOD} provide an optimal basis for describing the ensemble of short-time DFT modes at each frequency.  It is thus not surprising that traces of the wavepackets described by the first two SPOD modes can be observed in the short-time DFT modes and indeed in all of the different types of DMD modes.  However, none of these other DMD/DFT modes clearly isolate the coherent wavepacket structures.  Instead, they each show one of the many ways in which the coherent structures represented by the SPOD modes can coexist in one particular realization of the turbulent flow.  Clearly, the elementary coherent structures provided by SPOD are more useful for understanding and modeling the jet than are these random realizations provided by the other forms of DMD.

\subsubsection{Resolvent modes}

Fourth, we compute resolvent modes of the jet and compare them with the SPOD modes.  Several authors have computed resolvent modes for jets in recent years \citep{Garnaud:2013, Jeun:2016, Semeraro:2016, Semeraro:2016b, Schmidt:2017} and their properties have been shown to predict certain properties of the jet.  Our focus here is on making comparisons with SPOD, as suggested by our analysis in \S\ref{Sec:Resolvent}.

The jet is governed by the compressible Navier-Stokes equations, so the flow operator $\mathcal{A}$ in~(\ref{Eq:resolvent_LNS_time}) corresponds to these equations linearized about the mean flow computed from the LES data.  The output operator $\mathcal{C}$ selects the domain $\Omega = \{x,r\in[0,\,30]\times[0,\,6]\}$ in order to match the support of the LES snapshots used for the empirical methods.  The input operator $\mathcal{B}$ multiplies each component of $\vc{\eta}$ by the turbulent kinetic energy of the jet (again computed from the LES data).  This choice is motivated by the analysis of \cite{Towne:2017a} and emphasizes the forcing terms in regions of the jet where turbulent fluctuations are of high enough amplitude to interact non-linearly.  The same compressible energy inner product \citep{Chu:1965} used for the POD modes is adopted for both the input and output spaces.

The equations are discretized using fourth-order finite differences on a grid with $950\times250$ points in $x$ and $r$, respectively.  The computational domain includes the physical domain described above plus a surrounding sponge region that prevents waves from being reflected back into the computational domain at its boundaries.  The overall numerical scheme is the same as in \cite{Schmidt:2017a}; additional details are available in that reference.  As indicated by the above definition of $\Omega$, the sponge regions carry zero weight in the discretized inner products.  The numerical approximations of the leading resolvent output modes $\tilde{\td{U}}$ and singular values ${\text{\bf \textSigma}}$ are calculated from the eigenvalue decomposition  $\tilde{\td{R}}\tilde{\td{R}}^*=\tilde{\td{U}}{\text{\bf \textSigma}}^2\tilde{\td{U}}^*$ using a standard Arnoldi method (see Appendix~\ref{Sec:Res_comp} for a description of the preceding notation for the discretized equations).


\begin{figure}[!t]
\input{fig13.tex}
\centering
\includegraphics[trim=0cm 0.0cm 0cm 0.0cm, clip=true, width=0.95\textwidth]{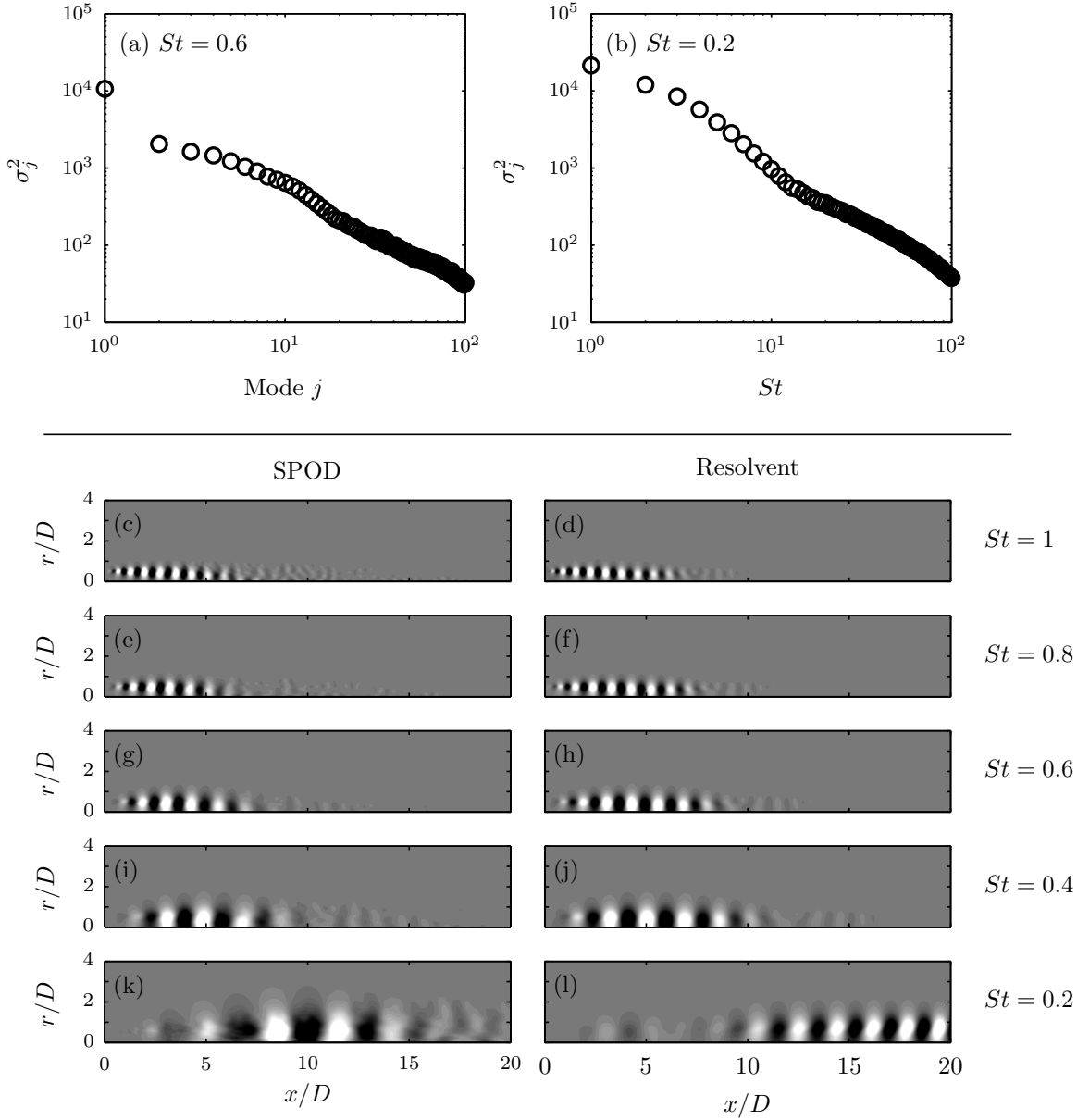}
\caption{Resolvent modes of the Mach 0.4 turbulent jet: gain as a function of mode number for (a) $St = 0.6$ and (b) $St = 0.2$; (c-l) Comparison between the leading SPOD mode (left column) and resolvent output mode (right column) at the five indicated frequencies.  The pressure field of each mode is shown.}
\label{fig:Jet_RES}
\end{figure}

The theory developed in \S\ref{Sec:Resolvent} provides a rigorous basis for comparing SPOD and resolvent modes.  Figure~\ref{fig:Jet_RES}(a,b) shows the gains of the first 100 resolvent modes for $St = 0.6$ and 0.2.  A large gain separation between the first and second modes is observed for $St = 0.6$ but not for $St = 0.2$.  This mirrors the behavior of the SPOD eigenvalues observed in Figure~\ref{fig:Jet_SPOD}(b).  The leading resolvent modes provide an accurate approximation of the leading SPOD modes at frequencies that exhibit this low-rank behavior.  For example, the pressure fields of the leading SPOD and resolvent output modes are compared for several frequencies in Figure~\ref{fig:Jet_RES}(c-l).  The low-rank behavior is observed in the resolvent gain and SPOD eigenvalue spectra for all but the lowest frequency shown in the figure, and for these frequencies the resolvent modes approximate the corresponding SPOD modes with striking accuracy.  The only discernible difference is that the resolvent-mode wavepackets tend to be slightly longer.  On the other hand, the leading resolvent mode at $St = 0.2$ is quite different from the corresponding SPOD mode.  Similarly, the suboptimal resolvent modes at all frequencies (not shown) do not have a one-to-one correspondence with the suboptimal SPOD modes.  

The theoretical connection between SPOD and resolvent modes enables us to glean useful information from these observations.  The close match between the leading SPOD and resolvent modes for $St \gtrsim 0.3$ implies that the wavepackets described by these modes are robust and insensitive to correlations within the nonlinear forcing field.  Conversely, the mismatch of lower-frequency and suboptimal modes indicates that it is necessary to account for correlated forcing in order to model the low-frequency jet dynamics as well as the wavepacket variability described by the suboptimal SPOD modes at all frequencies.   Leveraging these ideas to better understand the jet dynamics and construct accurate reduced order models is the subject of ongoing research \citep{Schmidt:2017, Towne:2017a}.


\section{Conclusions}
\label{Sec:Conclusion}

This paper explores a space-time formulation of POD for stationary flows called spectral POD and its relationship with three other modal decomposition techniques.  The main results of our analysis can be summarized by the following four statements about SPOD.

First, SPOD modes represent physically meaningful coherent structures in the sense that each mode evolves coherently in space and time -- the part of the flow described by a particular SPOD mode is perfectly correlated with the part of the flow described by that same mode at all times and entirely uncorrelated with the part of the flow described by all other modes at all times.  In contrast, modes obtained from a more common form of POD, which we call space-only POD, do not necessarily evolve coherently in time -- the part of the flow described by a particular space-only POD mode is not necessarily correlated with the part of the flow described by the same mode at a later time, nor is it necessarily uncorrelated with the part of the flow described by a different mode at a later time.  This essential difference can be traced back to the definition of the stochastic ensembles that form the starting-point of each method; SPOD uses time-dependent flow realizations and thus retains dynamical information, whereas space-only POD discards all such information by treating each time instance of the flow as an independent realization of a random process.

Second, SPOD modes are optimally averaged DMD modes obtained from an ensemble DMD problem for stationary flows.  The ensemble DMD problem is defined using the `exact DMD' framework of \cite{Tu:2013}, and our result holds under the assumption of linearly consistent and zero-mean data ensembles.  The first of these requirements corresponds to the condition under which DMD modes are expected to approximate Koopman modes, while the second assumption is justified by the zero growth rate of Koopman modes for stationary flows.  We show that any linear combination of the DFT modes of the data ensembles are eigenvectors of this ensemble DMD problem.  Spectral POD modes can be written as linear combinations of these DFT modes at each frequency, and are thus DMD modes.  Specifically, SPOD modes are \emph{optimally averaged} DMD modes that provide the best description of the statistical variability of the flow.  

Third, SPOD modes are identical to resolvent modes in the special case in which the resolvent mode expansion coefficients are uncorrelated.  This result is based on a statistical perspective on the resolvent-mode reconstruction of turbulent flows and on the observation that the resolvent-mode expansion coefficients are inherently statistical quantities for stationary turbulent flows.  Uncorrelated expansion coefficients are usually associated with white-noise forcing of the linear dynamics.  While the nonlinear perturbation terms of the Navier-Stokes equations are unlikely to be truly white for real flows, this has proven to be a reasonable first approximation for modeling a variety of flows.  Accordingly, resolvent modes obtained from these models can be understood as non-empirical approximations of SPOD modes under the assumption of white-noise nonlinear forcing.

Fourth, SPOD modes provide fundamental insight into how the resolvent-mode expansion coefficients should be chosen so that the expansion properly captures the flow statistics.  Specifically, we show the importance of properly accounting for the statistical nature of the correlated expansion coefficients and provide a method of computing these statistics using SPOD modes that leads to convergent approximations of the flow.   All previous methods for obtaining the expansion coefficients assumed them to be deterministic quantities that can be entirely described by their amplitude and phase.  We show that this results in a rank-one approximation of the second-order flow statistics, no matter how many resolvent modes are retained in the expansion.  Since the optimal rank-one approximation of the cross-spectral density tensor is given by the leading SPOD mode at each frequency, the quality of the approximation that can be achieved using resolvent modes with deterministic expansion coefficients is governed by the low-rank nature of the second-order flow statistics rather then the resolvent operator.  

These results are demonstrated using two example problems.  The first is the linearized Ginzburg-Landau equation, which provides a simple model of a convectively unstable flow susceptible to non-modal growth.  When forced with white noise, the SPOD modes computed from data and resolvent modes are nearly the same, as expected.  Space-only POD modes computed from the same data do not capture these underlying dynamics and do not represent structures that evolve coherently in time.  The Ginzburg-Landau equation is then forced with spatially coherent input, and the results are used to demonstrate the inability of resolvent-mode expansions with deterministic expansion coefficients to reconstruct the power spectral density of the solution.  In contrast, these statistics converge when proper statistical expansion coefficients are employed.  

The second example problem is a Mach 0.4 turbulent jet.  The SPOD modes, computed from an LES database, isolate different space-time scales of the jet and provide insight into the low-rank dynamics of large-scale coherent wavepacket structures and how they might be modeled.  In contrast, space-only POD modes obscure these low-rank wavepacket dynamics by jumbling together many different space-time scales in each mode.  The DMD modes (without mean subtraction) and long-time DFT modes (i.e., DMD modes with mean subtraction) are similar, while the short-time DFT modes that are used to compute SPOD modes reveal the variability present in the jet.  Each of these types of DMD modes represents one possible way in which the coherent wavepacket structures identified by SPOD can coexist in one of the many possible realizations of the turbulent jet, which highlights the advantage of SPOD for turbulent flows.  

Overall, our results show that SPOD combines the advantages of space-only POD and DMD for stationary flows.  Each SPOD mode represents a structure that is dynamic in the same sense as DMD modes, but that also accounts for and optimally describes the statistical variability of the turbulent flow.  Consequently, SPOD modes could provide improved robustness over space-only POD and DMD modes for modeling, estimation, and control of turbulent flows. 

Spectral POD also automatically solves two challenges in DMD that have spurred a number of variations of the method \citep{Rowley:2017}.  First, DMD modes have no inherent ranking, making it challenging to select a reduced set of modes for model reduction.  Spectral POD, in contrast, gives a ranked set of modes in each frequency bin defined by the DFT rather than a collection of modes at many slightly different frequencies over the same frequency range.  Figure~\ref{fig:Jet_DMD}(b) provides a clear illustration of this; SPOD replaces the assortment of DMD/DFT modes with a ranked set of modes at the frequency indicated by the blue square.  Second, DMD can be sensitive to noisy data.  Spectral POD automatically reduces the effect of noise as a consequence of the spectral estimation process, and convergence can always be improved by including more flow realizations.   

The main disadvantage of SPOD is that it typically requires more data than either space-only POD or DMD.  The spatial statistics required for space-only POD usually converge more rapidly than the space-time statistics required for SPOD.  Standard applications of DMD use a single realization of a flow; SPOD requires at least a handful of realizations that are typically obtained from a longer time-series.  Additionally, if the mean is not subtracted, the frequencies identified by DMD are not constrained to be evenly spaced as they are for SPOD.  This can be advantageous for laminar or transitional flows with sharp spectral peaks, since resolving peaks using evenly spaced points requires longer data series.  \cite{Arbabi:2017} recently proposed an alternative method for computing Koopman modes for flows with sharp spectral peaks via an iterative application of the DFT; a similar approach could be used to improve the performance of SPOD for these flows.


\section*{Acknowledgments}

A.T. gratefully acknowledges support from NASA grant no. NNX15AU93A and from the NNSA Predictive Science Academic Alliance Program II, grant no. DE-NA0002373.  O.S. acknowledges support from the German science foundation (DFG) grant no. 3114/1-1 and O.S and T.C. acknowledge support from ONR-N0014-11-1-0753 and ONR-N00014-16-1-2445.  The authors also thank Dr. Kevin K. Chen for supplying portions of the code used for the Ginzburg-Landau problem and Dr. Guillaume A. Br{\`e}s for providing the jet LES data.  The LES study was supported by NAVAIR SBIR project, under the supervision of Dr. John T. Spyropoulos. The main LES calculations were carried out on CRAY XE6 machines at DoD HPC facilities in ERDC DSRC.

\appendix
\section{Derivation of the SPOD eigenvalue problem}
\label{appA}

The SPOD eigenvalue problem defined by~(\ref{Eq:SPOD_eigSpectral}) was first derived by \cite{Lumley:1967, Lumley:1970}.  Here, we offer an alternative derivation that we find to be more straight forward.  Using~(\ref{Eq:SPOD_CrossCor_stationary_IFT}), the correlation function ${\tc{C}}(\vc{x},\vc{x}^{\prime}, t - t^{\prime})$ can be written as
\begin{equation}
{\tc{C}}(\vc{x},\vc{x}^{\prime}, t - t^{\prime})   =  \int \limits_{-\infty}^{\infty} \tc{S}(\vc{x},\vc{x}^{\prime}, f) e^{\ii 2 \pi f t} e^{-\ii 2 \pi f t^{\prime}} d f
\label{Eq:CrossCor_stationary_IFT2}
\end{equation}
Inserting this form into~(\ref{Eq:SPOD_eig}) leads to the following simplifications:
\begin{equation}
 \int \limits_{-\infty}^{\infty} \int \limits_{\Omega} \int \limits_{-\infty}^{\infty} \tc{S}(\vc{x},\vc{x}^{\prime}, f) e^{\ii 2 \pi f t} e^{-\ii 2 \pi f t^{\prime}} \tc{W} {\vc{\phi}} (\vc{x}^{\prime},t^{\prime}) d f d \vc{x}^{\prime} d t^{\prime} = 
\lambda  {\vc{\phi}} (\vc{x},t),
\label{Eq:EigProb_stationary2}
\end{equation}
\begin{equation}
 \int \limits_{\Omega} \int \limits_{-\infty}^{\infty} \tc{S}(\vc{x},\vc{x}^{\prime}, f)   \tc{W} \left[ \int \limits_{-\infty}^{\infty}  {\vc{\phi}} (\vc{x}^{\prime},t^{\prime}) e^{-\ii 2 \pi f t^{\prime}} d t^{\prime} \right] e^{\ii 2 \pi f t} d f d \vc{x}^{\prime}   = 
\lambda  {\vc{\phi}} (\vc{x},t),
\label{Eq:EigProb_stationary3}
\end{equation}
\begin{equation}
 \int  \limits_{\Omega} \int \limits_{-\infty}^{\infty} \tc{S}(\vc{x},\vc{x}^{\prime}, f)   \tc{W} {\hat{\vc{\phi}}} (\vc{x}^{\prime},f) e^{\ii 2 \pi f t}  d f d \vc{x}^{\prime}  = 
\lambda  {\vc{\phi}} (\vc{x},t),
\label{Eq:EigProb_stationary4}
\end{equation}
where $\hat{\vc{\phi}} (\vc{x},f)$ is the temporal Fourier transform of ${ \vc{\phi}} (\vc{x},t)$.

For the solution ansatz ${\vc{\phi}} (\vc{x},t) = {\vc{\psi}} (\vc{x},f^{\prime}) e^{\ii 2 \pi f^{\prime} t}$ we have ${\hat{ \vc{\phi}}} (\vc{x},f) = { \vc{\psi}} (\vc{x},f^{\prime}) \delta\left( f - f^{\prime}\right)$.  Substituting these into~(\ref{Eq:EigProb_stationary4}) gives
\begin{equation}
\int \limits_{\Omega} \int \limits_{-\infty}^{\infty} \tc{S}(\vc{x},\vc{x}^{\prime}, f)   \tc{W} \vc{\psi}(\vc{x},f^{\prime}) \delta \left( f - f^{\prime} \right) e^{\ii 2 \pi f t}  d f d \vc{x}^{\prime}  = 
\lambda   \vc{\psi}(\vc{x},f^{\prime}) e^{\ii 2 \pi f^{\prime} t}
\label{Eq:EigProb_stationary5}
\end{equation}
or
\begin{equation}
 \int \limits_{\Omega} \tc{S}(\vc{x},\vc{x}^{\prime}, f^{\prime})   \tc{W} \vc{\psi}(\vc{x},f^{\prime}) e^{\ii 2 \pi f^{\prime} t}   d \vc{x}^{\prime}  = 
\lambda   \vc{\psi}(\vc{x},f^{\prime}) e^{\ii 2 \pi f^{\prime} t}.
\label{Eq:EigProb_stationary6}
\end{equation}
Multiplying both sides of~(\ref{Eq:EigProb_stationary6}) by $e^{-\ii 2 \pi f^{\prime} t}$ gives the final SPOD eigenvalue problem~(\ref{Eq:SPOD_eigSpectral}).


\section{Proof of ensemble DMD result}
\label{Sec:DMDensembleProof}

In this appendix, we prove that the modes of the ensemble DMD problem described in \S\ref{Sec:DMD_ensemble} are the DFT modes of each realization for zero-mean, linearly consistent data.  We begin by writing~(\ref{Eq:DMD_A_def}) as
\begin{equation}
\label{Eq:DMD_AXeqY}
\td{A} \td{X} = \td{Y}.
\end{equation}
\cite{Tu:2013} showed that~(\ref{Eq:DMD_AXeqY}) follows from~(\ref{Eq:DMD_A_def}) if and only if $\td{X}$ and $\td{Y}$ are linearly consistent.  We therefore assume that the ensemble data matrices are linearly consistent and note that this is precisely the condition under which DMD modes are expected to approximate Koopman modes \citep{Tu:2013}.

Since we wish to show that the eigenvalues of $\td{A}$ are the DFT modes of each flow realization, it is helpful to write $\td{X}$ and $\td{Y}$ in terms of these DFT modes.  To do so, we use the matrix 
\begin{equation}
\label{Eq:DMD_F}
\td{F}_{M} \triangleq {\frac {1}{\sqrt {M}}}{\begin{bmatrix}1&1&1&1&\cdots &1\\1&z &z ^{2}&z ^{3}&\cdots &z ^{M-1}\\1&z ^{2}&z ^{4}&z ^{6}&\cdots &z ^{2(M-1)}\\1&z ^{3}&z ^{6}&z ^{9}&\cdots &z ^{3(M-1)}\\\vdots &\vdots &\vdots &\vdots &\ddots &\vdots \\1&z ^{M-1}&z ^{2(M-1)}&z ^{3(M-1)}&\cdots &z ^{(M-1)(M-1)}\end{bmatrix}},
\end{equation}
which when applied to a vector gives its DFT, i.e., $\hat{\vd{v}} = \td{F}_{M}\vd{v}$ for $\vd{v} \in \mathbb{R}^{M}$. The scalar $z = e^{-\mathrm{i} 2 \pi /M}$ is the primitive $M$-th root of unity and the $1/\sqrt{M}$ scaling makes the DFT unitary and identical to~(\ref{Eq:SPOD_comput_DFT}) for rectangular window weights $w_{j}=1$.  Other windows, which are known to be beneficial in practice \citep{Bendat:1990}, can be incorporated into the present analysis by using the windowed data to define each $\td{Q}^{(n)}$.  We wish to take the DFT of each \emph{row} of the matrices $\td{Q}^{(n)}$.  To do so, we must apply $\td{F}_{M}$ to the conjugate transpose of each $\td{Q}^{(n)}$ and apply a second conjugate transpose to the product to restore the original size of the matrix, $\hat{\td{Q}}^{(n)} = \left(\td{F}_{M}({\td{Q}}^{(n)})^{*}\right)^{*}  = \td{Q}^{(n)}\td{F}_{M}^{*}$.  The first column of $\hat{\td{Q}}^{(n)}$ gives the $k = 1$ term from~(\ref{Eq:SPOD_comput_DFT}) and so on.  Since $\td{F}_{M}$ is unitary, it follows that $\td{Q}^{(n)} = \hat{\td{Q}}^{(n)}\td{F}_{M}$.  Inserting this into~(\ref{Eq:DMD_XY_def_ensemble}) gives
\begin{subequations}
\label{Eq:DMD_XY_def_ensemble2}
\begin{alignat}{2}
\td{X} =& \left[ \hat{\td{Q}}^{(1)}\td{F}_{M}\td{T}_{\mathrm{X}} , \dots, \hat{\td{Q}}^{(N_{e})}\td{F}_{M} \td{T}_{\mathrm{X}}  \right] =&& \left[ \hat{\td{Q}}^{(1)}\hat{\td{T}}_{\mathrm{X}} , \dots, \hat{\td{Q}}^{(N_{e})} \hat{\td{T}}_{\mathrm{X}}  \right], \\
\td{Y} =& \left[ \hat{\td{Q}}^{(1)}\td{F}_{M} \td{T}_{\mathrm{Y}} ,  \dots, \hat{\td{Q}}^{(N_{e})}\td{F}_{M}\td{T}_{\mathrm{Y}}  \right]=&& \left[ \hat{\td{Q}}^{(1)} \hat{\td{T}}_{\mathrm{Y}} , \dots, \hat{\td{Q}}^{(N_{e})} \hat{\td{T}}_{\mathrm{Y}}  \right],
\end{alignat}
\end{subequations}
where $\hat{\td{T}}_{\mathrm{X},\mathrm{Y}} = \td{F}_{M} {\td{T}}_{\mathrm{X},\mathrm{Y}}$.  

Since we have subtracted the mean from each flow realization, the zero-frequency ($k=1$) DFT mode is identically zero, so the first column of $\hat{\td{Q}}^{(n)}$ is zero for all $n$.  Accordingly, we can remove the first column from each $\hat{\td{Q}}^{(n)}$ and the first row from each instance of $\hat{\td{T}}_{\mathrm{X}}$ and $\hat{\td{T}}_{\mathrm{Y}}$ without changing $\td{X}$ and $\td{Y}$.  This is the key step that enables the remaining manipulations required to prove our result.  We will denote the modified forms of $\hat{\td{Q}}^{(n)}$, $\hat{\td{T}}_{\mathrm{X}}$, and $\hat{\td{T}}_{\mathrm{Y}}$ as $\tilde{\td{Q}}^{(n)}$, $\tilde{\td{T}}_{\mathrm{X}}$, and $\tilde{\td{T}}_{\mathrm{Y}}$, respectively.  It should be noted that while $\hat{\td{T}}_{\mathrm{X}}$ and $\hat{\td{T}}_{\mathrm{Y}}$ were rectangular matrices, $\tilde{\td{T}}_{\mathrm{X}}$ and $\tilde{\td{T}}_{\mathrm{Y}}$ are square $(M-1) \times (M-1)$ matrices. Specifically,
\begin{subequations}
\label{Eq:DMD_TtildeXY}
\begin{align}
\tilde{\td{T}}_{\mathrm{X}} =& {\frac {1}{\sqrt {M}}}{\begin{bmatrix}1&z &z ^{2}&z ^{3}&\cdots &z ^{M-2}\\1&z ^{2}&z ^{4}&z ^{6}&\cdots &z ^{2(M-2)}\\1&z ^{3}&z ^{6}&z ^{9}&\cdots &z ^{3(M-2)}\\\vdots &\vdots &\vdots &\vdots &\ddots &\vdots \\ \,\,\,\,\,1\,\,\,\,\, & \,\,\,\,\, z ^{M-1} \,\,\,\,\, &z ^{2(M-1)}&z ^{3(M-1)}&\cdots &z ^{(M-1)(M-2)}\end{bmatrix}} , \\[4pt]
\tilde{\td{T}}_{\mathrm{Y}} =& {\frac {1}{\sqrt {M}}}{\begin{bmatrix}z &z ^{2}&z ^{3}& z^{4} &\cdots &z ^{M-1}\\z ^{2}&z ^{4}&z ^{6}&z^{8}&\cdots &z ^{2(M-1)}\\z ^{3}&z ^{6}&z ^{9}& z^{12}&\cdots &z ^{3(M-1)}\\\vdots &\vdots &\vdots &\vdots &\ddots &\vdots \\z ^{M-1}&z ^{2(M-1)}&z ^{3(M-1)}& z^{4(M-1)}&\cdots &z ^{(M-1)(M-1)}\end{bmatrix}}.
\end{align}
\end{subequations}
These matrices can be factored as
\begin{subequations}
\label{Eq:DMD_TtildeXY_factor}
\begin{align}
\tilde{\td{T}}_{\mathrm{X}} =& \, \td{F}_{M-1} \, \mathrm{diag}\left(\left[ 1, z, z^{2}, \dots, z^{M-2} \right] \right),  \label{Eq:DMD_TtildeXY_factor_a}  \\  
\tilde{\td{T}}_{\mathrm{Y}} =& \, \mathrm{diag}\left(\left[ z, z^{2}, \dots, z^{M-1} \right] \right)   \, \td{F}_{M-1} \, \mathrm{diag}\left(\left[ 1, z,  \dots, z^{M-2} \right] \right), \label{Eq:DMD_TtildeXY_factor_b} 
\end{align}
\end{subequations}
which will prove useful in what follows.

Using these modified matrices, the input and output data matrices can be written as
\begin{subequations}
\label{Eq:DMD_XY_def_ensemble3}
\begin{align}
\td{X} =& \left[ \tilde{\td{Q}}^{(1)}\tilde{\td{T}}_{\mathrm{X}} , \dots, \tilde{\td{Q}}^{(N_{e})} \tilde{\td{T}}_{\mathrm{X}}  \right], \\
\td{Y} =& \left[ \tilde{\td{Q}}^{(1)} \tilde{\td{T}}_{\mathrm{Y}} , \dots, \tilde{\td{Q}}^{(N_{e})}\tilde{\td{T}}_{\mathrm{Y}}  \right].
\end{align}
\end{subequations}
Inserting~(\ref{Eq:DMD_XY_def_ensemble3}) into~(\ref{Eq:DMD_AXeqY}) gives the expression
\begin{equation}
\label{Eq:DMD_eigprob}
\td{A}\left[ \tilde{\td{Q}}^{(1)}, \dots, \tilde{\td{Q}}^{(N_{e})}\right] \begingroup 
\setlength\arraycolsep{2pt} 
\renewcommand{\arraystretch}{1}
\begin{bmatrix}
\tilde{\td{T}}_{\mathrm{X}} &  & \\
 & \ddots &  \\
 &  & \tilde{\td{T}}_{\mathrm{X}} \end{bmatrix} \endgroup = \left[ \tilde{\td{Q}}^{(1)}, \dots, \tilde{\td{Q}}^{(N_{e})} \right] 
 \begingroup 
\setlength\arraycolsep{2pt} 
\renewcommand{\arraystretch}{1}
 \begin{bmatrix}
\tilde{\td{T}}_{\mathrm{Y}} &  & \\
 & \ddots &  \\
 &  & \tilde{\td{T}}_{\mathrm{Y}} \end{bmatrix}
 \endgroup.
\end{equation}
Since $\td{F}_{M-1}$ is unitary and powers of $z$ are non-zero, (\ref{Eq:DMD_TtildeXY_factor_a}) shows that $\tilde{\td{T}}_{\mathrm{X}}$ is the product of two invertible matrices and is therefore itself invertible.  As a result, ({\ref{Eq:DMD_eigprob}) can be written in the form given by~(\ref{Eq:DMD_eigprob_final}) with ${\text{\bf \textLambda}}_{\mathrm{DMD}} \triangleq \tilde{\td{T}}_{\mathrm{Y}}\tilde{\td{T}}_{\mathrm{X}}^{-1}$.  Finally, using~(\ref{Eq:DMD_TtildeXY_factor}), ${\text{\bf \textLambda}}_{\mathrm{DMD}}$ reduces to the form given by~(\ref{Eq:DMD_LAMBDA}).

\section{Discretized SPOD/resolvent-mode relations}
\label{Sec:Res_comp}

In practice, SPOD and resolvent modes are always approximated on discrete grids.  To help make the results of \S\ref{Sec:Resolvent} as accessible and useful as possible, in this appendix we write the discrete forms of some of the key equations relating these different types of modes.  The discretized forms of all continuous variables and operators will be represented by the corresponding upright symbols, e.g., $\tc{S}_{yy}$ becomes $\td{S}_{\vd{y}\vd{y}}$.  Time or frequency dependencies can be inferred from the continuous variables and these arguments are suppressed for brevity.

We begin by writing the discretized form of the linearized flow equations and output as
\begin{equation}
\label{Eq:disc_linEqs}
\frac{d \qvecd'}{dt} - \td{A}\qvecd' = \td{B}{\text{\bf \texteta}}
\end{equation} 
and
\begin{equation}
\label{Eq:disc_y}
\vd{y} = \td{C}\qvecd',
\end{equation} 
respectively (to be clear, $\td{A}$ is the discretization of $\mathcal{A}$, not the DMD matrix).  Note that the input and output do not need to be discretized on the same grid as the linearized flow operator and state vector, since the input and output matrices $\td{B}$ and $\td{C}$ can be used to interpolate from one grid to another.  We will assume that the the discretized output $\vd{y}$ is defined on the same grid as the flow data used to estimate SPOD modes.  

The inner products on the input and output spaces are approximated as $\langle \hat{\vd{y}}_{1}, \hat{\vd{y}}_{2}  \rangle_{ \vd{y}} = \hat{\vd{y}}_{2}^{*} \td{W}_{\vd{y}} \hat{\vd{y}}_{1}$ and $\langle \hat{{\text{\bf \texteta}}}_{1}, \hat{{\text{\bf \texteta}}}_{2}  \rangle_{ {\text{\bf \texteta}}} = \hat{{\text{\bf \texteta}}}_{2}^{*} \td{W}_{{\text{\bf \texteta}}} \hat{{\text{\bf \texteta}}}_{1}$, respectively.  It is important to note that the positive-definite Hermitian matrices $\td{W}_{\vd{y}}$ and $\td{W}_{{\text{\bf \texteta}}}$ must account for both the weight matrices and the numerical quadrature of the integrals in~(\ref{Eq:resolvent_ip_y}) and~(\ref{Eq:resolvent_ip_f}).

The discretized resolvent operator is
\begin{equation}
\label{Eq:disc_resolventOp}
\td{R} = \td{C}(\ii 2\pi f {\td{ I}} - \td{A})^{-1}\td{B}.
\end{equation} 
If $\td{W}_{\vd{y}} = \td{W}_{{\text{\bf \texteta}}} = \td{I}$, then the resolvent modes are given by the singular value decomposition 
\begin{equation}
\label{Eq:disc_resolvent_SVD_WeqI}
\td{R} = \td{U} {\text{\bf \textSigma}} \td{V}^{*}.
\end{equation}
The singular values appear within the diagonal positive-semidefinite matrix ${\text{\bf \textSigma}}$ and the input and output modes are contained in the columns of the orthonormal matrices $\td{V}$ and $\td{U}$, respectively.  When either of the weight matrices $\td{W}_{\vd{y}}$ or $\td{W}_{{\text{\bf \texteta}}}$ is not the identity, the resolvent modes can be recovered from the singular value decomposition of a weighted matrix,
\begin{equation}
\label{Eq:disc_resolvent_SVD_WneqI}
\tilde{\td{R}} = \td{W}_{\vd{y}}^{1/2} \td{R} \td{W}_{{\text{\bf \texteta}}}^{-1/2}  = \tilde{\td{U}} {\text{\bf \textSigma}} \tilde{\td{V}}^{*}.
\end{equation}
The singular values again appear on the diagonal of ${\text{\bf \textSigma}}$ and the input and output modes are contained in the columns of the matrices $\td{V} = \td{W}_{{\text{\bf \texteta}}}^{-1/2} \tilde{\td{V}}$ and $\td{U} = \td{W}_{\vd{y}}^{-1/2} \tilde{\td{U}} $, respectively.  The resolvent operator is recovered as 
\begin{equation}
\label{Eq:disc_resolvent_decomp}
\td{R} = \td{U} {\text{\bf \textSigma}} \td{V}^{*}\td{W}_{{\text{\bf \texteta}}}.
\end{equation}
The resolvent-mode expansion of the Fourier-transformed output is then $\hat{\vd{y}} = \td{U}{\text{\bf \textSigma}}{\text{\bf \textbeta}}$, where ${\text{\bf \textbeta}} = \td{V}^{*}\td{W}_{{\text{\bf \texteta}}}\hat{{\text{\bf \texteta}}}$.  The discretized cross-spectral density tensor is thus
\begin{equation}
\label{Eq:disc_Syy_resolvent}
\td{S}_{\vd{y}\vd{y}} = E\{ \hat{\vd{y} } \hat{\vd{y}}^{*} \} = \td{U} {\text{\bf \textSigma}} \td{S}_{{\text{\bf \textbeta}}{\text{\bf \textbeta}}} {\text{\bf \textSigma}} \td{U}^{*},
\end{equation}
where $\td{S}_{{\text{\bf \textbeta}}{\text{\bf \textbeta}}} = E\{{\text{\bf \textbeta}}{\text{\bf \textbeta}}^{*}\} = \td{V}^{*}\td{W}_{{\text{\bf \texteta}}}\td{S}_{{\text{\bf \texteta}}{\text{\bf \texteta}}} \td{W}_{{\text{\bf \texteta}}}  \td{V}$ and $\td{S}_{{\text{\bf \texteta}}{\text{\bf \texteta}}} = E\{ \hat{{\text{\bf \texteta}} } \hat{{\text{\bf \texteta}}}^{*} \}$.

Using~(\ref{Eq:disc_Syy_resolvent}) along with the data-based SPOD estimates discussed in \S\ref{Sec:SPOD_comp}, the output cross spectral density can be written in terms of SPOD or resolvent modes as
\begin{equation}
\label{Eq:disc_Syy_both}
\td{S}_{\vd{y}\vd{y}} = {\text{\bf \textPsi}} {\text{\bf \textLambda}} {\text{\bf \textPsi}}^{*} = \td{U} {\text{\bf \textSigma}} \td{S}_{{\text{\bf \textbeta}}{\text{\bf \textbeta}}} {\text{\bf \textSigma}} \td{U}^{*}
\end{equation}
This is the discrete equivalent of~(\ref{Eq:resolvent_Syy_4}), which is the central expression in \S\ref{Sec:Resolvent}.  If the resolvent-mode expansion coefficients are uncorrelated, then $\td{S}_{{\text{\bf \textbeta}}{\text{\bf \textbeta}}} = \td{I}$ and~(\ref{Eq:disc_Syy_both}) reduces to 
\begin{equation}
\label{Eq:disc_Syy_white}
\td{S}_{\vd{y}\vd{y}} = {\text{\bf \textPsi}} {\text{\bf \textLambda}} {\text{\bf \textPsi}}^{*} = \td{U} {\text{\bf \textSigma}}^{2} \td{U}^{*}
\end{equation}
and so ${\text{\bf \textSigma}}^{2} = {\text{\bf \textLambda}}$ and $\td{U} = {\text{\bf \textPsi}}$.
More generally, the expansion coefficients can be written in terms of SPOD modes as
\begin{equation}
\label{Eq:disc_Sbb_SPOD}
\td{S}_{{\text{\bf \textbeta}}{\text{\bf \textbeta}}} = {\text{\bf \textSigma}}^{-1} \td{G} {\text{\bf \textLambda}} \td{G}^{*} {\text{\bf \textSigma}}^{-1},
\end{equation}
where $\td{G} = \td{U}^{*}\td{W}_{\vd{y}}{\text{\bf \textPsi}}$ is the discrete equivalent of $\gamma_{ij}$ defined in~(\ref{Eq:resolvent_gamma}).  These equations are analogous to~(\ref{Eq:resolvent_Sbb_SPOD}) and~(\ref{Eq:resolvent_gamma}).  Equivalently, $\td{S}_{{\text{\bf \textbeta}}{\text{\bf \textbeta}}}$ can be conveniently written in terms of the data matrix $\hat{\td{Q}}$ used to define the discrete SPOD eigenvalue problem.  If we define $\td{E} = {\text{\bf \textSigma}}^{-1}\td{U}^{*}\td{W}_{\vd{y}}\hat{\td{Q}}$, then $\td{S}_{{\text{\bf \textbeta}}{\text{\bf \textbeta}}} = \td{E}\td{E}^{*}$.  

If the statistical vector ${\text{\bf \textbeta}}$ is replaced by a deterministic vector $\vd{b}$, then the approximation of the output cross-spectral density matrix is $\td{S}_{\vd{y}\vd{y}} = \left(\td{U} {\text{\bf \textSigma}} \vd{b}\right)\left(\td{U} {\text{\bf \textSigma}} \vd{b}\right)^{*}$,which is clearly a rank-one matrix.  The optimal deterministic expansion coefficients, i.e., the ones that reconstruct the leading SPOD mode, are given by the vector $\vd{b}^{\mathrm{opt}} = {\text{\bf \textSigma}}^{-1} \td{G}_{1} \lambda_{1}^{-1/2}$, where $ \td{G}_{1}$ is the first column of $\td{G}$ and $\lambda_{1}$ is the first entry of the diagonal matrix ${\text{\bf \textLambda}}$.

\bibliographystyle{jfm}
\bibliography{Towne_SPOD_bib}

\end{document}